\documentclass[english,pra, reprint, superscriptaddress, groupedaddress, showkeys, showpacs, nofootinbib, noraggedbottom, balancelastpage,longbibliography]{revtex4-2}
\usepackage[T1]{fontenc}
\usepackage[latin9]{inputenc}
\usepackage{babel}
\usepackage{array}
\usepackage{booktabs}
\usepackage{units}
\usepackage{multirow}
\usepackage{amsmath}
\usepackage{amssymb}
\usepackage{graphicx}
\usepackage{xcolor}
\usepackage{mathrsfs}

\makeatletter

\providecommand{\tabularnewline}{\\}

\definecolor{blueviolet}{rgb}{0.2, 0.2, 0.6}
\definecolor{webgreen}{rgb}{0,.5,0}
\definecolor{webbrown}{rgb}{.6,0,0}
\usepackage[pdftex,          
	bookmarks=false,          
	colorlinks=true, 
	urlcolor=webbrown, 
	linkcolor=webgreen, 
	citecolor=blueviolet,
	pdfstartpage=1,
	pdfstartview={FitH},  
	bookmarksopen=false
	]{hyperref}

\makeatother

\begin{document}

\global\long\def\bra{\langle}%
\global\long\def\ket{\rangle}%
\global\long\def\half{\frac{1}{2}}%
\global\long\def\pr{\prime}%
\global\long\def\kk#1{\left|#1\right\ket }%
\global\long\def\bbra#1{\left\bra #1\right|}%

\global\long\def\a{\alpha}%
\global\long\def\b{\beta}%
\global\long\def\g{\gamma}%
\global\long\def\d{\delta}%
\global\long\def\l{\lambda}%
\global\long\def\m{\mu}%
\global\long\def\n{\nu}%
\global\long\def\o{\omega}%
\global\long\def\s{\sigma}%
\global\long\def\D{\Delta}%
\global\long\def\O{\Omega}%
\global\long\def\P{\Phi}%
\global\long\def\T{\Theta}%

\def\prg#1{\paragraph{{\bf #1}}}%
\global\long\def\dg{\dagger}%
\global\long\def\tr{\text{Tr}}%
\global\long\def\id{1}%
\global\long\def\hc{\text{h.c.}}%
\global\long\def\diff{\text{d}}%

\global\long\def\A{\mathsf{A}}%
\global\long\def\C{\mathsf{C}}%
\global\long\def\DD{\mathsf{D}}%
\global\long\def\F{\mathsf{F}}%
\global\long\def\G{\mathsf{G}}%
\global\long\def\H{\mathsf{H}}%
\global\long\def\II{\mathsf{I}}%
\global\long\def\K{\mathsf{K}}%
\global\long\def\L{\mathsf{L}}%
\global\long\def\OO{\mathsf{O}}%
\global\long\def\PP{\mathsf{P}}%
\global\long\def\R{\mathsf{R}}%
\global\long\def\RP{\mathsf{RP}}%
\global\long\def\S{\mathsf{S}}%
\global\long\def\SO{\mathsf{SO}}%
\global\long\def\SU{\mathsf{SU}}%
\global\long\def\TT{\mathsf{T}}%
\global\long\def\U{\mathsf{U}}%
\global\long\def\W{\mathsf{W}}%
\global\long\def\X{\mathsf{X}}%
\global\long\def\Z{\mathsf{Z}}%

\global\long\def\aa{\hat{\alpha}}%
\global\long\def\bb{\hat{\beta}}%
\global\long\def\gg{\hat{\gamma}}%
\global\long\def\ph{\hat{n}}%
\global\long\def\k{\hat{\phi}}%
\global\long\def\rr{\hat{R}}%
\global\long\def\lh{\hat{L}^{2}}%
\global\long\def\lreg{\overleftarrow{L}}%
\global\long\def\lbold{\overleftarrow{\boldsymbol{L}}}%
\global\long\def\lu{\overleftarrow{X}}%
\global\long\def\ru{\overrightarrow{X}}%

\global\long\def\cs{a}%
\global\long\def\csr{S}%

\global\long\def\vh{\textbf{v}}%
\global\long\def\xh{\textbf{x}}%
\global\long\def\yh{\textbf{y}}%
\global\long\def\zh{\textbf{z}}%
\global\long\def\wh{\textbf{w}}%
\global\long\def\uh{\textbf{u}}%

\global\long\def\lb{\bar{\ell}}%
\global\long\def\pl{P_{\text{leak}}}%
\global\long\def\pk{P_{\text{ok}}}%
\global\long\def\om{\o_{\text{max}}}%
\global\long\def\lmax{\ell_{\text{max}}}%
\global\long\def\st{\frac{L}{2}}%
\global\long\def\SS{L}%
\global\long\def\quadgate{\textsc{quad}}%
\global\long\def\crot    {\textsc{crot}}%
\global\long\def\cphase  {\textsc{cphs}}%

\global\long\def\dd{\hat{D}}%
\global\long\def\sz{\hat{S}_{Z}}%
\global\long\def\sx{\hat{S}_{X}}%
\global\long\def\r {\hat{X}}%
\global\long\def\p {\hat{X}_{P}}%
\global\long\def\y {\hat{Y}}%
\global\long\def\zz{\hat{Z}}%
\global\long\def\xl{\overline{X}}%
\global\long\def\zl{\overline{Z}}%
\global\long\def\z{%
\text{\ooalign{\hidewidth\raisebox{0.2ex}{--}\hidewidth\cr$Z$\cr}}%
}

\global\long\def\one{\mathbf{1}}%
\global\long\def\onep{\mathbf{1}^{\prime}}%
\global\long\def\onepp{\mathbf{1}^{\prime\prime}}%
\def\twoo{\mathbf{2}}%
\def\two#1{\mathbf{2}_{#1}}%
\global\long\def\three{\mathbf{3}}%
\global\long\def\threep{\mathbf{3}^{\prime}}%
\global\long\def\four{\mathbf{4}}%
\global\long\def\five{\mathbf{5}}%

\newcommand{\edt}[1]{{\color{webbrown}{#1}}}%

\title{Robust encoding of a qubit in a molecule}
\author{Victor V. Albert,$^{1,2}$ Jacob P. Covey,$^1$ and John Preskill$^{1,2}$}
\affiliation{Institute for Quantum Information and Matter$^1$ and Walter Burke Institute for Theoretical Physics$^2$\\ California Institute of Technology, Pasadena CA 91125, USA}

\begin{abstract}
We construct quantum error-correcting codes that embed a finite-dimensional code space in the infinite-dimensional Hilbert state space of rotational states of a rigid body. These codes, which protect against both drift in the body's orientation and small changes in its angular momentum, may be well suited for robust storage and coherent processing of quantum information using rotational states of a polyatomic molecule. Extensions of such codes to rigid bodies with a symmetry axis are compatible with rotational states of diatomic molecules, as well as nuclear states of molecules and atoms. We also describe codes associated with general nonabelian compact Lie groups and develop orthogonality relations for coset spaces, laying the groundwork for quantum information processing with exotic configuration spaces.
\end{abstract}
\date{\today}
\maketitle

\section{Introduction}

Quantum systems described by continuous variables arise in many laboratory settings. For example, a microwave resonator in a superconducting circuit or the motional degree of freedom of a trapped ion can be viewed as a harmonic oscillator with an infinite-dimensional Hilbert space. Such continuous-variable systems have potential applications to quantum information processing. However, quantum information encoded in an oscillator can be easily damaged by ubiquitous noise sources such as dissipation and diffusive motion in phase space. 

Robustness against noise can be achieved more easily by encoding a protected finite-dimensional system within the infinite-dimensional Hilbert space of an oscillator. One method for doing so was proposed some years ago by Gottesman, Kitaev, and Preskill (GKP) \cite{Gottesman2001}. A GKP code is a quantum error-correcting code designed to protect against noise that slightly shifts the position or momentum of an oscillator. The ideal basis states for the code space are ``grid states'' supported on periodically spaced points in position or momentum space. By measuring the code's check operators, one can diagnose a shift error that may have occurred, without disturbing the encoded quantum information, and then correct the error (if the shift introduced by noise is not too large) by performing a compensating shift. These codes are expected to perform well against realistic noise, including dissipation, which typically acts locally in phase space \cite{Terhal2016,codecomp,Noh2018}. Construction of GKP grid states has recently been demonstrated experimentally \cite{Fluhmann2018,CampagneIbarcq2019}.

\begin{figure}[t]
\includegraphics[width=1\columnwidth]{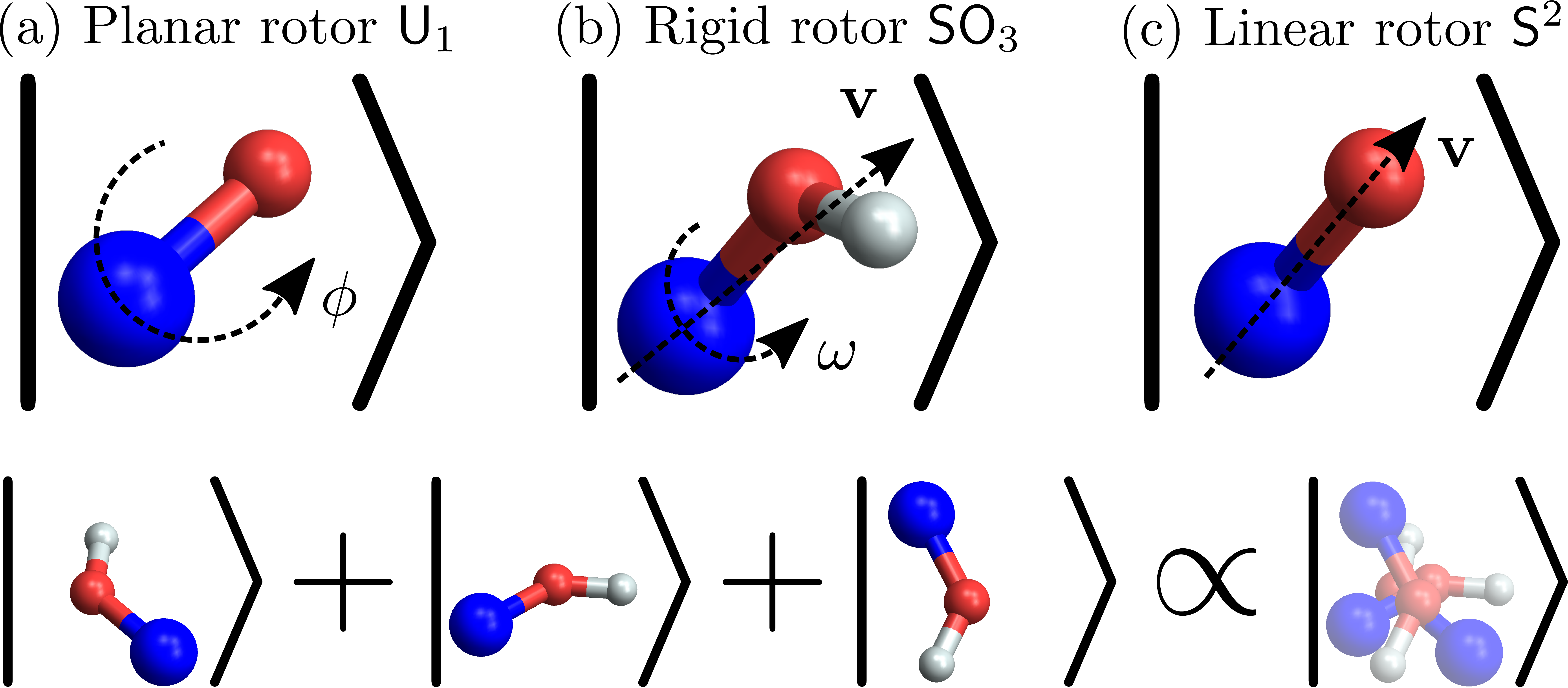}
\caption{
\label{fig:zero}\textsc{Rigid bodies.} A molecular code protects against errors in the orientation and angular momentum of a rigid body, which may be \textbf{(a)}
a planar rotor whose orientation is an element of the two-dimensional rotation group $\U_1$, \textbf{(b)} a rigid rotor whose orientation is an element of the 3-dimensional rotation group $\SO_3$, or \textbf{(c)} a linear rotor whose orientation is a point on the two-sphere $\S^2$. A basis state for the code, or \textit{codeword}, is a superposition of a finite number of orientations. 
}
\end{figure}

In this paper, we develop GKP-like codes that protect against, not noise that shifts the position and momentum of an oscillator, but rather noise that shifts the (continuous) orientation and (discrete) angular momentum of an asymmetric rigid body. GKP codes for 
objects that rotate about a fixed axis [Fig.~\ref{fig:zero}(a)] were already discussed in \cite{Gottesman2001}. In that case, the orientation of the object corresponds to an element of the two-dimensional rotation group $\U_1=\SO_2=\C_{\infty}$. New issues arise for an object that rotates freely in three dimensions [Fig.~\ref{fig:zero}(b-c)], with orientation described as an element of the 3-dimensional rotation group $\SO_3$ (for an object with no symmetries) or a point on the two-sphere $\S^2=\SO_3/\U_1$ (for an object with a symmetry axis). 

Our work is motivated by recent progress in trapping and coherently manipulating individual diatomic and polyatomic molecules \cite{Anderegg2019,Kozyryev2017a,Liu2019,Covey2018,Jamadagni2019}. Since we only consider a molecule's rotational degrees of freedom, for our purposes a molecule is equivalent to a rigid body. For that reason, we refer to quantum codes embedding a protected finite-dimensional subspace in the infinite-dimensional Hilbert space of a rigid body as \textit{molecular codes}.

The rigid rotor Hamiltonian describing molecular rotational motion
is inherently anharmonic; because the energy levels are unevenly spaced, transitions between levels can be individually addressed using microwave fields. 
Hence, proposals for storing quantum
information in molecules \cite{DeMille2002,Lee2005,Yelin2006,Rabl2007,Ortner2011,Kuznetsova2011,Blackmore2018,Ni2018,Hudson2018,Yu2019,Campbell2019,Sawant2019}
(see also \cite{Negretti2011,Cote2014}) typically pick out two low-lying
long-lived energy eigenstates as basis states for a qubit. One can also introduce an external electric field, and encode a qubit using the resulting ``pendular'' eigenstates \cite{Rost1992,Slenczka1994,Karra2016}. Other proposals have advocated using vibrational or spin degrees of freedom \cite{Tesch2002,Baldovi2015}.

Rigid rotor energy eigenstates, if spaced sufficiently far apart in angular momentum, provide protection against small jumps in angular momentum, but are unprotected against dephasing in the angular-momentum eigenstate basis resulting from fluctuations in the rotor's orientation. Our molecular codes, inspired by GKP codes, are designed to protect against both momentum kicks and orientational diffusion of a single molecule. Here we develop the theory of molecular codes and generalizations thereof. Laboratory realizations of these coding schemes that actually improve the coherence times of molecular qubits may still be far off, but we propose laying the foundations for molecular quantum error-correction as a challenging goal for the physicists and chemists of the NISQ era \cite{Preskill2018}.

Though our work is partially motivated by advances in molecular physics, the coding methods we use are best explained in an abstract group-theoretic framework, which we will summarize in the next section. In Sec.~\ref{sec:Realizations}, we enumerate a variety of physical settings, in molecular physics and beyond, where our code constructions may be applicable. The connection with rotational states of the three molecular rotors from Fig.~\ref{fig:zero}(a-c) is developed in greater detail in Secs.~\ref{sec:planar-rotor}-\ref{sec:Linear-rotor-codes}, respectively. Section~\ref{sec:A-qubit-on} discusses extensions to more abstract state spaces. Section \ref{sec:Conclusion} contains conclusions and ideas for future work. 

\section{Summary of our framework}

We describe a family of codes that generalize the GKP codes \cite{Gottesman2001}, which were initially formulated to encode a finite-dimensional system in the infinite-dimensional Hilbert space of a bosonic mode, or of many bosonic modes. Each code in our generalized GKP code family is associated with a nested sequence of groups
\begin{equation}
	\H\subset \K\subset\G\, .
\end{equation}
Here $\G$ is a continuous group of shifts in the position of a physical object. If no nontrivial subgroup of $\G$ leaves the object invariant, and any position can be reached by applying an element of $\G$ to a standard initial position, then we may regard the ``position eigenstates'' $\{|g\ket, g\in \G\}$ as a basis for the Hilbert space of the object.  The generalized GKP code is a subspace of this Hilbert space defined by two properties: (1) The discrete subgroup $\H$ of the continuous group $\G$ leaves any state in the code space invariant, (2) The subgroup $\K$ acts transitively on a basis for the code space.

For the standard GKP code, $\G$ is the abelian noncompact group $\R$, the group of translations in position space of a particle in one spatial dimension. The subgroup $\K$ is the infinite discrete group containing all translations of the particle by an integer multiple of $\a$, where $\a$ is a fixed real number. The subgroup $\H$ contains all translations by an integer multiple of $d\a$, where $d$ is the dimension of the code space. In this case, we may choose the basis for the code space to be (up to normalization)
\begin{align}
	\kk{\overline k} \propto \sum_{h\in \Z}\kk{q= (k+ hd)\a},
\end{align}
where $|q\ket$ is a position state of the oscillator and $k\in\{0,1,\cdots, d{-}1\}$. We refer to each such basis element of the code as a \textit{codeword}. Thus a translation of $q$ by $d\a$ leaves the codewords invariant, and a translation of $q$ by $\a$ permutes the codewords according to $k\to k{+}1$ modulo $d$. A shift in $q$ due to an error can be detected by measuring $q$ modulo $\a$. 

In addition to errors that shift the value of $q$, the GKP code also protects against errors that introduce $q$-dependent phases. Phase errors which are diagonal in the $q$ basis are described by functions on $\R$. Such functions can be Fourier-expanded using irreducible representations (irreps) of $\R$, labeled by the momentum $p$. The irreps that preserve the code space are those with $p$ an integer multiple of $\frac{2\pi}{d\a}$, and those that act trivially on the code space have $p$ an integer multiple of $\frac{2 \pi}{ \a}$.

For a generalized GKP code, the detectable position shifts are labeled by elements of the coset space $\G/\K$, and the ``logical'' position shift errors that preserve the code space are labeled by elements of $\K/\H$. Undetectable logical phase errors correspond to representations of $\G$ which represent the subgroup $\H$ trivially, but represent $\K$ nontrivially.

\begin{figure}[]
\includegraphics[width=0.95\columnwidth]{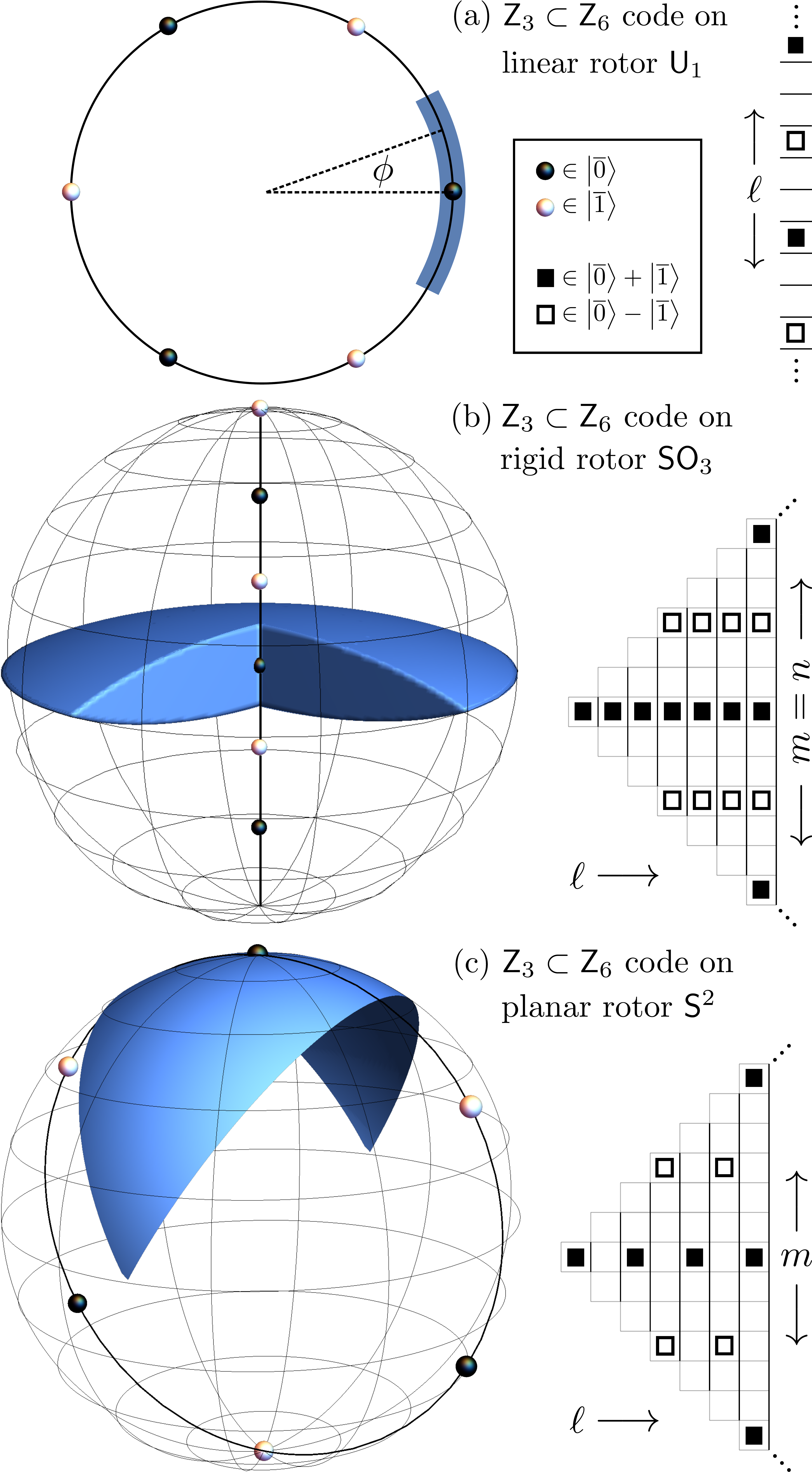}\caption{
\label{fig:Zd-rotors}
\textsc{Codeword constructions.} \textbf{(a)}
Left panel: Sketch of the planar rotor state space $\protect\U_{1}$. Black/white points represent the positions present in the two codewords (\ref{eq:rotor-qubit1}) of the $\protect\Z_{3}\subset\protect\Z_{6}$ GKP rotor code. Correctable shifts $(-\frac{\pi}{6},\frac{\pi}{6}]$
are highlighted in blue. Right panel: $\protect\U_{1}$ angular momentum ladder $\ell\in\protect\Z$. Black/white squares represent momentum states present in the logical-$X$ codewords (\ref{eq:rotor-qubit2}). \textbf{(b)} Sketch of the same features of
the $\protect\Z_{3}\subset\protect\Z_{6}$ molecular code (\ref{eq:Z3}).
Left panel: Position space is drawn as a ball of radius $\pi$ with antipodal
points identified, and each $\protect\SO_{3}$ rotation by angle $\protect\o$
around axis $\protect\vh\in\protect\S^{2}$ corresponds to the vector
$\protect\o\protect\vh$ on the ball. The set of correctable rotations is in blue, but part of it is cut out to show that it contains the origin (meaning that small rotations around any axis are correctable). Right panel: Momentum space is a 3D square pyramid with height labeled by $\ell$ and base by $|m|,|n|\protect\leq\ell$.
We plot only the $m=n$ part, where the 
codewords (\ref{eq:Z3inSO3inMomentumXBasis}) have support. \textbf{(c)}
Sketch of similar features of the $\protect\Z_{3}\subset\protect\Z_{6}$
linear rotor code (\ref{eq:ZN-on-S2-codewords}). Left panel: the blue spherical lune contains all points that are closer to the enclosed black point than to any other black or white point. Right panel: Momentum space is a 2D pyramid with base $|m|\leq\ell$, showing states participating in the logical-$X$ codewords (\ref{eq:S2-logical-X}).
}

\end{figure}

In Sec.~\ref{sec:planar-rotor}, we illustrate the concepts underlying generalized GKP codes by discussing the example of a planar rotor. In this case $\G$ is $\U_1$, the infinite compact group of rotations in a two-dimensional plane, $\K$ is the finite subgroup of $\U_1$ containing rotations by an angle which is an integer multiple of $\frac{2\pi}{d N}$, and $\H$ is the subgroup of $\K$ containing rotations by an angle which is an integer multiple of $\frac{2\pi}{N}$. Here $N,d$ are positive integers, and $d$ is the dimension of the code space. This code can correct a rotation of the planar rotor by any angle less than $\frac{\pi}{dN}$, and can correct a shift in angular momentum by any integer less than $N/2$. The structure of this code, for the case $N=3$ and $d=2$, is depicted in Fig.~\ref{fig:Zd-rotors}(a).

While these planar rotor codes were already introduced in \cite{Gottesman2001}, generalized GKP codes where $\G$ is nonabelian have not been previously discussed to our knowledge. In Sec.~\ref{sec:Rigid-rotor-codes}, we introduce \textit{molecular codes}, which can protect an asymmetric rigid body from rotational shift errors and angular momentum kicks. In this case $\G$ is $\SO_3$, the infinite compact group of proper rotations in 3D space. The finite subgroups $\H\subset \K\subset \SO_3$ can be chosen in various ways. By choosing $\H=\Z_N\subset \K = \Z_{dN}$ to be discrete cyclic groups of rotations about one axis (for chemists, $\Z_N=\C_N$), we obtain codes that can correct small rotations of the body about \textit{any} axis, and can also correct momentum kicks that change the total angular momentum of the body by $\delta \ell < N/2$. For a pictorial representation of this code in the case $N=3$, $d=2$, see Fig.~\ref{fig:Zd-rotors}(b).

We also discuss examples where $\H$ and $\K$ are finite nonabelian subgroups of the rotation group.  Guided by the stabilizer formalism, we show that for each molecular code there is a Hamiltonian which has the code as its ground space. Each ideal codeword is not normalizable, a superposition of an finite number of position eigenstates, but there are normalizable approximate codewords which maintain good error-correcting properties.

We generalize the code construction further in Sec.~\ref{sec:A-qubit-on}, where we allow $\G$ to be any finite group, compact Lie group, or sufficiently well-behaved non-compact group. This formulation provides a unified treatment that encompasses molecular codes ($\G=\SO_{3}$),  CSS codes ($\G=\Z_{D}^{\times n}$) and GKP codes for qudits
($\Z_{D}$), planar rotors ($\U_{1}$ or $\Z$), and oscillators ($\R$).

In Sec.~\ref{sec:Linear-rotor-codes}, we discuss the \textit{linear rotor}, a rigid body with a symmetry axis, such as a heteronuclear diatomic molecule. For this case, the quantum codes we construct are not generalized GKP codes as defined above, because the position basis states of the linear rotor are indexed not by elements of a group, but rather by points in the coset space $\SO_3/\U_1=\S^2$.  Codewords of a linear rotor code are uniform superpositions of antipodal points on $\S^2$, which lie in the same \textit{orbit} of $\H$ acting on $\S^2$, where $\H$ is a finite subgroup of $\SO_3$. See Fig.~\ref{fig:Zd-rotors}(c) for the case $\H=\Z_3$ and codespace dimension two. 

The linear rotor codes can also correct small rotations about any axis, and analyzing correction of momentum kicks follows closely the corresponding discussion for molecular codes. However, for correction of \textit{combinations} of rotations and momentum kicks, there are complications which arise because each $\SO_3$ rotation acting on $\S^2$ has fixed points.

Coset spaces arise in both generalized GKP codes and linear rotor codes, but for different reasons. In GKP codes, position basis states are in one-to-one correspondence with elements of the group $\G$, and the position shifts detected by the code are labeled by elements of $\G/\K$. In linear rotor codes, the position basis states themselves are in one-to-one correspondence with elements of the coset space $\SO_3 / \U_1$. Since coset spaces play a central role in both settings, we formulate position and momentum bases, shift operators, and orthogonality relations for general $\G/\H$ in Appx.~\ref{appx:GH}. These are applicable to $\H$-symmetric molecules when $\G=\SO_3$ (see Sec.~\ref{subsec:other-systems}), and may be of independent interest for general $\G$.

\section{Experimental realizations\label{sec:Realizations}}

Before proceeding to discuss code constructions in more detail, in this section we briefly mention some of the physical settings where these constructions might apply. The rotational states of a molecule provide  one such setting, where the orientations of a molecule correspond to elements of $\SO_3$ (in the case of an asymmetric polyatomic molecule) or $\S^2$ (in the case of a heteronuclear diatomic molecule). In addition, other physical systems, including atomic or molecular hyperfine, vibrational, and electronic states, as well as atomic ensembles and levitated nanoparticles, realize similar configuration spaces.

\subsection{Molecular rotors}\label{subsec:molecular-rotors}

GKP codes were realized experimentally \cite{Fluhmann2018,CampagneIbarcq2019} nearly 20 years after the initial proposal \cite{Gottesman2001}, and full-fledged error correction for molecular qubits may still be many years away [\citealp{Patterson2018}, Sec.~V.D]. Nevertheless, significant steps toward the realization of molecular codes may be feasible during the NISQ era \cite{Preskill2018} as the technology for trapping and controlling molecules \cite{Shuman2010,Anderegg2017,Truppe2017,Collopy2018,Moses2016,Koch2018} continues to advance.

Laser cooling and trapping techniques have recently enabled several
seminal advances for diatomic polar molecules, namely: the creation
of low-entropy arrays in an optical lattice \cite{Moses2015,Reichsollner2017},
trapping and imaging in tweezer arrays \cite{Anderegg2019,Liu2019} and magnetic traps \cite{Fortagh2007,Caldwell2019},
and even the first quantum degenerate gas of polar molecules \cite{DeMarco2019}. Coherence times of $\sim100$ ms to $\sim1$ second in angular momentum states of diatomic polar molecules have already been observed in several experiments
\cite{Neyenhuis2012,Park2017,Seesselberg2018}. 

Laser cooling and quantum control of polyatomic molecules continues
to be a rapidly-progressing field \cite{Kozyryev2017,Kozyryev2017a,Patterson2018,Ivanov2019}.
Further, the possibility of angular momentum state-resolved detection
has recently been considered \cite{Patterson2018,Covey2018}.  
In addition, quantum gates of optically-trapped symmetric top molecules have recently been analyzed~\cite{Yu2019}. Symmetric top molecules also hold promise for simulating quantum magnetism~\cite{Wall2013,Wall2015}. Moreover, specific classes of polyatomic linear polar molecules that feature more than one optically active metal atom have recently been proposed for laser cooling and trapping~\cite{ORourke2019}. Prospects for cooling other complex polyatomic molecules have also been analyzed~\cite{Kozyryev2016}.

Here we highlight a few techniques that could help realize aspects of our codes in real systems.

\prg{Rotational states}
The codewords for our codes can be expressed as coherent superpositions of several different molecular orientations. Alternatively, each codeword can be expressed as a coherent superposition of eigenstates of angular momentum (\textit{a.k.a.} ``rotational states''). When discussing experimental realization of the codes, the basis of rotational states is far more convenient than the position-eigenstate basis, because rotational states can be directly addressed using experimental tools.

For the case of a planar rotor, with configuration space $\U_1$, the rotational basis states $\{|\ell\ket\}$ transform as one-dimensional irreducible representations of $\U_1$; for the case of a polyatomic molecule, with configuration space $\SO_3$, the basis states $\{|_{mn}^{\ell}\ket\}$ correspond to matrix elements of irreducible representations of $\SO_3$; and for the case of a diatomic molecule, with configuration space $\S^2$, the basis states $\{|_{m}^{\ell}\ket\}$ correspond to spherical harmonics. In molecular physics \cite{browncarrington,krebs_book,Demtroder2005,PaulE.S.W}, $\ell$ corresponds to the total angular momentum of the rotor, $m$ is the $\zh$-component of the angular momentum in the lab frame, and $n$ is the $\zh$-component in the rotor frame. 
How codewords are expressed as linear combinations of rotational states is illustrated in the right panels of Fig.~\ref{fig:Zd-rotors}(a-c), respectively, for three simple rotor codes.

\prg{Microwave dressing}
One can try to stabilize the codewords using polychromatic microwave dressing. Note that we consider single molecules and neglect any effects due to their interaction. The inherent rigid-rotor Hamiltonian for $\SO_3$ and $\S^2$ is diagonal in the angular momentum basis, with eigenvalues $\ell(\ell+1)$.\footnote{\label{fn:spherical-top}The actual Hamiltonian depends on the molecule's moments of inertia \cite{browncarrington,krebs_book,Demtroder2005,PaulE.S.W}. We use the ``spherical top'' Hamiltonian to simplify the analysis.} Therefore, transitions between states with different momenta are individually addressable (unlike, e.g., the transitions of a harmonic oscillator Hamiltonian). Selection rules for the internal indices $n,m$ are dictated by the polarization of the
microwave field. Thus, the energy and polarization of a microwave field
can be tuned to couple two angular momentum states that are neighbors in the angular momentum pyramid. That is, the value of $\ell$, $m$, or $n$ can change by 1 unit in a single-photon transition.

However, the angular momentum states making up each codeword are widely spaced in the internal indices.  For example, in the case of the $\SO_{3}$ code depicted in Fig.~\ref{fig:Zd-rotors}(b), the codewords have support only on $\{|_{mn}^{\ell}\ket\}$ states such that $m=n$ is an integer multiple of 3. 
For $\S^{2}$ {[}Fig.~\ref{fig:Zd-rotors}(c){]},
a similar pattern emerges, except that for even $\ell$ the rotational state $|_{m}^{\ell}\ket$ is populated only if $m$ is an even multiple of $3$, while, for odd $\ell$, $\{|_{m}^{\ell}\ket\}$ is populated only if $m$ is an odd multiple of $3$. 

Because a single microwave tone couples states that differ by just 1 unit of $m$ or $\ell$, a sequence of virtual transitions induced by multiple pulses would be needed to couple states with more widely separated values of $m$ or $\ell$. For example, coupling states with $|\delta m|=3$ requires
a 3-photon transition that is sufficiently detuned from the two
intermediate states, and coupling states with $\delta\ell=2$ requires
a two-photon transition sufficiently detuned from the one intermediate
state. We outline a scheme to generate these states in Appendix~\ref{appx:Microwave-dressing}. This scheme requires many pulses, but it is on par with previously proposed molecular
dressing schemes \cite{Gorshkov2013,Manmana2013,Sundar2018}.

We have neglected rotational-state-dependent trapping effects, which are prominent in optical dipole traps~\cite{Gorshkov2011,Neyenhuis2012,Yan2013,Rosenband2018}. These effects will be negligible when considering a single molecule in the motional ground state of the trap, whose intensity can be robustly stabilized. In this case, the unique Stark shift for each rotational state due to the trap simply requires an updated microwave frequency catalog for all transitions. However, this spread in polarizability poses a practical problem when considering many molecules, since one must ensure that they all experience the same optical intensity. Accordingly, alternative trapping schemes may be more appropriate for the applications proposed in this work. Magnetic micro-traps~\cite{Fortagh2007} are compatible with electronic spin doublet or triplet molecules such as CaF, SrF, YbF, or YO. Radio-frequency electric traps are compatible with molecular ions~\cite{Hudson2018,Patterson2018}. Such trapping potentials are substantially less dependent on the rotational state of the molecule since they couple to magnetic dipoles and electric monopoles, respectively.

More generally, one can consider engineering the desired pulses to generate states or correct errors via established optimal-control schemes
\cite{DeVivie-Riedle2007,Koch2018}. It has been shown that one can control the planar \cite{Boscain2012,Magann2018}, linear \cite{Zhdanov2011,Boscain2014,Tehini2019}, and even rigid \cite{Boscain2019} rotors, and it would be useful to extend these and other efforts \cite{Ma2018,Zhang2019} to stabilizing the required code subspace.

\prg{Crystal fields}
In a class of quantum error-correcting codes called \textit{stabilizer codes}, the code space is the simultaneous eigenspace with eigenvalue 1 of a set of commuting Pauli operators, which are called \textit{check operators}. A special subclass of stabilizer codes are the CSS codes, for which each check operator can be chosen to be either \textit{$Z$-type} or \textit{$X$-type}; the $Z$-type operators $\{\sz^{(i)}\}$ are diagonal in the computational basis, and the $X$-type operators $\{\sx^{(j)}\}$ permute the computational basis states. The code subspace may be regarded as the degenerate ground space of the Hamiltonian
\begin{equation}
H_{\text{code}}=-\sum_{i}\sz^{(i)} -\sum_{j}\sx^{(j)}\,.\label{eq:stab}
\end{equation}

Our molecular codes are not stabilizer codes, but as we explain in Sec.~\ref{sec:Rigid-rotor-codes}, the code space is the degenerate ground space of a Hamiltonian which is a sum of $Z$-type and $X$-type terms. Here the $X$-type check operator rotates the molecule, while the $Z$-type check operator is diagonal in the position basis but alters the total angular momentum. Just like its oscillator
counterpart {[}\citealp{Gottesman2001}, Sec.~XIII{]}, this molecular Hamiltonian is gapless, but ground states of an approximate
gapped version would be close to the approximate codewords
we introduce in Sec.~\ref{subsec:Approximate-codewords}.

The $\sz$ check operators are momentum kicks which couple well-separated
angular momentum states $\{|_{mn}^{\ell}\ket\}$ for $\SO_{3}$ or $\{|_{m}^{\ell}\ket\}$
for $\S^{2}$. For example, $\sz$  for
a linear rotor code based on the octahedral group is a superposition
of octopole ($\ell=4$) spherical harmonics; see Eq.~(\ref{eq:S2-stab-T-in-O}). Such harmonics are in principle present in a general interaction with a bath \cite{Lemeshko2017}. However, simple laser, DC, or microwave fields produce only $\ell\leq2$
harmonics {[}\citealp{krebs_book}, Chs.~4 and 7{]}.

One way to generate the required higher value of $\ell$ is to put the molecule into
a crystal lattice. For rotor codes based on a discrete subgroup
$\K\subset\SO_{3}$, $\sz$ is the lowest-$\ell$ function that is
symmetric under $\K$. Thus, putting the rotor into a $\K$-symmetric
lattice yields a background field whose dominant term is exactly $\sz$.
For example, putting a linear rotor into an octahedrally symmetric
lattice yields a background potential \cite{Devonshire1936,Kiljunen2005}
that is exactly the $\sz$ (\ref{eq:S2-stab-T-in-O}) required for
the octahedral code. This potential is minimized at those orientations
of the rotor that are superposed to construct the codewords; in fact, these degenerate minima were noticed earlier in an experimental context \cite{Pandey1986}.
Similarly, embedding into a two-dimensional square lattice yields the appropriate $\sz$
(\ref{eq:ZN-on-S2-Sz}) for a linear-rotor version of the planar rotor
code introduced in Sec.~\ref{sec:planar-rotor}. To access subgroups
of $\SO_{3}$ forbidden in crystals, one could consider embedding
a molecule in a quasicrystal.

Crystal symmetries can enforce only the $\sz$ check operator condition; the $\sx$ check operator condition must be imposed by some other means. 
The $\sx$ operators are trigonometric functions of the angular momentum
operators $\lbold$ for $\SO_{3}$ or $\hat{\boldsymbol{L}}$ for
$\S^{2}$. These are not naturally available, as the rigid
rotor Hamiltonian (\ref{eq:L2-action-on-lmn-states}) and its generalizations$^{\ref{fn:spherical-top}}$
contain terms that are at most bilinear in the angular momentum components.
However, there are other terms in the full rotor-in-lattice Hamiltonian {[}\citealp{Barker1975},
Eq.~(7.2){]}, and, akin to superconducting circuit schemes \cite{Leghtas2014},
one might engineer the molecule's environment (for example, by embedding the molecule in a liquid Helium nanodroplet \cite{Shepperson2017}) to arrange for adiabatic elimination to provide the required $\sx$ terms.

\prg{Nuclear spin coupling}

If an error causes the molecule to rotate slightly, we recover from the error by applying a compensating small rotation.  The desired rotation can be executed by turning on a Hamiltonian which is linear in the angular momentum. But since the natural rigid rotor Hamiltonian is quadratic, this linear term is not so easily realized in the laboratory.

One way to provide a Hamiltonian term which is linear in the molecule's angular momentum is to couple the rotational states of the molecule to nuclear spin states via nuclear spin-rotation interactions [\citealp{browncarrington}, Eq.~(1.32)]\cite{Park2017,Ospelkaus2010}
\begin{equation}\label{eq:quadrupole}
H_\text{nsr}=\boldsymbol{I} \cdot \lbold\,,  
\end{equation}
where $\boldsymbol{I}$ is the nuclear spin. The nuclear spin can serve as a convenient ancilla system, and the orientation of the molecule can be controlled by manipulating the nuclear spin.
Similar approaches have been applied to solid-state systems in which electronic spins are coupled to nuclear spins~\cite{Unden2016}. 
This is roughly analogous to using a superconducting Josephson-junction device coupled to a bosonic mode for manipulating the states of a bosonic error-correcting code. 

We also need to correct momentum kicks by applying unitary operations that change the value of $\ell$. Operations which shift the occupation number of a cavity can be applied by coupling the cavity to a 3-level atom \cite{Oi2013} or by using linear optics \cite{Radtke2017}. Similar schemes could shift the value of $\ell$ for a $\U_1$ rotor. Extensions of such schemes may be helpful for controlling the rotational states of higher-dimensional rotors.

\begin{table}
\begin{tabular}{ccc}
\toprule 
~Space $\X$~ & Group $\H$ & Quotient space $\X/\H$\tabularnewline
\midrule
$\R$ & $\Z$ & Wigner-Seitz unit cell $\U_{1}$\tabularnewline
\midrule
\multirow{7}{*}{$\SO_{3}$} & $\Z_{N}$ & lens space $\L_{2N,1}$\tabularnewline
 & dihedral $\DD_{N}$ & prism space\tabularnewline
 & tetrahedral $\TT$ & octahedral space\tabularnewline
 & octahedral $\OO$ & truncated cube space\tabularnewline
 & ~~~icosahedral $\II$~~~ & ~Poincar\'e dodecahedral space~\tabularnewline
 & $\U_{1}$ & two-sphere $\S^{2}$\tabularnewline
 & $\OO_{2}$ & projective plane $\RP^{2}$\tabularnewline
\midrule 
$\S^{2}$ & $\Z_{N},\DD_{N},\TT,\OO,\II$ & spherical two-orbifold\tabularnewline
\bottomrule
\end{tabular}

\caption{\label{t:quaotient_spaces} 
Quotient spaces mentioned in this work
\cite{postnikov,orbifold} (see also {[}\citealp{jose_book}, Sec.~3.8{]}\cite{Wulker2019,thurston_book,weeks_book}). Spaces associated with $\SO_3$ characterize  rotational states of various molecules (see Sec.~\ref{subsec:other-systems}).
$\protect\Z_N=\C_N$ is the order-$N$ cyclic group, $\protect\DD_{N}$ is the order-$2N$ dihedral group, $\protect\U_{1}=\protect\SO_{2}=\C_{\infty}$
is the circle group, and $\protect\OO_{2}=\protect\SO_{2}\rtimes\protect\Z_{2}=\DD_{\infty}$
is the group of planar rotations and reflections. Some of these spaces are shown in Figs.~\ref{fig:Zd-rotors} and \ref{fig:TinSO3}.
}
\end{table}

\subsection{Spin systems}
\label{subsec:atomic-ensembles}

Certain combinations of spins offer another platform for simulating the linear rotor space
$\S^{2}$ and quotient spaces $\SO_{3}/\H$ from Table~\ref{t:quaotient_spaces}.
We list three manifestations: $L$ spin-$1/2$ systems in a totally symmetric spin state,
$L$ spin-$N/2$ systems in a totally symmetric state, and a pair of
spin-$L/2$ systems. In the limit of large $L$, each of these systems provides a useful approximation to one of the spaces of interest. While the first two cases are usually studied in the context of atomic ensembles, the third case can easily arise in the nuclear spin manifold of an atom or a molecule.

\prg{Many small spins}
$L$ spin-1/2 particles in a totally symmetric spin state have a total angular momentum of $L/2$. The $L\to \infty$ limit of this large collective spin is sometimes said to be a semiclassical limit, meaning that the spin-$L/2$ object behaves like a continuous classical spin when $L$ is large.
An intuitive way to understand
this limit is to consider the spin-coherent states
\begin{equation}
|\vh\ket_{\text{SC}}=\left(e^{-i\varphi/2}\cos{\textstyle \frac{\vartheta}{2}}\kk{_{1/2}^{1/2}}+e^{i\varphi/2}\sin{\textstyle \frac{\vartheta}{2}}\kk{_{-1/2}^{1/2}}\right)^{\otimes \SS}
\end{equation}
for $\vh=(\vartheta,\varphi)\in\S^{2}$ \cite{Arecchi1972,perelomov_book}.
These states are not orthogonal; instead they form an overcomplete frame for the collective spin's $(\SS+1)$-dimensional Hilbert space, with overlap $|\bra\vh|\vh^{\pr}\ket_{\text{SC}}|^{2}=(\frac{1+\vh\cdot\vh^{\pr}}{2})^{\SS}$.
As $\SS\to\infty$, the states become orthogonal
and correspond to the position states $|\vh\ket$ of $\S^{2}$ (Table~\ref{t:GoverH}.B). For finite $L$, superpositions of these spin-coherent states can be approximate codewords  for a linear rotor code. 

Numerous manifestations of entangled ensembles of many spin-1/2 atoms have recently been demonstrated~\cite{Haas2014,Strobel2014,Lucke2014,McConnell2015}, and the current status of the field is summarized in Ref.~\cite{Pezze2018}.

\prg{Many medium spins}

Any pure state of a spin-$1/2$ system is invariant under a continuous $\U_1$ subgroup of the rotation group $\SO_3$; each pure state corresponds to a point on the Bloch sphere, and a rotation about the axis aligned with that Bloch vector leaves the state invariant. In contrast, there are pure states in higher-spin representations for which the subgroup which preserves the state is a nontrivial \textit{discrete} subgroup $\H$ of $\SO_3$. 
For example, the spin-$2$ state $|\TT\ket\propto|_{-2}^{2}\ket+\sqrt{2}|_{1}^{2}\ket$
is invariant under the tetrahedral subgroup $\TT$. Therefore, applying $\SO_3$ rotations to $|\TT\ket$ generates a manifold of states $\{\,|\cs\ket_{\TT}\}$, where the label $a$ is a point in the coset space $\cs\in\SO_{3}/\TT$.  
The spin-coherent states $\{\,|\cs\ket_{\TT}^{\otimes L}\}$, obtained by taking a tensor product of many identical elements of this manifold, approximate the position-basis states of $\SO_3/\TT$ in the limit of large $L$. This idea can be generalized: spin-coherent states $\{\,|\cs\ket_{\H}^{\otimes L}\}$ approximate the position-basis states for the coset space $\SO_3/\H$, if $|\cs\ket_{\H}$ is a higher-spin state with invariance group $\H$.

The above $\TT$-symmetric and similar $\H$-symmetric
states {[}\citealp{Kobayashi2012}, Table~2{]} --- examples of Perelomov coherent states \cite{perelomov_book} --- have been used
as a mean-field ansatz for the ground space of spin-$N$
Bose-Einstein condensates \cite{Stamper-Kurn2013,Barnett2006}. We will use such coherent states to extract error syndrome information for molecular codes (see Sec.~\ref{subsec:msmnts}).

\prg{Two large spins}

Instead of using only the symmetric subspace, one can consider the
entire space of a pair of spin-$\SS/2$ systems. Per the addition rules {[}\citealp{VMH},
Ch.~8{]},
\begin{equation}
{\textstyle \SS/2\otimes \SS/2}=0\oplus1\oplus\cdots\oplus \SS\,,\label{eq:twolargespins}
\end{equation}
the $(L+1)^2$ orthonormal basis states for this system can be chosen to be the angular-momentum eigenstates $\{|_{m}^{\ell}\ket\}$, with $\ell \le L$ and $|m|\le \ell$. These are precisely the rotational states of a linear rotor, except for the truncation $\ell \le L$. Formally, then, the state space of a pair of spin-$L/2$ systems matches the state space on $\S^2$ in the limit $L\to \infty$.
Since the normalizable approximate codewords of the linear rotor code are necessarily truncated for large $L$ anyway, these approximate codewords can be accurately realized using a pair of spin-$L/2$ systems for sufficiently large $L$ (see Sec.~\ref{subsec:Approximate-codewords}).

If one instead considers two different spins $L/2$ and $L^\pr/2$, one obtains a different band of $\S^2$ momentum states. While developing codes for such band-limited subspaces is outside the scope of this work, it is possible that our coding strategies may also be useful there.

As a concrete experimental platform for large-spin systems, we can consider nuclear spin spaces of molecules or single atoms. Diatomic molecules such as NaCs \cite{Ni2018} offer exactly the band-limited subspaces mentioned above. Concerning single atoms, Lanthanide species such as dysprosium (Dy), holmium (Ho), and erbium (Er) have large total spin manifolds in the their ground states due to their large nuclear spins and many unpaired electrons in their f-shells. Accordingly, such atoms have already attracted attention for the possibility of scaling up quantum computing by collectively encoding in multilevel atoms~\cite{Brion2007,Brion2008,Saffman2008}. Ho in particular has the largest hyperfine ground space of any atom, with 128 ground states~\cite{Saffman2008}. Laser cooling and trapping techniques are well established for Dy~\cite{Youn2010}, Ho~\cite{Miao2014}, and Er~\cite{McClelland2006}, as well as other lanthanides. Moreover, quantum degenerate gases of Dy~\cite{Lu2011,Lu2012} and Er~\cite{Aikawa2012,Aikawa2014} are widely used for novel quantum simulations based on their large magnetic dipole moments.

\subsection{Other systems}
\label{subsec:other-systems}

\prg{Planar rotors}
Several systems have the configuration space of the planar rotor. The system depicted in Fig.~\ref{fig:zero}(a) is a diatomic molecule confined to rotate in a two-dimensional plane, but one can also consider a two-ion crystal \cite{Urban2019}. Other possibilities include the phase difference between two superconductors on either side of a Josephson junction \cite{girvinbook} and orbital angular momentum of light \cite{Nicolas2014}.

One can also embed the first few angular momentum states of the planar rotor in the linear and rigid rotors. For fixed angular momentum $L$, the linear rotor subspace $\{|^{L}_m\ket\}$ with $|m|\leq L$ is equivalent to the band-limited subspace $\{|\ell\ket,|\ell|\leq L\}$ of the planar rotor.

\prg{Symmetric molecules}

A molecule with symmetry
group $\H$ has an orientation state space parameterized by $\SO_{3}/\H$
(see Table~\ref{t:quaotient_spaces}). For example, the methane molecule CH$_3$ has the tetrahedral symmetry group $\TT$, and the alkaline earth monomethoxide (MOCH$_3$) family --- potentially useful for quantum computing \cite{Yu2019} --- has symmetry group $\Z_3$. This is also the relevant symmetry group of Posner molecules, postulated to have potentially useful quantum effects \cite{Fisher2015,YungerHalpern2019}.
The symmetry group of the fullerene molecule is the icosahedral group $\II$, and the 3-manifold $\SO_{3}/\II$ has an exotic shape that was once proposed as a model for the geometry of the universe \cite{Luminet2003,weeks_book}. It is interesting that such exotic spaces are readily accessible in relatively simple molecules. Completely asymmetric and $\U_1$-symmetric molecules correspond, respectively, to rigid and linear rotors from Sec.~\ref{subsec:molecular-rotors}.

More generally, if a group $\G$ acts transitively on the states of a quantum system, and the subgroup $\H$ of $\G$ leaves the states invariant, then the configuration space of the system is $\G/\H$. In Appendix \ref{appx:GH}, we develop mathematical tools for parameterizing the position eigenstates and the dual momentum states of such a system, including orthogonality/completeness
relations, and a Poisson summation formula \cite{Justel2018}.

\prg{Electronic states}
One can consider embedding certain spaces from Table~\ref{t:quaotient_spaces} in the electronic eigenstates of single atoms. The eigenstates of hydrogen offer a platform for a band-limited subspace of the linear rotor
$\S^{2}$, and even the space $\SU_{2}$ (closely related to the rigid rotor $\SO_3$; see Appx.~\ref{appx:voronoi}). Let us label the atom's eigenstates by $\kk{\nu,~^\ell_m}$, where $0\leq|m|\leq\ell < \nu$ and the energy $E_{\nu,\ell,m}\propto 1/\nu^2$. For fixed energy $\nu=L$, the manifold of states is the same subspace of $\S^2$ as that obtained by combining two large spins in Eq.~(\ref{eq:twolargespins}). If we instead consider all values of $\nu,\ell,m$, we obtain $\SU_2$ by an appropriate unitary transformation, related to writing the hydrogen atom in parabolic coordinates \cite{BANDER1966}.

\prg{Vibrational states}
One can also consider using vibrational states of atoms or molecules to encode quantum information \cite{Tesch2002}. As control over vibrational states improves, it may be possible to implement bosonic error-correcting codes \cite{codecomp}. Position-state subspaces of harmonic oscillators also yield the two rotational spaces of interest. For example, considering position states $|x,y,z\ket$ of three oscillators with $x^2+y^2+z^2$ constant yields $\S^2$. With four oscillators, one obtains $\SU_2$. To simulate $\S^2$ using momentum states, one can take all Fock states of two oscillators with even total occupation number.

\prg{Levitated nanoparticles}
The codewords of our $\SO_3$ and $\S^2$ codes are coherent superpositions of different possible orientations for a rigid body. Though we have emphasized the potential applications to atoms and  molecules, the same ideas can be applied to any quantized 3-dimensional rigid body that can be coherently manipulated. 
While there is a size limitation due to decoherence,
we are on the cusp of entering the quantum regime for levitated nanoscale
particles of helium \cite{Childress2017}, vaterite \cite{Arita2013},
diamond (alone \cite{Hoang2016} or doped \cite{Neukirch2015}), and
silicon \cite{Reimann2018,Ahn2018,Delic2019}, to name a few. Nanoparticles
may seem to be unlikely candidates for quantum computing, but it would be interesting nonetheless to try to stabilize quantum superpositions of their orientational states (cf.~\cite{Stickler2018b}).

\section{Error correction basics \\ ~~~~~~~ for the planar rotor\label{sec:planar-rotor}}

The goal of error correction is to encode quantum information into a cleverly-chosen subspace (the \textit{code}) such that it is possible to recover said information from errors caused by physical noise. Before proceeding to discuss codes which protect against noise acting on a 3-dimensional rigid body, we will review a simpler case which was previously considered in \cite{Gottesman2001}: encoding a finite-dimensional system in the infinite-dimensional Hilbert space of a \textit{planar rotor}. By discussing this case we can introduce the key concepts underlying our code constructions in a familiar mathematical setting. The interested reader can consult \cite{gottesman_thesis}[\citealp{preskillnotes}, Ch.~7] for other introductory material on quantum error correction.

The position-basis eigenstates for a planar rotor are in one-to-one correspondence with the elements of the two-dimensional rotation group $\U_1=\SO_2=\C_{\infty}$. Equivalently, these are the position eigenstates for a particle moving on a circle; the basis elements may be denoted $\{|\phi\ket, \phi\in[0,2\pi)=\U_1\}$, with continuum normalization $\bra \phi|\phi^\pr\ket = \delta(\phi-\phi^\pr)$. A dual basis is provided by the angular-momentum eigenstates (\textit{a.k.a.} ``rotational states'') $\{|\ell\ket,\ell\in \Z\}$, where $\bra\phi|\ell\ket =\frac{1}{\sqrt{2\pi}} e^{i\ell \phi}$ and hence $\bra \ell | \ell^\pr \ket = \delta_{\ell \ell^\pr}$. 

Noise might rotate the system, applying an operator
\begin{equation}
   \r_{\phi^{\pr}}=e^{-i\phi^{\pr}\hat{L}}=\int_{\U_1} \diff\phi~|\phi+\phi^{\pr}\ket\bra\phi|\,;\label{eq:U1:position-shift}
\end{equation}
alternatively, noise might kick the angular momentum, applying some power of the kick operator
\begin{equation}
	\zz=e^{i\hat{\phi}}=\sum_{\ell\in\Z}|\ell+1\ket\bra\ell|\,.\label{eq:U1:momentum-shift}
\end{equation}
In fact, we can expand an arbitrary noise channel $\mathcal{E}$ acting on the density operator $\rho$ of the planar rotor in terms of a complete basis of operators, where each element of the basis is a product of an $\r_\phi$ operator and an $\ell$th power of the $\zz$ operator:
\begin{align}
   \mathcal{E}(\rho)=\int_{\U_{1}^{\times2}}\diff\phi \diff\phi^{\pr}\sum_{\ell,\ell^{\pr}\in\Z}\mathcal{E}_{\phi\phi^{\pr}}^{\ell\ell^{\pr}}\r_{\phi}\zz^{\ell}\,\rho\,\zz^{\ell^{\pr}\dagger}\r_{\phi^{\pr}}^{\dagger}.\label{eq:U1-channel}
\end{align}
Above, the expansion coefficients $\mathcal{E}_{\phi\phi^{\pr}}^{\ell\ell^{\pr}}$ are such that $\cal{E}$ is a channel. Our goal is to encode a finite-dimensional logical system in the infinite-dimensional Hilbert space of the rotor, where this logical system is protected against any error $\r_\phi \zz^\ell$ where both $\phi$ and $\ell$ are sufficiently small. In other words, if $\rho$ consists of states in the logical (\textit{a.k.a.} code) subspace, and if $\cal{E}$ is expanded using only such \textit{correctable} $\r_\phi \zz^\ell$, then we will be able recover the original $\rho$ from $\cal{E}(\rho)$. Otherwise, recovery may not be possible, and logical information stored in $\rho$ may become corrupted.

\subsection{A protected qubit}
\label{subsec:U1-protected-qubit}

For example, the two orthonormal basis states of a protected qubit can be chosen to be [see Fig.~\ref{fig:Zd-rotors}(a)] 
\begin{subequations}
\label{eq:rotor-qubit1} 
\begin{align}
\kk{\overline{0}}&={\textstyle \frac{1}{\sqrt{3}}}\left(\kk{\phi=0}+\kk{{\textstyle \phi=\frac{2\pi}{3}}}+\kk{{\textstyle \phi=\frac{4\pi}{3}}}\right),\\\kk{\overline{1}}&={\textstyle \frac{1}{\sqrt{3}}}\left(\kk{{\textstyle \phi=\frac{\pi}{3}}}+\kk{{\textstyle \phi=\pi}}+\kk{{\textstyle \phi=\frac{5\pi}{3}}}\right).
\end{align}
\end{subequations}
Both basis states are eigenstates with eigenvalue 0 of $\hat \phi$ modulo 
$\frac{\pi}{3}$. Suppose that $|\overline \psi\ket$ is an arbitrary state in the code space spanned by $|\overline 0\ket$ and $|\overline 1\ket$. If an error occurs which causes $\phi$ to shift by $\delta \phi\in( -\frac{\pi}{6},\frac{\pi}{6} )$, we can unambiguously diagnose the error by measuring $\hat \phi$ modulo $\frac{\pi}{3}$. Once $\delta \phi$ is known, we can correct the error by applying a unitary transformation that shifts $\phi$ by $-\delta \phi$, restoring the state of the rotor to the initial undamaged state $|\overline \psi\ket$.

Alternatively, we may expand the basis states of the code in the angular-momentum eigenstate basis, finding
\begin{subequations}
\label{eq:rotor-qubit2}
 \begin{align}
{\textstyle \frac{1}{\sqrt{2}}}\left(\kk{\overline{0}}+\kk{\overline{1}}\right)&={\textstyle \sqrt{\frac{3}{\pi}}}\sum_{s\in\Z}\kk{\ell=6s},\\{\textstyle \frac{1}{\sqrt{2}}}\left(\kk{\overline{0}}-\kk{\overline{1}}\right)&={\textstyle \sqrt{\frac{3}{\pi}}}\sum_{s\in\Z}\kk{\ell=6s+3}\,.
\end{align}
\end{subequations}
Both basis states are eigenstates with eigenvalue 0 of $\hat L$ modulo 3. Suppose an error occurs which causes the angular momentum to shift by $\delta \ell \in\{ -1,0,1\}$. We can unambiguously diagnose the error by measuring $\hat L$ modulo 3. Once $\delta\ell$ is known, we can correct the error by applying a unitary transformation that shifts $\ell$ by $-\delta \ell$. Furthermore (see below), $\hat \phi$ modulo $\frac{\pi}{3}$ and $\hat L$ modulo 3 are compatible observables that can be measured simultaneously. Therefore, we can correct any combination of shifts in $\phi$ and $\ell$, as long as the shift in $\phi$ is no larger than $\frac{\pi}{6}$ and the shift in $\ell$ is no larger than $1$.

The code basis states in Eqs.~(\ref{eq:rotor-qubit1}-\ref{eq:rotor-qubit2}) are not normalizable and therefore unphysical. However, we may replace the position eigenstates in Eq.~(\ref{eq:rotor-qubit1}) by narrow wave packets; then the sum over $s$ in Eq.~(\ref{eq:rotor-qubit2}) is modulated by a broad envelope function. In that case, the code states are physical, and the nice error-correction properties we noted still hold, up to negligibly small corrections. 

Our main task in this paper will be to generalize this code construction, in various directions. For that purpose, it will be convenient to have other ways to describe the code. Our first alternative description uses the stabilizer language \cite{gottesman_thesis,grasslbook}. 

\prg{Stabilizer formalism}

A \textit{stabilizer code} may be characterized as the simultaneous eigenspace with eigenvalue 1 of a set of commuting unitary operators, called the \textit{stabilizer generators}. For the code specified by Eqs.~(\ref{eq:rotor-qubit1}-\ref{eq:rotor-qubit2}), we may choose these operators to be
\begin{equation}
	\sz\equiv\zz^{6}=e^{i6\hat{\phi}},\quad\sx\equiv\r_{\frac{2\pi}{3}}=e^{-i\frac{2\pi}{3}\hat{L}}\,.
\end{equation}
To check that these operators commute, recall the relation $e^{i\hat{\phi}}\hat{L}e^{-i\hat{\phi}}=\hat{L}-1$, and the identity $\r e^{\a\hat{L}}\r^{\dg}=e^{\a\r\hat{L}\r^{\dg}}$ for any unitary $\r$ and scalar $\a$. $\sz$ and $\sx$ are the code's \textit{check operators}, which we can measure to diagnose errors. Note that measuring $\sz$ is equivalent to measuring $\hat \phi$ modulo $\frac{\pi}{3}$ and that measuring $\sx$ is equivalent to measuring $\hat L$ modulo 3, as we asserted earlier, and that we can perform these measurements simultaneously because $\sz$ and $\sx$ commute.

Furthermore, we note that the operators
\begin{equation}
   \zl\equiv\zz^{3}=e^{i3\hat{\phi}},\quad\xl=\r_{\frac{\pi}{3}}=e^{-i\frac{\pi}{3}\hat{L}}\,,
\end{equation}
also commute with the stabilizer generators, which means that these are \textit{logical operators} which preserve the code space. We see also that $\zl$ and $\xl$ anticommute, and that they square to the identity on the code space, where $\sz=\sx=1$. Thus $\zl$ and $\xl$ may be regarded as the logical Pauli operators acting on the encoded qubit, where $\zl$ is diagonal in the basis $\{|\overline 0\ket,|\overline 1\ket\}$ and $\xl$ is diagonal in the conjugate basis $\{\frac{1}{\sqrt{2}}\left(|\overline 0\ket\pm |\overline 1\ket\right)\}.$

\prg{CSS construction}
We may also describe our protected qubit using the language of \textit{Calderbank-Shor-Steane (CSS) codes} \cite{gottesman_thesis,grasslbook}. In the CSS construction, a quantum error-correcting code is built from a classical error-correcting code $\K$ and a subcode $\H\subset \K$. 

In the case of the protected qubit with basis states (\ref{eq:rotor-qubit1}), the code $\K$ is a 6-state system embedded in the infinite-dimensional Hilbert space of the rotor, with the 6 states corresponding to 6 equally spaced angular positions of the rotor, rotated by $\phi = \frac{2\pi}{6}k, k \in \{0,1,\cdots, 5\}$, relative to a standard reference orientation. This classical system is protected against errors that shift the rotor slightly, rotating it through an angle $\delta \phi\in (-\frac{\pi}{6},\frac{\pi}{6})$. 
The subcode $\H$ has three states, with orientations $\phi = \frac{2\pi}{3}k, k \in \{0,1,2\}$, and protects against rotations which are twice as large: $\delta \phi\in(-\frac{\pi}{3},\frac{\pi}{3})$. In the associated quantum code, each of the basis states~(\ref{eq:rotor-qubit1}) is a uniform superposition of all the elements of a \textit{coset} of $\H$ in $\K$, the trivial coset (the elements of $\H$) for the basis state $|\overline 0\ket$, and the nontrivial coset for the basis state $|\overline 1\ket$. The protection of this qubit against shifts of the rotor is inherited from the corresponding property of the classical code $\K$.

There is a dual description of this quantum code, making use of the angular momentum basis of the rotor rather than its position basis. The classical code $\H^\perp$, dual to $\H$, contains all angular momentum eigenstates where $\ell$ is an integer multiple of 3. These two classical codes are dual in the sense that the representations of the group $\U_{1}$ contained in $\H^\perp$ represent the elements of $\H$ trivially. Similarly, the classical code $\K^\perp$ dual to $\K$ contains all angular momentum eigenstates where $\ell$ is an integer multiple of 6, those representations which represent $\K$ trivially. Evidently, $\K^\perp$ is a subcode of $\H^\perp$. For the quantum code, each basis state in Eq.~(\ref{eq:rotor-qubit2}) is a uniform superposition of all the elements of a coset of $\K^\perp$ in $\H^\perp$, the trivial coset for the basis state $\frac{1}{\sqrt{2}}\left(|\overline 0\ket + |\overline1\ket\right)$, and the nontrivial coset for the basis state $\frac{1}{\sqrt{2}}\left(|\overline 0\ket - |\overline1\ket\right)$. The classical code $\H^\perp$ protects against shifts of the angular momentum by $\delta \ell\in\{-1,0, 1\}$, and the quantum code inherits this property. 

Viewed as an abstract group, the code $\K$ is the subgroup $\Z_6$ of $\U_1$, and $\H$ is the subgroup $\Z_3\subset \Z_6$. The construction can be easily generalized to $\K = \Z_{dN}$ and $\H=\Z_N$, where $d$ and $N$ are positive integers, in which case the quantum code is $d$-dimensional. In the stabilizer language, this more general code has stabilizer generators
\begin{equation}
   \sz=\zz^{dN},\quad\sx=\r_{\frac{2\pi}{N}}\,.
\end{equation}
Its logical operators,
\begin{equation}
   \zl=\zz^{N},\quad\xl=\r_{\frac{2\pi}{dN}}\,,\label{eq:U1-logicals}
\end{equation}
are generalized Pauli operators, obeying the Heisenberg-Weyl commutation relation $\zl\xl=e^{i\frac{2\pi}{d}}\xl\zl$. This quantum code protects against position shifts by $\delta\phi$ with $|\delta\phi|<\frac{\pi}{dN}$ and momentum kicks by $\delta\ell$ with $|\delta\ell| \leq (N-1)/2$ (for odd $N$). Note the tradeoff: increasing $N$ improves the protection against angular momentum kicks, but weakens the protection against rotations. 

\prg{Partial Fourier transform}
There is yet another way to describe the code construction, using the notion of a \textit{partial Fourier transform}, which will be helpful as we seek further  generalizations. Recall that the position and angular-momentum bases for the planar rotor are related by Fourier transforming:
\begin{subequations}
\begin{align}
   |\ell\ket&=\int_{-\pi}^{\pi}\diff\phi|\phi\ket\bra\phi|\ell\ket={\textstyle \frac{1}{\sqrt{2\pi}}}\int_{-\pi}^{\pi}\diff\phi|\phi\ket e^{i\ell\phi},\\|\phi\ket&=\sum_{\ell\in\Z}|\ell\ket\bra\ell|\phi\ket={\textstyle \frac{1}{\sqrt{2\pi}}}\sum_{\ell\in\Z}|\ell\ket e^{-i\ell\phi}.
\end{align}
\end{subequations}
It is  useful to imagine that the above integral over $\phi$ is carried out in two steps. We write $\phi= a + \frac{2\pi}{N} h$, where $a\in (-\frac{\pi}{N},\frac{\pi}{N}]$ and $h\in \{0,1, \cdots , N{-}1\}$; then integrating $\phi$ from $-\pi$ to $\pi$ is equivalent to integrating $a$ from $-\frac{\pi}{N}$ to $\frac{\pi}{N}$, and summing $h$ from 0 to $N{-}1$. Likewise, we can do the sum over $\ell$ in two steps; we write $\ell = Ns + \l$, where $s\in \Z$ and $\l \in \{0,1, \cdots, N{-}1\}$, and we separate the infinite sum over $s$ from the finite sum over $\l$. When we speak of a ``partial Fourier transform'', we mean performing one of these two steps without the other. 

By performing the sum over $h$ but not the integral over $a$, we obtain a new orthonormal basis
\begin{align}
\kk{\cs;\l}&\equiv{\textstyle \frac{1}{\sqrt{N}}}\sum_{h\in\Z_{N}}e^{i\frac{2\pi}{N}\l h}\kk{\phi=\cs+{\textstyle \frac{2\pi}{N}h}}\notag\\&=e^{-i\l\cs}{\textstyle \sqrt{\frac{N}{2\pi}}}\sum_{s\in\Z}e^{-iNs\cs}\kk{\ell=Ns+\l}\,,
\end{align}
with normalization
$
  \bra a;\l|a';\l'\ket = \delta(a-a')\delta_{\l\l'}
$.
From now on, the presence of a semicolon inside a ket declares that ket to be an element of this basis.

This $\{|a;\l\ket\}$ basis is convenient for our purposes because shifts in position or angular momentum affect only one of the two indices. A shift in angular momentum by $\delta\ell$
acts on the basis according to
\begin{equation}
	|\cs;\l\ket\to\kk{\cs;\left(\l+\delta\ell\right)\text{ mod }N}\,
\end{equation}(up to a phase), shifting $\l\to\l+\delta\ell$ modulo $N$. A shift in position by ${\delta\phi}$
shifts $\cs\to\cs+{\delta\phi}$ modulo $\frac{2\pi}{N}$,
\begin{equation}
    |\cs;\l\ket\to\kk{\textstyle{\left(\cs+{\delta\phi}\right)\text{ mod }\frac{2\pi}{N};\l}}\,.
\end{equation}

To recover our previous code construction, we choose $d$ basis states $\{|\overline k \ket\}$ with $\l=0$ and $a = \frac{2\pi}{dN} k$, finding
\begin{align}
\kk{\overline{k}}=\kk{{\textstyle \frac{2\pi}{dN}k;0}}\notag&={\textstyle \frac{1}{\sqrt{N}}}\sum_{h\in\Z_{N}}\kk{{\textstyle \phi=\frac{2\pi}{dN}k+\frac{2\pi}{N}h}}\notag\\&={\textstyle \sqrt{\frac{N}{2\pi}}}\sum_{s\in\Z}e^{-i\frac{2\pi}{d}sk}|\ell=Ns\ket\,.\label{eq:zakforZ3}
\end{align}
If an error occurs in which $|\delta \ell| \le (N-1)/2$ (for odd $N$) and $|\delta \phi| < \frac{\pi}{dN}$, we diagnose the error by performing a measurement which determines the value of $\l$ and also the value of $a$ (mod $\frac{2\pi}{dN}$). 
Then the value of $a$ unambiguously identifies the shift in $\phi$ and the value of $\l$ unambiguously identifies the shift in $\ell $. Once known, these shifts can be corrected to recover the initial undamaged code states. 

The orientation label $\phi$ of the planar rotor can be viewed as the element of the group $\U_{1}$ describing the rotation which reaches $\phi$ starting from a standard initial orientation.  The basis $\{|a;\l\ket\}$ for the rotor's Hilbert space reflects a decomposition of $\U_{1}$ which may be written symbolically as 
\begin{equation}
\U_{1}\cong\U_{1}/\Z_{N}\times\widehat{\Z_{N}}\,.\label{eq:u1factorization}
\end{equation}
That is, $a$ labels an element of $\U_{1}/\Z_{N}$ (a coset of $\Z_{N}$ in $\U_{1}$), and $\l$ labels an element of $\widehat{\Z_{N}}$ (an irreducible representation of $\Z_{N}$). Our error-correction procedure makes use of a finer decomposition:
\begin{equation}
\U_{1}\cong\U_{1}/\Z_{dN}\times \Z_{dN}/\Z_N \times \widehat{\Z_{N}}\,.
\label{eq:u1-further-factorization}
\end{equation}
The correctable rotation error is an element of $\U_{1}/\Z_{dN}$, the correctable angular momentum kick is an element of $\widehat{\Z_{N}}$, and code basis states correspond to elements of $\Z_{dN}/\Z_N$.
We will use similar decompositions in our constructions of quantum codes for more general groups.

\subsection{Gates, recovery \& initialization}
\label{subsec:U1-Gates-recovery-init}

To use the above codes for quantum computation on multiple encoded rotors, we need to initialize in the code subspace, execute quantum gates, and perform the measurement-based error-correction described above. For these tasks, we need operators other than the Pauli-type operators $\r_{\phi}$ (\ref{eq:U1:position-shift}) and $\zz^\ell$ (\ref{eq:U1:momentum-shift}). As is typical of quantum codes, there is an ``easy'' subset of all possible operators that aid us in the above tasks in a reasonably fault-tolerant manner. For $\U_1$-rotors, such \textit{normalizer} or \textit{symplectic} operations are generated by certain quadratic functions of the rotors' positions and momenta \cite{Bermejo-Vega2014,Bermejo-Vega2016}.

\prg{Symplectic operations}

Single-rotor symplectic operations include unitary operators generated
by Hamiltonians that are polynomials in angular momentum of at most
degree two. The quadratic-phase operator $\quadgate_{\varphi}=e^{-i\varphi\hat{L}\left(\hat{L}+1\right)/2}$
(with angle $\varphi\in\U_{1}$) maps 
\begin{equation}
\zz\to\r_{\varphi}\zz\,,
\end{equation}
while commuting with position shifts $\r_{\phi}$ (also generated by $\hat{L}$). The analogous two-rotor ``conditional-phase'' operator, $\cphase_{\varphi}=e^{-i\varphi\hat{L}\otimes\hat{L}}$ \{cf. \cite{Grimsmo2019}, Eq.~(23)\},
commutes with $\r_{\phi}\otimes\id$ and $\id\otimes\r_{\phi}$, but
maps
\begin{equation}
\hat{Z}\otimes\id\to\hat{Z}\otimes\hat{X}_{\varphi}\quad\text{and}\quad\id\otimes\hat{Z}\to\hat{X}_{\varphi}\otimes\hat{Z}\,.
\end{equation}

Another operation is the conditional rotation, 
\begin{equation}
\crot\equiv e^{-i\hat{\phi}\otimes\hat{L}}=\int_{\U_{1}}\diff\phi|\phi\ket\bra\phi|\otimes\r_{\phi}\,,\label{eq:CNOT-gate-3}
\end{equation}
shifting the position of the second rotor by $\phi$, conditioned
on the first rotor being at position $\phi$. This maps
\begin{equation}
\r_{\phi}\otimes\id\to\r_{\phi}\otimes\r_{\phi}\quad\text{and}\quad\id\otimes\zz\to\zz^{\dg}\otimes\zz\,,
\end{equation}
while acting trivially on $\zz\otimes\id$ and $\id\otimes\r_{\phi}$.

The $\quadgate$ and $\cphase$ operations can be realized by
turning on Hamiltonians quadratic in angular momenta for a specified
amount of time {[}cf. Eq.~(\ref{eq:quadrupole}){]}. The $\crot$
operation however cannot be obtained from the ``Hamiltonian'' $H=\hat{\phi}\otimes\hat{L}$,
because such an $H$ would not be invariant under $2\pi$-rotations
of the first rotor, and therefore would not be single-valued. (A similar
problem plagues the ``Hamiltonian'' $\hat{\phi}$, present in the exponent of $\zz$, while $\hat{\phi}^2$ is not single-valued even when exponentiated.) To produce such an operator in the lab, one can consider
adapting implementations of the related oscillator phase operator
to rotors \cite{Oi2013,Radtke2017} (see Sec.~\ref{subsec:molecular-rotors}).

\prg{Logical gates}

The above symplectic operations, for certain $\varphi$, perform logical Clifford operations on the encoded qudits. The gate $\quadgate_{\frac{2\pi}{dN^{2}}}$
performs a logical qudit rotation mapping $\zl\to\xl\zl$
(up to a phase), while $\cphase_{\frac{2\pi}{dN^{2}}}$ and $\crot$
act as entangling gates. 

In the case of a logical qubit ($d=2$) with logical operators $\zl=\zz^{N}$
and $\xl=\r_{\frac{\pi}{N}}$ (\ref{eq:U1-logicals}), the symplectic
operations producing the above logical transformations act on the
rotor positions $\phi_{1,2}$ and momenta $\ell_{1,2}$ as follows:
\begin{equation}\label{eq:phi-ell-drift}
\begin{array}{lcll}
\quadgate_{\frac{\pi}{N^{2}}} & : & \phi\to\phi-\frac{\pi}{N^{2}}\ell+c\,\,\,\,\,\,\,\, & \ell\to\ell\phantom{{\displaystyle \half}}\\
\cphase_{\frac{\pi}{N^{2}}} & : & \phi_{1}\to\phi_{1}-\frac{\pi}{N^{2}}\ell_{2} & \ell_{1}\to\ell_{1}\phantom{{\displaystyle \half}}\\
 &  & \phi_{2}\to\phi_{2}-\frac{\pi}{N^{2}}\ell_{1} & \ell_{2}\to\ell_{2}\phantom{{\displaystyle \half}}\\
\crot & : & \phi_{1}\to\phi_{1} & \ell_{1}\to\ell_{1}-\ell_{2}\phantom{{\displaystyle \half}}\\
 &  & \phi_{2}\to\phi_{2}+\phi_{1} & \ell_{2}\to\ell_{2}\,,\phantom{{\displaystyle \half}}
\end{array}
\end{equation}
with constant $c=\frac{\pi}{2}\frac{N-1}{N^{2}}$ (cf. {[}\citealp{Gottesman2001},
Sec.~IX{]}). We have assumed in Eq.~(\ref{eq:phi-ell-drift}) that $\phi$ and $\ell$ simultaneously have definite values, which makes sense for an encoded state assuming that $\phi$ and $\ell$ are sufficiently small. These transformations do not amplify correctable position
and momentum shifts into uncorrectable ones, and a small overrotation or underrotation in the implementation of one of the logical gates introduces only correctable errors, not logical errors. In this sense, the logical gates are fault-tolerant.

The above symplectic operations do not provide a universal set of
logical operations. One way to upgrade to such a set is to include
unitaries generated by the logical operators $\{\xl,\zl\}$ themselves.
The gates $e^{i\varphi(\xl+\hc)}$ and $e^{i\varphi^{\pr}(\zl+\hc)}$
allow for arbitrary single-qudit rotations, while $e^{i\varphi^{\pr\pr}(\xl\otimes\xl+\hc)}$
allows for arbitrary logical $XX$-rotations. Such gates are however
not fault-tolerant, as fluctuations in the $\varphi$'s will produce
undetectable errors. One can also consider using Hamiltonians that
are cubic (or higher) in angular momenta.

\prg{Diagnosis \& Recovery}

A shift in the position by $\delta\phi$ and momentum by $\delta\ell$
maps logical states $|\overline{k}\ket\to|{\textstyle \frac{2\pi}{dN}k+\delta\phi;\delta\ell}\ket$
(up to a phase).
To diagnose the errors we need to measure $\hat \phi$ mod $\frac{2\pi}{dN}$ and $\hat L$ mod $N$. Once this \textit{error syndrome} is known, we can undo the damage by applying  $\r_{\delta\phi}^{\dg}$
(\ref{eq:U1:position-shift}) and $\zz^{\delta\ell\dg}$
(\ref{eq:U1:momentum-shift}) to the corrupted logical states.

To measure $\hat{\phi}$ mod $\frac{2\pi}{dN}$, we need an ancilla that can resolve all possible values of this syndrome, while revealing no information about the protected encoded state.
One way to extract the syndrome is to encode the ancilla using the same code that protects the data \cite{Steane1997}. Specifically, we may prepare an ancillary rotor in the logical-$X$ eigenstate
$|\overline{0}_{X}\ket$, a uniform superposition of the position eigenstates $\{|\phi= \frac{2\pi}{dN}k'\rangle, k' = 0, 1, \dots dN{-}1\}$, which is therefore invariant under the rotation $\phi \to \phi +\frac{2\pi}{dN}$.
Applying the $\crot$ gate (\ref{eq:CNOT-gate-3}) to a noisy logical state and a noiseless ancilla
yields
\begin{equation}
\crot\kk{{\textstyle \frac{2\pi}{dN}k+\delta\phi;\delta\ell}}\otimes\kk{\overline{0}_X}=\kk{{\textstyle \frac{2\pi}{dN}k+\delta\phi;\delta\ell}}\otimes
\hat X_{\delta \phi} |\bar 0_X\rangle.
\end{equation}
The ancilla can then be measured in the $\{|\phi\rangle\}$ basis, and the measured value modulo $\frac{2\pi}{dN}$ determines the shift $\delta\phi$. If the ancilla is noisy or the measurement is imperfect, then the extracted value of $\delta\phi$ is likewise noisy; nevertheless, if a fresh supply of ancilla rotors is continuously available, this recovery procedure will with high likelihood prevent small displacements of the data rotor from accumulating to produce an uncorrectable logical error. 

To measure $\hat{L}$ mod $N$, we need an ancilla that
can resolve the $N$ values of the syndrome. In this case, we could initialize an ancilla rotor in the state $|\phi=0\rangle$, and apply $\cphase_{\frac{2\pi}{N}}$ to the data and ancilla rotors. This gate rotates the ancilla by $\frac{2\pi}{N}\delta \ell$, and the value of $\delta \ell$ can therefore be extracted by measuring the ancilla in the position basis. Since the syndrome takes discrete values, some noise resilience is built into the procedure --- $\delta \ell$ is determined by rounding off the measured value of $\phi$ to the nearest multiple of $\frac{2\pi}{N}$.

Since we only need to resolve a discrete number of momentum syndrome
values, a discretized version of the above scheme using a qu$N$it
ancilla works just as well. Let $\{|h_{z}\ket,h\in\Z_{N}\}$ be the
position states of the qunit, and initialize the qunit in the state
$|0_{z}\ket$. Then apply the entangling gate
\begin{equation}
\cphase^{\pr}\equiv\sum_{\ell\in\Z}|\ell\ket\bra\ell|\otimes{\cal X}^{\ell}\,,
\end{equation}
where ${\cal X}$ satisfies ${\cal X}|h_{z}\ket=|h+1_{z}\ket$ (modulo
$N$) and ${\cal X}^{N}$ is the identity. This yields
\begin{equation}
\cphase^{\pr}\kk{{\textstyle \frac{2\pi}{dN}k+\delta\phi;\delta\ell}}\otimes\kk{0_{z}}=\kk{{\textstyle \frac{2\pi}{dN}k+\delta\phi;\delta\ell}}\otimes\kk{\delta\ell_{z}}\,,
\end{equation}
and measuring the qu$N$it in the position basis then reveals the syndrome.

\prg{Initialization}

The above error-correction procedures can equivalently be used to
initialize in certain logical states. For example, consider one rotor initialized
in $|\phi=0\ket$, coupled to an ancillary qu$N$it initially in $|0_{z}\ket$.
Applying $\cphase^{\pr}$ yields
\begin{equation}
\cphase^{\pr}\,\kk{\phi=0}\otimes\kk{0_{z}}\propto\sum_{\l\in\Z_{N}}|0;\l\ket\otimes|\l_{z}\ket\,.
\end{equation}
Measuring the ancilla in the $|h_{z}\ket$ basis to obtain
a particular $\l=\l_{\star}$ collapses the rotor state to $|0;\l_{\star}\ket$.
Applying a momentum kick $\zz^{\l_{\star}\dg}$ then produces the
logical state $|\overline{0}\ket=|0;0\ket$, thereby completing the
initialization. Analogous initialization schemes use the position syndrome
measurement.

\section{Molecular codes}
\label{sec:Rigid-rotor-codes}

By a ``molecular code'' we mean a finite-dimensional subspace of the infinite-dimensional Hilbert space of a rigid body in three dimensions (\textit{a.k.a.} a ``rigid rotor''). To define a basis for this infinite-dimensional Hilbert space, we imagine fixing a coordinate system in the laboratory, pinning the body's center of mass, and specifying the orientation of the body relative to a standard initial configuration in this fixed coordinate system. For a molecule with no symmetries, the possible orientations are in one-to-one correspondence with the elements of the 3D special orthogonal group $\SO_{3}$; thus we may choose the ``position'' basis $\{|R\ket , R\in\SO_{3}\}$. This correspondence between group elements and orientations of the body follows the same logic as in our discussion of the planar rotor in Sec.~\ref{sec:planar-rotor}, where we identified position-basis eigenstates with elements of $\U_{1}$. 

If the body has symmetries, using a group element to specify the orientation becomes redundant, and the position basis should be refined accordingly. For example, if there is an axis of symmetry (as for a diatomic molecule composed of two distinct nuclei), the body is invariant under the $\U_{1}$ subgroup of rotations about the symmetry axis, and the possible orientations are in one-to-one correspondence with the coset space $\SO_{3}/\U_{1}$, which is equivalent to the two-sphere $\S^2$. If in addition the axis of symmetry has no preferred direction, so the body is invariant under reflections that invert the axis (as for a diatomic molecule composed of two identical nuclei), then the position eigenstates are labeled by the elements of the real projective space $\SO_{3}/\mathsf{O}_{2} =\mathsf{RP}^{2}$. In this section, we assume the body has no symmetries; the case of a body with a symmetry axis (\textit{a.k.a.}  a ``linear rotor'') is discussed in Sec.~\ref{sec:Linear-rotor-codes}.

An \textit{active} rotation of the body described by $\SO_{3}$ element $S$ alters the orientation of the body according to $R \to SR$; that is, the group element describing the orientation of the body is left multiplied by $S$. This rotation is represented by the unitary operator $\ru_{S}$, which acts on the Hilbert space according to
\begin{equation}
	\ru_{S}: |R\ket \to |SR\ket. 
\end{equation}
We will also consider \textit{passive} rotations, rotations of the coordinate system in the laboratory, which act on the position basis according to
\begin{equation}
	\lu_{S}: |R\ket \to |R S^{-1}\ket. 
\end{equation}

As in our discussion of the $\U_{1}$ case, there is a Fourier-conjugate basis of angular momentum states (\textit{a.k.a.} ``rotational states''), defined for $\SO_{3}$ by \begin{subequations}
\label{eq:SO3-Fourier-basis}
\begin{align}
	|_{mn}^{\ell}\ket & ={\displaystyle \int_{\SO_{3}}}\diff R {\textstyle \sqrt{\frac{2\ell+1}{8\pi^{2}}}}D_{mn}^{\ell}(R)|R\ket,\\
|R\ket & ={\displaystyle \sum_{\ell\geq0}\sum_{|m|,|n|\leq\ell}}{\textstyle \sqrt{\frac{2\ell+1}{8\pi^{2}}}}D_{mn}^{\ell\star}(R)|{}_{mn}^{\ell}\ket .   
\end{align}
\end{subequations}
The $\{D_{mn}^{\ell}\}$ denote matrix elements of the angular momentum $\ell$ irreducible representation of $\SO_3$, obeying
\begin{equation}
\label{eq:Wigner-multiplication}
	D_{mn}^\ell(SR) = \sum_p D_{mp}^\ell(S)D_{pn}^\ell(R)
\end{equation}
and $D_{mn}^{\ell}(R^{-1})=D_{nm}^{\ell\star}(R)$, with normalization\footnote{\label{fn:Z-gauge}We are free to choose any orthonormal basis we like for the irrep $D^\ell$. When an explicit form for $D^\ell_{mn}$ is needed, we use the complex-conjugated Wigner $D$-functions from {[}\citealp{VMH}, Sec.~4.3{]}.}
\begin{align}\label{eq:SO3-orthonormality}
 {\displaystyle \int_{\SO_{3}}}\diff R D_{mn}^{\ell\star}(R)D_{m^{\pr}n^{\pr}}^{\ell^{\pr}}(R) = {\textstyle \frac{8\pi^{2}}{2\ell+1}\d_{\ell\ell^{\pr}}} \d_{mm^{\pr}}\d_{nn^{\pr}}.   
\end{align}
The integral is with respect to the invariant Haar measure $\diff R$ on $\SO_{3}$, here normalized so that the volume of the group $\int_{\SO_3} \diff R = 8\pi^2$. 

The elements of the conjugate basis transform under active and passive rotations according to
\begin{subequations}
\label{eq:SO3-position-shifts}
\begin{align}
\ru_{R}|{}_{mn}^{\ell}\ket&={\displaystyle \sum_{p}}D_{pm}^{\ell\star}(R)|{}_{pn}^{\ell}\ket,\label{eq:SO3-position-shifts-1}\\
 \lu_{R}|{}_{mn}^{\ell}\ket&={\displaystyle \sum_{p}}D_{pn}^{\ell}(R)|{}_{mp}^{\ell}\ket.\label{eq:SO3-position-shifts-2}
\end{align}
\end{subequations}
These and other useful properties of the $D_{mn}^{\ell}$ matrices are summarized in Table~\ref{t:G}. 

The functions $D_{mn}^\ell(R)$ are a complete basis for functions which map the group to complex numbers. Hence an arbitrary function $f(R)$ on the group can be Fourier expanded in this basis:
\begin{subequations}
    \begin{align}
f(R) & ={\displaystyle \sum_{\ell\geq0}\sum_{|m|,|n|\leq\ell}}{\textstyle \sqrt{\frac{2\ell+1}{8\pi^{2}}}}f_{mn}^\ell D_{mn}^{\ell}(R)\, ,\\
   f_{mn}^\ell  & ={\displaystyle \int_{\SO_{3}}}\diff R {\textstyle \sqrt{\frac{2\ell+1}{8\pi^{2}}}}D_{mn}^{\ell\star}(R)f(R)\, .
\end{align}
\end{subequations}
We'll use the notation $\hat f$ for the operator associated with the function $f(R)$ that is diagonal in the $\{|R\ket\}$ basis:
\begin{equation}
	\hat f = \int_{\SO_{3}}\diff R \, |R\ket f(R) \bra R|\,.
\end{equation}
Fourier expanding this operator, we obtain
\begin{equation}
	\hat f  ={\displaystyle \sum_{\ell\geq0}\sum_{|m|,|n|\leq\ell}}{\textstyle \sqrt{\frac{2\ell+1}{8\pi^{2}}}}f_{mn}^\ell \dd_{mn}^{\ell}\, .\\
\end{equation}

An arbitrary operator $E$ acting on the Hilbert space of the rotor can be expanded in terms of an operator basis, in which each element of the basis is a diagonal operator followed by an active rotation. After Fourier expanding the diagonal operator, $E$ has the expansion
\begin{align}
	E = \int_{\SO_{3}}\diff S \sum_{\ell\geq0}\sum_{|m|,|n|\leq\ell} E_{mn}^\ell(S) \ru_{S} \dd_{mn}^\ell\,,
\end{align}
where we have absorbed an $\ell$-dependent numerical factor into the coefficient $E_{mn}^\ell(S)$. This is the analog of the expansion of an operator acting on a qudit in terms of the Pauli operator basis. Therefore, a completely positive noise channel $\mathcal{E}$ acting on a state $\rho$ of the rigid rotor has an expansion of the form
\begin{align}
    \mathcal{E}(\rho)=&\int_{\SO_{3}}\diff S\int_{\SO_{3}}\diff S^{\pr}\sum_{\ell,m,n}\sum_{\ell^{\pr},m^{\pr},n^{\pr}}\mathcal{E}_{mn}^{\ell}\!~_{m^{\pr}n^{\pr}}^{\ell^{\pr}}(S,S^{\pr})\notag\\ &~~~~~~~~~~~~~~~~~~~\ru_{S}\dd_{mn}^{\ell}\rho\dd_{m^{\pr}n^{\pr}}^{\ell^{\pr}\dagger}\ru_{S^{\pr}}^{\dagger}\,
	.\label{eq:SO3-channel}
\end{align}
Our goal is to construct a code that allows us to recover successfully from any error of the form
\begin{align}\label{eq:error-term}
	\rho \to \ru_{S} \dd_{mn}^\ell \rho  \dd_{m^\pr  n^\pr }^{\ell^\pr\dagger}\ru_{ S^\pr }^\dagger,
\end{align}
where the position shifts $S, S^\pr $ and momentum kicks $\ell,\ell^\pr$ are sufficiently small. Using this code, we can recover from the noise channel $\mathcal{E}$ with high fidelity if $\ru_{S} \dd_{mn}^\ell \rho  \dd_{m^\pr  n^\pr }^{\ell^\pr\dagger}\ru_{ S^\pr }^\dagger$ has most of its support on small values of $S, S^\pr ,\ell,\ell^\pr$.

To determine how well these codes protect against physical rigid-rotor noise models \cite{Ramakrishna2005,Zhong2016,Stickler2016,Schmidt2015,Papendell2017,Stickler2018,Stickler2018a}, one would expand the noise operators in terms of position and momentum shifts and estimate the probability of an uncorrectable error. A similar analysis for code states of harmonic oscillators rather than rotors {[}\citealp{codecomp}, Sec.~VII{]} shows that physically relevant noise is typically correctable with high probability. If the noise acts locally in phase space, our rotor codes should perform well.

\subsection{$\protect\Z_{N}\subset\protect\Z_{2N}$ codes} \label{subsec:abelian-mol-codes}

We want to construct a finite-dimensional code subspace for the rigid rotor which is protected against small shifts in position and in angular momentum.
For this purpose, we will specify a discrete subgroup $\H\subset \G=\SO_3$ and consider the basis defined by the corresponding partial Fourier transform. In this section,  we assume $\H$ is $\Z_{N}$, an abelian group of rotations about the $\zh$-axis; the case where $\H$ is nonabelian is discussed in Secs.~\ref{subsec:dihedral-codes}-\ref{subsec:Nonabelian-subgroup-codes}. 

As for the case of $\G=\U_1$ from Sec.~\ref{subsec:U1-protected-qubit}, our code construction makes use of a basis defined by a partial Fourier transform associated with the subgroup $\Z_N$. The elements of this basis are
\begin{equation}
    \kk{\csr\Z_{N};\l}\equiv{\textstyle \frac{1}{\sqrt{N}}}\sum_{h\in\Z_{N}}e^{i\frac{2\pi}{N}\l h}\ru_{\csr}\kk{{\textstyle \frac{2\pi}{N}}h,\zh}\,;
    \label{eq:zakforZ3inSO3}
\end{equation}
here, $\kk{\o, \zh}$ denotes the position eigenstate $\kk{R}$, where $R\equiv R_{\o,\zh}$ is a rotation by angle $\o \in [0,2\pi)$ about the $\zh$-axis. Note that the index $\l \in \{0,1,\cdots, N-1\}$ indicates the irreducible representation of $\Z_N$ according to which the state $\kk{\csr\Z_{N};\l}$ transforms. From now on, the presence of a semicolon inside a ket declares that ket to be an element of this basis.

The rotation $S$ is a representative of a coset in  the \textit{lens space} $\SO_3/\Z_N$. Coset representatives are not unique, because of the freedom to multiply $S$ on the right by an element of $\Z_N$ without modifying the coset. We label each coset using the representative $S$ that is as close as possible to the identity rotation. We call the set of such representatives the \textit{fundamental Voronoi cell}, denoting it by $\F_{\SO_3/\Z_N}$ (see Appx.~\ref{appx:voronoi}). It is shown in blue in Fig.~\ref{fig:Zd-rotors}(b) for $N=6$. 

\begin{figure}
\includegraphics[width=1\columnwidth]{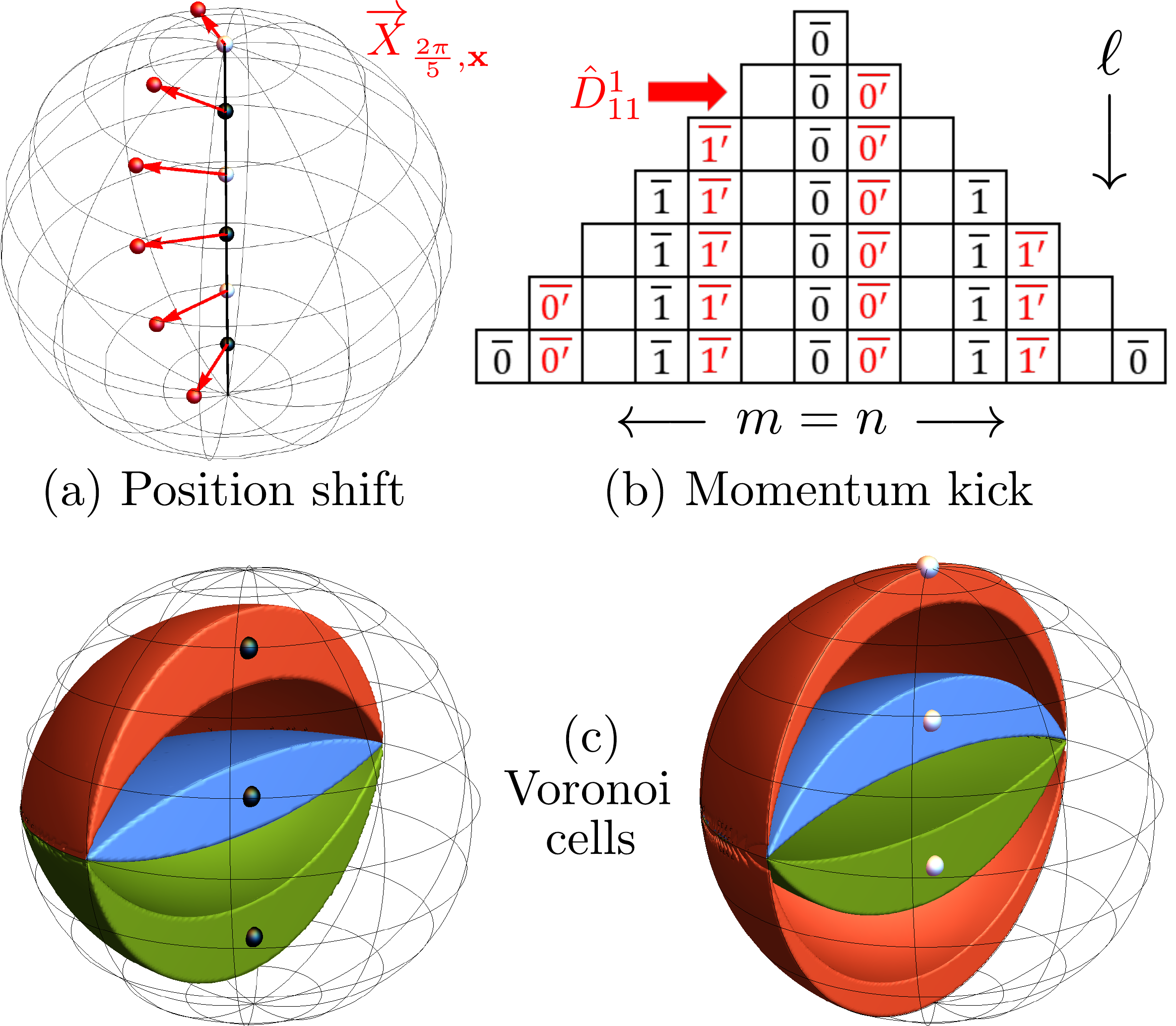}
\caption{
\label{fig:ladder}
{\textsc{Error and recovery.}}
\textbf{(a)} Effect of a typical position shift on the codeword group elements. The group $\SO_3$ is represented as a 3-ball with antipodal points identified. The three group elements which are superposed in the codeword $|\overline 0\ket$ ($|\overline 1\ket$) (\ref{eq:Z3}) are indicated as black (white) balls. 
Upon the indicated representative position shift, each group element is mapped to a group element corresponding to the red ball linked by a red arrow.
{\bf (b)} Sketch of the $|_{mm}^{\ell}\protect\ket$ angular momentum pyramid for $0\protect\leq|m|\protect\leq\ell$, where nonzero components of the logical-$X$ eigenstates $|\overline 0\ket_X$ and $|\overline 1\ket_X$ (\ref{eq:Z3inSO3inMomentumXBasis}) are marked in black  for the case $N=3$. Under the momentum kick $\dd_{11}^{1}$ (\ref{eq:SO3_momentum_shifts}), these codewords are mapped to error states $|\overline{0^{\pr}}\protect\ket_{X}$ and $|\overline{1^{\pr}}\protect\ket_{X}$, whose components are marked in red. 
{\bf (c)} Voronoi cells of group elements corresponding to codeword $|\overline 0\ket$ (left) and $|\overline 1\ket$ (right) are indicated as a red, blue, or green region.
}
\end{figure}

More generally, the $N$ images of $\F_{\SO_3/\Z_N}$ under the action of passive rotations $R\in\Z_N$ are called \textit{Voronoi cells} of $R$. These cells are disjoint, and together they cover $\SO_3$ in what is known as a Voronoi tiling. The six cells for $N=6$ are shown in groups of three in Fig.~\ref{fig:ladder}(c), right and left panels. Voronoi tilings exist for all discrete subgroups $\H\subset\SO_3$; fundamental cells $\F_{\SO_3/\H}$ for various $\H$ are bounded in orange and blue in Fig.~\ref{fig:TinSO3}.

\prg{Codewords}
As in Sec.~\ref{subsec:U1-protected-qubit}, code basis states are associated with the elements of the coset space $\Z_{dN}/\Z_N$, where $\Z_{dN}$ is a larger group of rotations about the $\zh$-axis that contains $\Z_N$. Here, for simplicity, we will assume that $d=2$, but the generalization to other values of $d$ is straightforward. Then the codewords are 
\begin{subequations}
\label{eq:Z3}
\begin{align}
\kk{\overline{0}} & ={\textstyle \frac{1}{\sqrt{N}}}\sum_{h\in\Z_{N}}\kk{{\textstyle \frac{2\pi}{N}}h,\zh}\label{eq:Z3zero}\, ,\\
\kk{\overline{1}} & ={\textstyle \frac{1}{\sqrt{N}}}\sum_{h\in\Z_{N}}\kk{{\textstyle \frac{2\pi}{N}}h+{\textstyle \frac{\pi}{N}},\zh}\, .
\end{align}
\end{subequations}
The state $|\overline 0\ket$ is the uniform superposition of elements of $\Z_N$, while $|\overline1\ket$ is the uniform superposition of the elements of the nontrivial coset of $\Z_N$ in $\Z_{2N}$, elements displaced from $\Z_N$ by $R_{\frac{\pi}{N},\zh}$, the $\frac{\pi}{N}$ rotation about the $\zh$-axis. In terms of the partially Fourier-transformed basis Eq.~(\ref{eq:zakforZ3inSO3}), we may express the codewords as
\begin{align}
    \label{eq:ZN-in-SO3-logicals-in-Zak-basis}
    \kk{\overline{0}}=|\Z_{N};0\ket\,\quad\text{and}\quad \kk{\overline{1}}=|R_{\frac{\pi}{N},\zh}\Z_{N};0\ket\,.
\end{align}

Using Eq.~(\ref{eq:SO3-Fourier-basis}), we can also express these codewords in the angular-momentum eigenstate basis. Since $D_{mn}^\ell(R)$ is a diagonal matrix for any rotation $R$ about the $\zh$-axis,$^{\ref{fn:Z-gauge}}$
\begin{equation}\label{eq:Dmn-diagonal}
	D_{mn}^\ell(\o,\zh) = \delta_{mn} e^{im\o},
\end{equation}
we easily compute (for $r\in\{0,1\}$)
\begin{align}
|\overline{r}\ket & =\sum_{\ell\geq0}{\textstyle \sqrt{\frac{N\left(2\ell+1\right)}{8\pi^{2}}}}\sum_{|pN|\leq\ell}\left(-1\right)^{pr}\kk{_{pN,pN}^{\ell}}\,.\label{eq:Z3inSO3inMomentumBasis}
\end{align}
We see that, when expanded in the $|{}_{mn}^\ell\ket$ basis, the only states that occur with nonzero coefficients are those for which $m=n$ is an integer multiple of $N$. This property will ensure that the codewords are well protected against sufficiently small angular momentum kicks. 

It is also useful to express the codewords in the logical-$X$ basis, defining
$|\overline r\ket_X = \frac{1}{\sqrt{2}}\left(|\overline 0\ket +(-1)^r|\overline 1\ket\right)$ and finding
\begin{align} 
	\kk{\overline 0}_X &= \sum_{\ell\geq0}{\textstyle \sqrt{\frac{N\left(2\ell+1\right)}{4\pi^{2}}}}\sum_{|2pN|\leq\ell}\kk{_{2pN,2pN}^{\ell}}\,,\label{eq:Z3inSO3inMomentumXBasis}\\
	\kk{\overline 1}_X &=  \sum_{\ell\geq N}{\textstyle \sqrt{\frac{N\left(2\ell+1\right)}{4\pi^{2}}}}\sum_{|(2p+1)N|\leq\ell}\kk{_{(2p+1)N,(2p+1)N}^{\ell}}.\notag
\end{align}
The codeword $|\overline 0\ket_X$, expanded in the $|{}_{mn}^\ell\ket$ basis, includes only basis states in which $m=n$ is an even multiple of $N$, and the codeword $|\overline 1\ket_X$ includes only basis states with $m$ an odd multiple of $N$. The $|{}_{mn}^\ell\ket$'s which occur with nonzero coefficients are indicated schematically in Fig.~\ref{fig:Zd-rotors}(b) for $N=3$, with black squares indicating basis states in the expansion of $|\overline 0\ket_X$ and white squares indicating basis states in the expansion of $|\overline 1\ket_X$. 

\prg{Position shifts}
We will correct an error of the form Eq.~(\ref{eq:error-term}) in two steps. In the first step, we diagnose and reverse the shift in the position basis $\ru_S:|R\ket\to |SR\ket$. After the position shift is corrected, we can proceed to correct the momentum kick $\dd_{mn}^\ell$.

For any coset space $\G/\H$, we can label the cosets using coset representatives chosen from the fundamental Voronoi cell $\F_{\G/\H}$. Then the action of $\G$ on the cosets is described by the \textit{induced representation} \cite{Arovas,so3engbook,Preskill1991} 
\begin{equation}
\ru_{R}\kk{\csr\H;0}=\kk{RST\H;0}\, ,\label{eq:inducedrepZninSO3}
\end{equation}
where $T$ is a compensating element of $\H$ chosen to ensure that $RST\in \F_{\G/\H}$. (If $RS\in \F_{\G/\H}$, then no compensating element of $\H$ is needed.)

The effect of a representative rotation on the group elements making up the codewords is shown in Fig.~\ref{fig:ladder}(a). For correcting position shifts acting on the codewords, the relevant coset space is $\SO_3/\Z_{2N}$. We quotient out $\Z_{2N}$ rather than $\Z_N$, because both $|\overline 0\ket$ and $|\overline1\ket$ are superpositions of elements of $\Z_{2N}$; hence an element of $\SO_3/\Z_{2N}$ characterizes the shift away from the code space induced by $\ru_{R}$ in Eq.~(\ref{eq:inducedrepZninSO3}), without revealing any information that distinguishes $|\overline 0\ket$ from $|\overline 1\ket$. 
We divide the basis elements $\{\kk{\csr\Z_{N};\l}\}$ (\ref{eq:zakforZ3inSO3}) into two disjoint subsets, where each subset is parametrized by an element of $\F_{\SO_3/\Z_{2N}}$ rather than an element of $\F_{\SO_3/\Z_{N}}$, as follows:
\begin{align}
\kk{\tilde{S}\Z_{N};\l}&=\ru_{\tilde{S}}\kk{\Z_{N};\l}\,\text{~and~}\label{eq:ZN-two-sets-of-bases}\\
\kk{\tilde{S}R_{\frac{\pi}{N},\zh}\Z_{N};\l}&=\ru_{\tilde{S}}\kk{R_{\frac{\pi}{N},\zh}\Z_{N};\l}\,\notag,
\end{align}
where $\l \in \{0,1,\cdots, N-1\}$ as before, but now $\tilde S \in \F_{\SO_3/\Z_{2N}}$. Since each element of $S\in \F_{\SO_3/\Z_{N}}$ can be uniquely expressed as either $S= \tilde S$ or $S= \tilde SR_{\frac{\pi}{N},\zh}$, for $\tilde S\in \F_{\SO_3/\Z_{2N}}$, this is the same basis as described earlier, just with a different labeling than before. The first set is the set of all states obtained by acting with a rotation $\tilde S \in \F_{\SO_3/\F_{2N}}$ on the logical zero codeword; the second set is obtained similarly from logical one. (If we wished to construct a $d$-dimensional codespace, we would divide the basis into $d$ disjoint subsets, with each subset parametrized by an element of $\F_{\SO_3/\Z_{dN}}$.)

To diagnose the error, we measure the value of $\tilde S$, and then apply $\ru_{\tilde S}^\dagger$ to attempt to correct the effect of the position shift. If the actual shift error is $\ru_{R}$, where $R$ is contained in the $\F_{\SO_3/\Z_{2N}}$, this recovery procedure successfully corrects the position-shift error, mapping
\begin{subequations}
\label{eq:ZN-on-SO3-position-correction}
    \begin{align}
    \ru_{R}\kk{\Z_{N};\l}&\to\kk{\Z_{N};\l}\\\ru_{R}\kk{R_{\frac{\pi}{N},\zh}\Z_{N};\l}&\to\kk{R_{\frac{\pi}{N},\zh}\Z_{N};\l}\,.
\end{align}
\end{subequations}
Subsequent momentum-kick correction, described below, would then complete the recovery, correctly mapping the corrupted states back to their respective codewords (\ref{eq:ZN-in-SO3-logicals-in-Zak-basis}).
However, if $R$ is not contained in $\F_{\SO_3/\Z_{2N}}$, then there may be an uncorrected logical error which interchanges $|\Z_{N};\l\ket$ and $|R_{\frac{\pi}{N},\zh}\Z_{N};\l \ket$ on the right-hand side of Eq.~(\ref{eq:ZN-on-SO3-position-correction}).

Thus the code protects against any position-shift error $\ru_R$ for $R\in \F_{\SO_3/\Z_{2N}}$. This set of correctable rotations is indicated by the blue region in Fig.~\ref{fig:Zd-rotors}(b). How large an angular rotation can be tolerated depends on the axis of rotation, and can be determined by analyzing the geometry of the fundamental cell $\F_{\SO_3/\Z_{2N}}$ (see Appx.~\ref{appx:voronoi}).
We find that a rotation by angle $\o$ about an axis with polar angle $\T$ is contained within $\F_{\SO_3/\Z_{2N}}$ for 
\begin{equation}
|\o|<\om(\T)\equiv\left|2\cot^{-1}\left(\cos\T~\cot\frac{\pi}{4N}\right)\right|\,.\label{eq:omega-max}
\end{equation}
The maximum correctable rotation angle $\om(\T)$ is smallest for rotations about the $\zh$-axis, where $\om(0)= \frac{\pi}{2N}$, as for the planar rotor code discussed in Sec.~\ref{subsec:U1-protected-qubit}. The largest correctable rotation angles occur when the rotation axis is in the equatorial plane, where $\om(\frac{\pi}{2})= \pi$. Thus, \textit{any} rotation about such an axis is correctable, unless the rotation angle is precisely $\pi$. The relative volume occupied by correctable rotations in $\SO_3$ is $\frac{1}{2N}$.

\prg{Momentum kicks}

We have now described how to correct the position shift $\ru_R$ in Eq.~(\ref{eq:error-term}). Next, we need to understand how to contend with a momentum kick 
\begin{equation}
\dd_{mn}^{\ell}\equiv\int_{\SO_{3}}\diff R \,|R\ket D_{mn}^{\ell}\left(R\right) \bra R|\, \label{eq:SO3_momentum_shifts}
\end{equation}
acting on the code space.

We can compute the action of $\dd_{mn}^\ell$ on the codewords using Eq.~(\ref{eq:Dmn-diagonal}), finding
\begin{subequations}
    \label{eq:hat-D-on-codewords}
\begin{align}
	\dd_{mn}^{\ell}\kk{\overline{0}}&=\delta_{mn}\kk{\Z_{N};\l=m\,(\textrm{mod}\,N)}\,,\\
	\!\!\!\!\!\!\!\!\!\!\!\!\!\!\!\!\!\!\!\!\!\!\!\!\!\!\!\!\!\!\!\!\!\!\!\!\!\!\!\!\!\!\!\!\!\!\!\!\!\dd_{mn}^{\ell}\kk{\overline{1}}&=\delta_{mn}e^{i\frac{\pi}{N}m}\kk{R_{\frac{\pi}{N},\zh}\Z_{N};\l=m\,(\textrm{mod}\,N)}.
\end{align}
\end{subequations}
After the noise acts on the encoded state, and the position-shift error has been corrected, we measure the value of $\l$, the syndrome for the momentum-shift error. The key thing to notice is that, while $\l $ determines the value of $m$ (mod $N$), the codeword-dependent phase $e^{i\frac{\pi}{N}m}$ in Eq.~(\ref{eq:hat-D-on-codewords}) depends on the value of $m$ (mod $2N$). In fact, for any value of $\ell\ge N$, the operator $\dd_{NN}^\ell$ is a nontrivial logical operator that preserves the code space and flips the relative phase of $|\overline 0\ket$ and $|\overline 1\ket$.

Once the value of $\l$ is known, we attempt recovery by applying the unitary operator $U_m$ with the action
\begin{align}\label{eq:momentum-shift-recovery}
    U_{m}:\,&\kk{\Z_{N};m}\to\kk{\Z_{N};0}\,,\\&\kk{R_{\frac{\pi}{N},\zh}\Z_{N};m}\to e^{-i\frac{\pi}{N}m}\kk{R_{\frac{\pi}{N},\zh}\Z_{N};0}\,\notag,
\end{align}
where $m$ is chosen to be the integer with minimal absolute value such that $m = \l$ (mod $N$). For example, we can choose $U_m$ to have the same action as $\dd_{mm}^{N\dagger}$ on the position eigenstates $|R_{\o, \zh}\ket$, namely 
\begin{equation}
\label{eq:recovery-m}
    U_{m}\kk{R_{\o,\zh}}=e^{-im\o}\kk{R_{\o,\zh}}.
\end{equation}
How $U_m$ acts on $|R\ket$ when $R$ is not a rotation about the $\zh$-axis can be chosen arbitrarily. We note though, that this extended operator can be diagonal in the $\{|R\ket\}$ basis. That is, the recovery operation after a momentum kick can be a phase shift that depends on the orientation of the rotor, and has the action  Eq.~(\ref{eq:recovery-m}) when the rotor's orientation differs from the standard reference orientation by a rotation about the $\zh$-axis. 

Notice that $m$ (mod $2N$) is unambiguously determined by $\l =m$ (mod $N$) for any $m$ satisfying $|m| < N/2$. Therefore, the damage to the codewords caused by the action of $\dd_{mn}^\ell$ will be corrected successfully when $|m| < N/2$. Since $|m| \le \ell$, we conclude that the code protects against any momentum kick $\dd_{mn}^\ell$ such that
\begin{equation}
\ell <  N/2 \,.\label{eq:correctableshiftsZNinSO3}
\end{equation}
For $\ell \ge m > N/2$, however, a logical error may occur.

It is also instructive to consider how the momentum kick $\dd_{mn}^\ell$ acts on the basis of angular momentum eigenstates. We observe that 
\begin{align}\label{eq:3j-integral}
&\!\!\!\!\!\!\! \bra _{MN}^L| \dd_{mn}^\ell    |_{m^\pr  n^\pr }^{\ell^\pr}\ket 
 = \int_{\SO_3} \diff R  \bra _{MN}^L| R\ket D_{mn}^\ell(R)\bra R   |_{m^\pr  n^\pr }^{\ell^\pr}\ket \\
&\!= {\textstyle\frac{\sqrt{(2L+1)(2\ell^\pr+1)}}{8\pi^2 }} \int_{\SO_3} \diff R ~ D_{MN}^{L\star}(R)D_{mn}^\ell(R) D_{m^\pr  n^\pr }^{\ell^\pr}(R).\notag
\end{align}
The group integral (\ref{eq:3j-integral}) can be expressed in terms of Clebsch-Gordan coefficients [see Table~\ref{t:G}.G]. For our purposes, what's noteworthy is that selection rules for addition of angular momenta require the integral to vanish unless
\begin{align}
 |\ell^\pr-\ell|\le    L\le \ell^\pr+\ell\, , \quad M= m^\pr+m\, , \quad N = n^\pr+n\, .\label{eq:CG}
\end{align}
As indicated in Eq.~(\ref{eq:Z3inSO3inMomentumBasis}), if the codewords are expanded in the $\{|_{m^\pr  n^\pr }^{\ell^\pr}\ket\}$ basis, then $m^\pr = n^\pr $ is an integer multiple of $N$ for all states that occur with nonzero coefficients. For $\ell < N/2$, the momentum kick $\dd_{mn}^\ell$ changes the value of $m^\pr $ and $ n^\pr $ by less than half the spacing between successive multiples of $N$. Therefore, these shifts in $m^\pr $ and $ n^\pr $ can be unambiguously identified and corrected. However, it is simpler to understand how the recovery procedure works in detail using the expansion of the codewords in the the position basis $\{|R\ket\}$  [as in Eq.~(\ref{eq:momentum-shift-recovery})] rather than the angular momentum basis (detailed in Appx.~\ref{appx:Approximate-states}).

The effect of an angular momentum kick $\dd_{1,1}^1$ is visualized in Fig.~\ref{fig:ladder}(b) for the case $N=3$. Recalling Eq.~(\ref{eq:hat-D-on-codewords}), this kick shifts the code space to a subspace with $\l = 1$, and $\dd_{-1,-1}^1$ shifts the code space to a subspace with $\l = 2$. In either case, measuring $\l$ points to a unique error with $\ell \le 1$ which can then be corrected. However, $\dd_{2,2}^2$ also maps the code space to the same subspace with $\l = 2$ as $\dd_{-1,-1}^1$, imparting a \textit{different} codeword-dependent phase; according to Eq.~(\ref{eq:hat-D-on-codewords}), the codeword $|\overline 1\ket$ acquires the relative phase $\exp(-i\frac{\pi}{3})$ when $\dd_{-1,-1}^1$ acts on the codespace, and the relative phase $\exp(i\frac{2\pi}{3})$ when $\dd_{2,2}^2$ acts on the codespace. Therefore, if the $\dd_{2,2}^2$ error occurs, and is misdiagnosed as a $\dd_{-1,-1}^1$ error, a nontrivial logical error results when recovery is attempted.

\prg{Logical operators}

The unitary active rotation $\ru_{\frac{\pi}{N},\zh}$, acting on the code basis states in Eq.~(\ref{eq:Z3}), has the effect of interchanging $|\overline 0\ket$ and $|\overline 1\ket$. It can be regarded as the logical Pauli operator $\xl$ acting on the code space. This operation can similarly be performed by the passive rotation $\lu_{\frac{\pi}{N},\zh}$ [since the codewords (\ref{eq:Z3}) consist of position states forming an abelian group].
In other words, we can rotate the molecular frame or the lab frame to perform this operation. Active and passive rotations always commute, $\ru_R \lu_S=\lu_S\ru_R$. We will use this fact to infer the momentum kick syndrome $\l$ using passive rotations, without interfering with position shifts $\ru_R$.

We have already noted that $\dd_{NN}^\ell$ (for any $\ell\ge N$), acting on the code basis states, preserves $|\overline 0\ket$ and flips the phase of $|\overline 1\ket$. Thus its action on the code space is equivalent to the logical Pauli operator $\zl$. However, $\dd_{NN}^\ell$ is not unitary as an operator acting on the full Hilbert space of the rotor. Why isn't $\dd_{NN}^\ell$ unitary? Recall that $\dd_{mn}^\ell$ is diagonal in the $\{|R\ket\}$ basis, with eigenvalues $\{D_{mn}^\ell(R)\}$. The trouble is that for rotations  that are not about the $\zh$-axis, $\{D_{mm}^\ell(R)\}$ does not have modulus 1, and therefore cannot be an eigenvalue of a unitary operator. Specifically, if we parametrize $R_{\a\b\g}$ using Euler angles in the $ZYZ$ convention, where $\a\in [0,2\pi)$, $\b \in [0,\pi]$, $\g\in [0,2\pi)$,
\begin{align}\label{eq:ZninSO3gates}
D_{NN}^N(\a,\b,\g)= e^{iN(\a +\g)} \cos^{2N}(\b/2)\, , 
\end{align}
which has modulus less than 1 for nonzero $\b$.

To formulate an implementation of $\zl$ that is achievable in the laboratory, we should find a unitary extension of its logical action to the full Hilbert space. As for the recovery operation described earlier, we can choose this logical $\zl $ to be a phase shift that depends appropriately on the orientation of the rotor. 

One way to produce logical gates involves turning on the Hamiltonian $\dd_{NN}^N+\hc$ to perform a logical $Z$-axis rotation with angle proportional to the time the Hamiltonian is turned on. This provides unitary logical $Z$-gates, and analogous two-qubit $ZZ$-gates can be performed via the Hamiltonian $\dd_{NN}^N\otimes\dd_{NN}^N+\hc$. However, such gates are subject to over- or under-rotation errors. 


\prg{Check operators}

Once we have operators whose action on the code space matches that of the logical Pauli operators $\xl$ and $\zl$, we can square these operators to define the check operators for the code. Then the code space can be said to be the simultaneous eigenspace with eigenvalue 1 of these operators.

Recall that $\xl$ is a position shift defined as either left or right multiplication by $R_{\frac{\pi}{N},\zh}$, corresponding to either an active or a passive rotation, respectively.
We choose our ``$X$-type'' stabilizer to be a passive rotation, yielding the unitary operator
\begin{align}
	\sx = \left(\lu_{\frac{\pi}{N},\zh}\right)^2 =\lu_{\frac{2\pi}{N},\zh}\,.
\end{align}
The condition $\sx = 1$ requires the codewords to be invariant under a position shift by $R_{\frac{2\pi}{N},\zh}$. The additional benefit of using a passive rotation is that \textit{all} of the partial Fourier-transformed states (\ref{eq:zakforZ3inSO3})
are eigenstates of $\sx$,
\begin{equation}
\lu_{\frac{2\pi}{N},\zh}\kk{\csr\Z_{N};\l}=e^{i\frac{2\pi}{N}\l}\kk{\csr\Z_{N};\l}\,.\label{eq:ZN-rightX-eigenstates}
\end{equation}
This way, the syndrome $\l$ can be extracted via a projective measurement onto eigenspaces of $\sx$.

As we've noted, $\dd_{NN}^N$ is not unitary, but nevertheless we may square it to obtain a (nonunitary) $Z$-type check operator
\begin{align}
 & \sz = (\dd_{N,N}^N)^2 = \dd_{2N,2N}^{2N} \\\notag
  &= \int \sin\b {\diff\a \diff\b \diff\g}\, |R_{\a\b\g}\ket e^{i2N(\a +\g)} \cos^{4N}(\b/2)\bra R_{\a\b\g}|\, .
	\end{align}
The eigenspace of $\sz$ with eigenvalue 1 contains rotations about the $\zh$-axis ($\b = 0$), by angle $\o = \a + \g = \frac{\pi}{N} h$, where $h\in \{0,1, \cdots, 2N-1\}$. The only states that satisfy these conditions and that are also invariant under $\sx$ are the states in the code space.

To check that $\sx$ and $\sz$ are really commuting operators, we observe that
\begin{align}
	\lu_S \dd_{mn}^\ell \lu_S^\dagger &= \int_{\SO_3} \diff R  |RS^{-1}\ket D_{mn}^\ell(R) \bra RS^{-1}| \notag\\
	&= \int_{\SO_3} \diff R  |R\ket D_{mn}^\ell (RS )\bra R|\, ,
\end{align}
where we have used the invariance of the Haar measure to obtain the second equality. Furthermore, if $S$ is a rotation about the $\zh$-axis, then, recalling the $ZYZ$ Euler-angle convention used here, right multiplication by $S$ merely changes the third angle:
\begin{equation}
	R_{\a,\b,\g} R_{\o,0, 0} = R_{\a,  \b, \g + \o}\,.
\end{equation}
Using this and Eq.~(\ref{eq:ZninSO3gates}),
\begin{align}
    D_{NN}^{N}\left(\a,\b,\g+{\textstyle \frac{\pi}{N}}\right)=-D_{NN}^{N}\left(\a,\b,\g\right),
\end{align}
we see that $\dd_{NN}^N $ and $\lu_{\frac{\pi}{N}, \zh}$ anticommute, not only acting on the code space, but also acting on the whole Hilbert space of the rotor. Correspondingly, $\sz = (\dd_{NN}^N)^2 $ and $\sx = (\lu_{\frac{\pi}{N}, \zh})^2$ commute, and thus can be simultaneously diagonalized.

\subsection{Measurement \& Initialization}
\label{subsec:msmnts}

Generalizing our discussion for the $\U_1$ rotor from Sec.~\ref{subsec:U1-Gates-recovery-init}, we outline procedures for extracting the momentum ($\l\in\Z_{N}$)
and position ($\tilde{S}\in\F_{\SO_{3}/\Z_{2N}}$) shift values using ancilla
systems. These procedures also allow us to perform logical state initialization.

\prg{Momentum syndromes}
There are $N$ different possible values
of $\l$, so a one-shot measurement requires an ancilla with at least $N$
orthogonal states. We use a qu$N$it ancilla with $Z$-eigenstates
$|h_{z}\ket$, $X$-eigenstates $|h_{x}\ket$ ($h\in\Z_{N}$), and Pauli operator $\cal{X}$ satisfying $\mathcal{X}|h_z\ket=|{h+1}_z\ket$ (modulo $N$). 

To measure $\l$, we initialize the qu$N$it in $|0_{z}\ket$ and entangle it with the rigid rotor by applying the gate 
\begin{subequations}
    \begin{align}
        \cphase&=\sum_{\ell\leq0}\sum_{|m|,|n|\leq\ell}|_{mn}^{\ell}\ket\bra_{mn}^{\ell}|\otimes{\cal X}^{n}\\&=\sum_{h\in\Z_{N}}\lu_{\frac{2\pi}{N}h,\zh}^{\dg}\otimes\kk{h_{x}}\bbra{h_{x}}\,,
    \end{align}
\end{subequations}
where the second line is obtained using Eqs.~(\ref{eq:SO3-position-shifts-2}) and  (\ref{eq:Dmn-diagonal}). This gate shifts the ``position'' of the qu$N$it by $n$, conditioned on the rotor having angular momentum $\zh$-component $n$ in the rotor frame.
Applying this to a partially Fourier-transformed basis state (\ref{eq:zakforZ3inSO3}) and using Eq.~(\ref{eq:ZN-rightX-eigenstates}) yields
\begin{equation}
\cphase~\kk{\csr\Z_{N};\l}\otimes\kk{0_{z}}=\kk{S\Z_{N};\l}\otimes|\l_{z}\ket\,.
\end{equation}
The value of $\l$ has thus been mapped onto the ancilla, and a subsequent $Z$-basis measurement of the ancilla allows us to extract $\l$. 

\prg{Position syndromes}
To diagnose position shifts without disturbing the logical information,
we have to use an ancilla to measure the syndrome $\tilde{S}\in\F_{\SO_{3}/\Z_{2N}}$
--- ``an $\SO_{3}$ rotation modulo $\Z_{2N}$''. In order to perfectly resolve all possible $\tilde{S}$ in one shot, the ancilla
needs to admit an orthonormal set of position states parameterized
by $\F_{\SO_{3}/\Z_{2N}}$. Such a set is exactly the set of orientations
of a $\Z_{2N}$-symmetric rigid body (see Sec.~\ref{subsec:other-systems}).
However, coupling an asymmetric molecule to a symmetric one is difficult.
Below, we show how to approximate the required position states using
generalized spin-coherent states of a finite-dimensional spin (see Sec.~\ref{subsec:atomic-ensembles}). 

We use a spin-$\SS$ ancilla $\{|_{s}^{\SS}\ket\,,\,|s|\leq\SS\}$,
which admits an irrep of $\SO_{3}$ with corresponding rotation matrices
$D^{\SS}(R)$. In order to ``mod out'' the $\Z_{2N}$ rotation,
we initialize the ancilla in any state $|\Z_{2N}\ket$
whose maximal invariant subgroup is $\Z_{2N}$, i.e.,
\begin{equation}
    \left|\bbra{\Z_{2N}}D^{\SS}\left(T\right)\kk{\Z_{2N}}\right|\,\,\,\begin{cases}
\,\,\,=1 & T\in\Z_{2N}\\
\,\,\,<1 & \text{otherwise}
\end{cases}\,.
\label{eq:spin-coh-condition}
\end{equation}
Such states exist for any $\SS\geq N$ {[}\citealp{bacry}, Table~10.1{]} (see also \cite{Barnett2006});
for example, the $\SS=N$ family of states $\cos\eta\kk{_{N}^{N}}+\sin\eta\kk{_{-N}^{N}}$
satisfies the above for any $\eta\in(0,\pi/4)$.

Any rotation $R\in\SO_{3}$ can be written as $R=\tilde{S}T$, with $\tilde{S}\in\F_{\SO_{3}/\Z_{2N}}$ and $T\in\Z_{2N}$. Therefore, applying
rotations to $|\Z_{2N}\ket$ and using Eq.~(\ref{eq:Wigner-multiplication}) yields the following set of generalized spin-coherent states $|\tilde{S}\ket_{\Z_{2N}}$,
\begin{align}
    D^{L}(R)\kk{\Z_{2N}}&=D^{L}(\tilde{S})D^{L}(T)\kk{\Z_{2N}}\notag\\&\propto D^{\SS}(\tilde{S})\kk{\Z_{2N}}\equiv|\tilde{S}\ket_{\Z_{2N}}
\end{align}
parameterized by $\tilde{S}\in\F_{\SO_{3}/\Z_{2N}}$.

To map the syndrome onto the ancilla, we use the conditional rotation (cf. {[}\citealp{Schmidt2016},~Eq.~(4){]})
\begin{equation}
    \crot=\int_{\SO_{3}}\diff R \,|R\ket\bra R|\otimes D^{\SS}(R)\,.
    \label{eq:ZN-crot}
\end{equation}
This gate applies a rotation $D(R)$ on the ancilla, conditioned on the
rotor being in the state $|R\ket$. When applied to the specified ancillary state, this gate maps $\tilde{S}\in\F_{\SO_3/\Z_{2N}}$ onto the ancilla while ignoring the logical state index. Applying it to the two sets of basis states from Eq.~(\ref{eq:ZN-two-sets-of-bases}), we have
\begin{align}
    \crot\,\kk{\tilde{S}\Z_{N};\l}\otimes|\Z_{2N}\ket&=\kk{\tilde{S}\Z_{N};\l}\otimes|\tilde{S}\ket_{\Z_{2N}}\notag\\
    \crot\,\kk{\tilde{S}R_{\frac{\pi}{N},\zh}\Z_{N};\l}\otimes|\Z_{2N}\ket&=\kk{\tilde{S}R_{\frac{\pi}{N},\zh}\Z_{N};\l}\otimes|\tilde{S}\ket_{\Z_{2N}}.
\end{align}

Unfortunately, a projective measurement in the overcomplete  $|\tilde{S}\ket_{\Z_{2N}}$ set of the states will not yield $\tilde S$ exactly, since the spin-coherent states are not orthogonal for any finite $\SS$ {[}\citealp{perelomov_book}, Sec.~2.3{]}. However, they
approach orthogonality in the limit $\SS\to\infty$, meaning
that a sufficiently large spin should be able to resolve points in
$\F_{\SO_{3}/\Z_{2N}}$ to desired accuracy.

\prg{Initialization}
The $\crot$ gate can also be used to initialize in the logical-$X$ state $|\overline{0}\ket_X\propto\sum_{T\in\Z_{2N}}|T\ket$ (\ref{eq:Z3inSO3inMomentumXBasis}). Say the rotor instead starts in the lowest-momentum state $|_{00}^{0}\ket$ --- a state outside of the codespace. Then, application of the gate yields (up to normalization)
\begin{equation}
    \crot~\kk{_{00}^{0}}\otimes\kk{\Z_{2N}}\propto\int_{\F_{\SO_{3}/\Z_{2N}}}\!\!\!\!\!\!\diff\tilde{S}\sum_{T\in\Z_{2N}}|\tilde{S}T\ket\otimes|\tilde{S}\ket_{\Z_{2N}}.
\end{equation}
A projective measurement obtaining some $\tilde S$ followed by a rotation $\ru_{\tilde S}^\dg$ yields the desired logical state.

\subsection{Normalizable codewords
\label{subsec:Approximate-codewords}}

Our logical codewords are not normalizable and therefore unphysical. To obtain normalizable states we may regulate the sum over the angular momentum $\ell$ in Eq.~(\ref{eq:Z3inSO3inMomentumBasis}) by introducing a broad envelope function which decays sufficiently rapidly for large $\ell$. In position space, this corresponds to replacing the position eigenstate $|\o,\zh\ket$ by a sharply peaked normalizable wavepacket which approximates $|\o,\zh\ket$. The protection against position shifts is mildly impaired due to this spreading of the codewords in position space. On the other hand, these approximate codewords still have support on angular momentum states such that $\ell$ is an integer multiple of $N$, and therefore the code continues to detect momentum kick operators $\dd^\ell_{mm}$ with $\ell < N$ and $|m|>0$. However, the kicks $\dd^\ell_{00}$ no longer leave the codewords invariant [as in Eq.~(\ref{eq:hat-D-on-codewords})], inducing a slight $\ell$-dependent distortion that can cause a logical error.

The oscillator GKP codes can be regulated  \cite{Gottesman2001} by applying the damping function $\exp(-\half\D^2\ph)$ to the ideal codewords, where $\ph$ is the number operator and $\D > 0$ is the damping strength {[}\citealp{codecomp}, Eq.~(7.12){]}. 
For molecular codes with configuration space $\SO_{3}$, we may use the damping function $\exp(-\half\D^2\lh)$ instead, where $\lh$ is the total angular momentum operator which satisfies
\begin{equation}
\lh|_{mn}^{\ell}\ket=\ell\left(\ell+1\right)|_{mn}^{\ell}\ket\,,\label{eq:L2-action-on-lmn-states}
\end{equation}
and generates orientational diffusion {[}\citealp{so3engbook}, Sec.~16.6{]}. The approximate codewords are thus (for $r\in\{0,1\}$)
\begin{align}
\!\!\kk{\tilde{r}} & =e^{-\half\D^2\lh}|\overline{r}\ket/\sqrt{\bra\overline{r}|e^{-\D^2\lh}|\overline{r}\ket}\label{eq:ZN-in-SO3-approxiamte-codewords}\\
 & =\sum_{\ell\geq0}{\displaystyle \sqrt{\frac{N\left(2\ell+1\right)e^{-\D^2\ell(\ell+1)}}{8\pi^{2}\bra\overline{r}|e^{-\D^2\lh}|\overline{r}\ket}}}\sum_{|pN|\leq\ell}\left(-1\right)^{pr}\kk{_{pN,pN}^{\ell}}.\nonumber 
\end{align}
Asymmetric diffusion is also possible, yielding an additional damping
term $e^{-\D^{\pr 2}p^{2}}$.

\prg{Average momentum}

The expectation value of the total angular momentum is infinite for
ideal codewords, but finite for approximate codewords. The square-root
of the expectation value of $\lh$,
\begin{equation}
\lb\equiv\bbra{\tilde{r}}\lh\kk{\tilde{r}}^{1/2}\sim\sqrt{\frac{3}{2\D^{2}}-\frac{1}{4}}\,,\label{eq:lbar}
\end{equation}
determines the average momentum of the approximate codewords $|\tilde{r}\ket$.
In Appx.~\ref{appx:Approximate-states}, we detail the calculation
that obtains the above $r$-independent result, valid in the $\D\to0$
limit. While the average photon number is proportional to the total
oscillator energy, energy for the rigid rotor is proportional to $\lb^{\,2}$.
For our normalizable codewords, the energy scales identically with
$\D$ as that for the normalizable GKP states \cite{Gottesman2001}.
The variance of $\lb$, $\s_{\lb}^2=\bbra{\tilde{r}}\hat{L}^{4}\kk{\tilde{r}}^{1/2}-\lb^{\,2}=O\left(\lb^{\,2}\right)$, is also similar to GKP states (for the latter, photon number moments satisfy a geometric distribution \cite{codecomp}).

\prg{Approximate correctability}

The spreading of the basis
states in position space gives rise to an intrinsic error in the approximate
code; the basis states $|\tilde{0}\ket$ and $|\tilde{1}\ket$
are imperfectly distinguishable even in the absence of noise. To quantify the probability of intrinsic memory error, we estimate
$\pl$ --- the probability that the approximate logical zero state
leaks into the union of Voronoi cells associated with the other logical
codeword {[}for $N=3$, these are the three cells in the right panel
of Fig.~\ref{fig:ladder}(c){]}. In Appx.~\ref{appx:Approximate-states},
we find in the $\D\to0$ limit,
\begin{align}
\pl & \sim\csc\left(\frac{\pi}{2N}\right)\,\frac{\D}{\sqrt{\pi}}\,\exp\left[-\left(\frac{\pi}{2N\D}\right)^{2}\right]\,.\label{eq:prob-of-leakage}
\end{align}
The right-hand side is exponentially suppressed with $1/\D^{2}$:
similar to GKP codes \cite{Gottesman2001}, a gentle smearing in position
space does not significantly affect the codes' performance. More concretely,
$\pl\approx10^{-3}$ requires an average momentum of $\lb\approx5.4$,
with 99\% of the approximate codewords supported on the $\ell\leq10$
momentum subspace. A more stringent $\pl\approx10^{-6}$ corresponds
to $\lb\approx8.1$, and requires $\ell\leq15$ to support 99\% of
the codewords. While these are reasonable numbers for the angular
momentum, we also have to keep in mind that, in contrast to the oscillator, the rotor energy is proportional to $\lb^{\,2}$.

As we have noted, the approximate codewords are supported on values
of $\ell$ which are integer multiples of $N$; therefore, angular
momentum kicks with $\ell < N$ and $|m|>0$ are detectable. However, the kicks $\dd^\ell_{00}$ slightly distort the codewords, leading to  the potential for a logical-$X$ error. To quantify this, we calculate the matrix element
$ \bra\tilde{0}|\dd_{00}^{\ell}|\tilde{1}\ket $ for $\ell < N$ in Appx.~\ref{appx:Approximate-states}. Numerically, as $\D\to0$, this element is suppressed exponentially with $1/\D^2$ for all cases tested. In said appendix, we estimate its asymptotic behavior, showing that its dependence on $\D$ is 
similar to $\pl$ (\ref{eq:prob-of-leakage}),
\begin{equation}
	\bra\tilde{0}|\dd_{00}^{\ell}|\tilde{1}\ket\approx2\left(2\ell+1\right)\exp\left[-\left(\frac{\pi}{2N\D}\right)^{2}\right]~\label{eq:momentum-distortion}.
\end{equation}
The ``$\approx$'' indicates that this asymptotic estimate is supported, to some extent, by numerical evidence.

\subsection{Dihedral molecular codes}
\label{subsec:dihedral-codes}

For molecular codes, $\G$ is the rotation group $\SO_{3}$, and up
until know we have considered the case where $\H,\K$ are abelian,
namely $\H=\Z_{N}$ and $\K=\Z_{dN}$ (for a code space of dimension
$d$). We may also construct codes for which $\K$, and perhaps also
$\H$, are nonabelian subgroups of $\SO_{3}$. Position correction
proceeds similarly as for the abelian molecular codes: one measures
values in the coset space $\SO_{3}/\K$ and applies a rotation to
map back into the codespace. Picking nonabelian subgroups allows for
more uniform correctable rotation sets $\F_{\SO_{3}/\K}$ than the
saucer-like $\F_{\SO_{3}/\Z_{2N}}$ from Fig.~\ref{fig:Zd-rotors}(b).
Detectable and correctable momentum kicks $\ell$ can be read off
by successive use of branching formulas, i.e., restricting $\SO_{3}$-irreps
$D^{\ell}$ to $\K$ and decomposing the resulting matrix into irreps
of $\K$, and then further restricting and decomposing into $\H$-irreps.
Here we describe $\DD_{N}\subset\DD_{2N}$ molecular codes, where
$\DD_{N}$ is the dihedral group.

\prg{Codewords}

The group $\Z_{N}$, containing rotations about the $\zh$-axis by
angle $\o=\frac{2\pi}{N} h$ with $h\in\{0,1,\cdots,N-1\}$, can be extended
to the dihedral group $\DD_{N}$ by adding the $\o=\pi$ rotation
around the $\xh$-axis. This dihedral group has $2N$ elements, the
original $N$  rotations contained in $\Z_N$ and also $N$ \textquotedblleft reflections\textquotedblright{}
--- rotations by $\pi$ about $N$ equally spaced axes on the equator.
In terms of Euler angles, the
rotations in $\DD_{N}$ are the elements $\{\frac{2\pi}{N}h,0,0\}$ of $\SO_3$, and the reflections are the elements $\{\frac{2\pi}{N}h,\pi,0\}$,
for $h\in\{0,1,\cdots,N-1\}$.

Here we consider an extension of the $\Z_{N}\subset\Z_{dN}$ codes
to $\DD_{N}\subset\DD_{dN}$. For the $d=2$ case, the coset space
$\DD_{2N}/\DD_{N}$ contains two cosets: the trivial coset (the $\DD_N$ subgroup of $\DD_{2N}$) and the nontrivial coset, which is obtained by multiplying all elements of $\DD_N$ by the rotation with Euler angles $\{\frac{\pi}{N},0,0\}$.
The logical codewords
are 
\begin{subequations}
\begin{align}
\kk{\overline{0}}&={\textstyle \frac{1}{\sqrt{2N}}}\sum_{T\in\DD_{N}}\kk T\\\kk{\overline{1}}&={\textstyle \frac{1}{\sqrt{2N}}}\sum_{T\in\DD_{N}}\kk{R_{\frac{\pi}{N}00}T}\,.
\end{align}\label{eq:DNcodewords}
\end{subequations}

\begin{figure}
\includegraphics[width=\columnwidth]{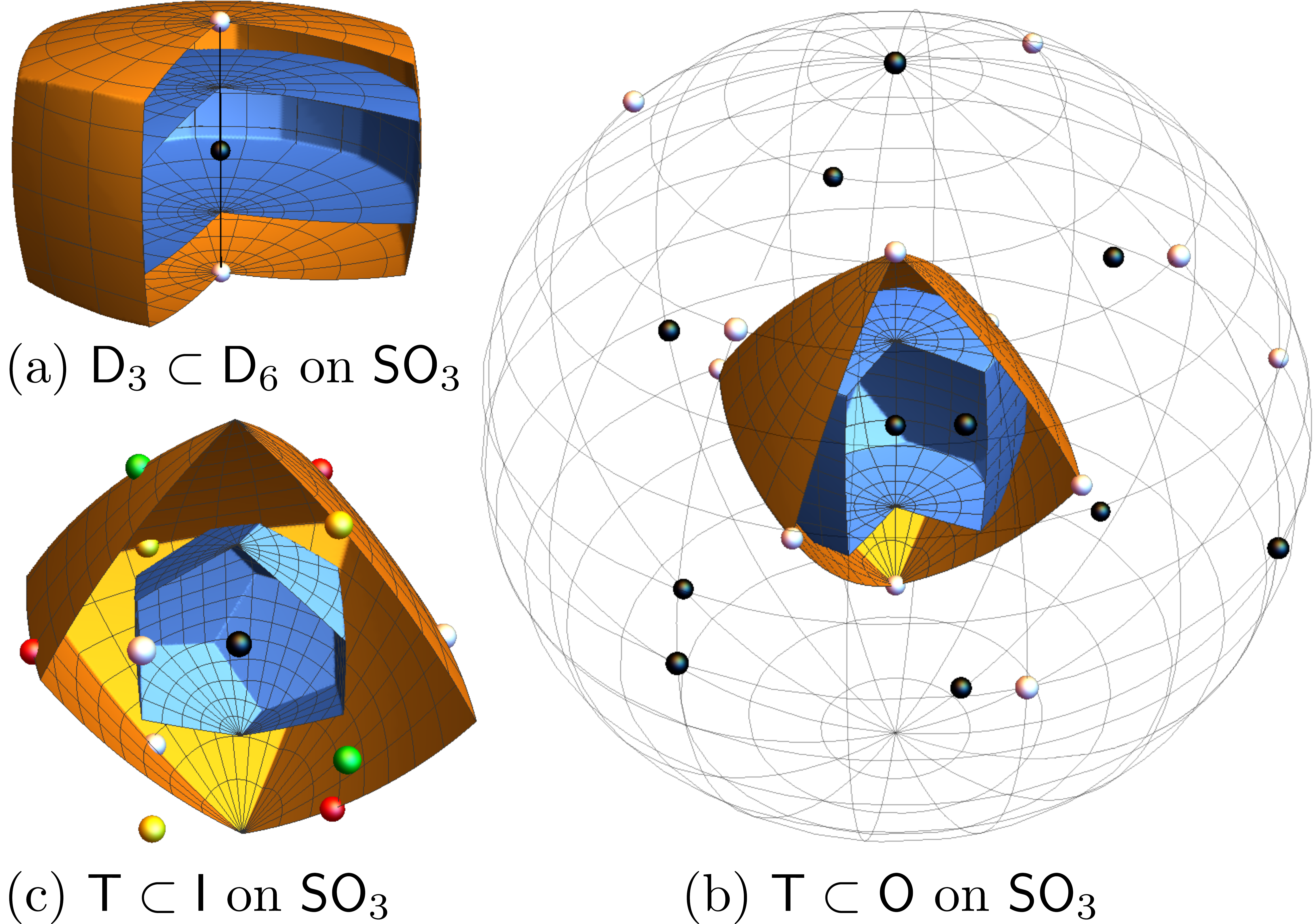}\caption{\label{fig:TinSO3}\textsc{Nonabelian subgroup codes.} {\bf (a)} Sketch
of the prism spaces $\protect\SO_{3}/\protect\DD_{3}$ (orange), $\protect\SO_{3}/\protect\DD_{6}$ (blue), and group elements representing the two logical codewords within $\protect\SO_{3}/\protect\DD_{3}$ for the $\protect\DD_3\subset\protect\DD_6$
dihedral code from Sec.~\ref{subsec:dihedral-codes}. Similar
sketches for the quotient spaces (see Table~\ref{t:quaotient_spaces})
and codewords of \textbf{(b)} the $\protect\TT\subset\protect\OO$
and \textbf{(c)} the $\protect\TT\subset\protect\II$ codes from Sec.~\ref{subsec:Nonabelian-subgroup-codes}.
} \end{figure}

This code family has much in common with the $\Z_{N}\subset\Z_{2N}$
code; in particular, it can correct momentum kicks with $ \ell < N/2$. The space of correctable position shifts --- the
prism space $\SO_{3}/\DD_{2N}$ [Table~\ref{t:quaotient_spaces} and Fig.~\ref{fig:TinSO3}(a)]
--- gets flatter with increasing $N$. Thus, as for the $\Z_{N}\subset\Z_{2N}$ codes, there is a tradeoff: as $N$ increases, the
code protects against larger momentum kicks, but at the cost of weakened performance against rotation errors. 

\prg{Partial Fourier transform}

For the $\DD_N\subset \DD_{2N}$ code, the partially Fourier-transformed basis generalizing Eq.~(\ref{eq:zakforZ3inSO3}) consists of pairs of basis states of the form
\begin{subequations}
\label{eq:DN-error-states}
\begin{align}
\kk{S\DD_{N};_{\m\n}^{\l}}&=\frac{1}{\sqrt{2N}}\sum_{T\in\DD_{N}}\z_{\m\n}^{\l}(T)\kk{ST}\\
\!\!\!\!\kk{SR_{\frac{\pi}{N}00}\DD_{N};_{\m\n}^{\l}}&=\frac{1}{\sqrt{2N}}\sum_{T\in\DD_{N}}\z_{\m\n}^{\l}(T)\kk{SR_{\frac{\pi}{N}00}T}\,.
\end{align}
\end{subequations}
Now $S$ is an element of the fundamental Voronoi cell $S\in\F_{\SO_{3}/\DD_{2N}}$, $\l$ labels an irreducible representation (irrep) of $\DD_N$, and $\z_{\m\n}^{\l}(T)$ denotes the matrix elements of that representation, evaluated for the $\DD_N$ element $T$. (The nonabelian group $\DD_N$ has both one-dimensional and two-dimensional irreps.) The codewords Eq.~(\ref{eq:DNcodewords}) correspond to $|\overline 0\ket = |\DD_N;{}_{00}^{\one}\ket$ and $|\overline 1 \ket = |R_{\frac{\pi}{N}00}\DD_N;{}_{00}^{\one}\ket$, where $\l=\one$ is the trivial irrep. The basis states $|S\DD_{N};_{\m\n}^{\l}\ket$ span all states that can be reached when a correctable error acts on the codeword $|\overline 0\ket$; $S$ is the rotation error and $_{\m\n}^{\l}$ indexes the momentum kick. Similarly, the basis states $|SR_{\frac{\pi}{N}00}\DD_{N};_{\m\n}^{\l}\ket$ span all states that can be reached when a correctable error acts on the codeword $|\overline 1\ket$.

\prg{Position shifts}
Our recovery consists of first correcting position shifts by measuring
$S$ and applying $\ru_{S}^{\dg}$ to map all error states into the
subspace $\{|\DD_{N};_{\m\n}^{\l}\ket,|R_{\frac{\pi}{N}00}\DD_{N};_{\m\n}^{\l}\ket\}$
for all $_{\m\n}^{\l}$. To extract $S$, we can readily adapt the procedure described in Sec.~\ref{subsec:msmnts}. This entails initializing an ancillary system in a $\DD_{2N}$-invariant state, performing the $\crot$ gate (\ref{eq:ZN-crot}), and measuring the ancilla in a basis of generalized spin-coherent states parameterized by $\F_{\SO_3/\DD_{2N}}$. A similar scheme can be used for state initialization.

\prg{Momentum kicks}

To see how the momentum kick operators $\dd^\ell_{mn}$ affect the codewords, we need to understand how the irreducible representation (irrep) of $\SO_3$ with angular momentum $\ell$ decomposes into irreps of $\DD_{2N}$ and $\DD_N$. When $N$ is odd, the group $\DD_N$ has two one-dimensional irreps, the trivial representation (which we denote by $\one$), and a nontrivial representation (denoted by $\onep$) which represents rotations by $+1$ and reflections by $-1$. In addition, there are $(N-1)/2$ two-dimensional irreps, which we denote by $\two{k}$ with $k \in\{1, 2, \cdots, (N-1)/2\}$. We can characterize a representation according to how the generators of $\DD_N$, the rotation $R_{\frac{2\pi}{N}00}$ and the reflection $R_{\frac{2\pi}{N}\pi0}$, are represented. All $\two{k}$ irreps represent the reflection by the $2\times 2$ matrix
\begin{align}\label{eq:DN-irrep-reflection}
\z^{\two{k}}(R_{\frac{2\pi}{N}\pi0})=\begin{pmatrix}0 & 1\\
1 & 0
\end{pmatrix},\end{align}
while the rotation is represented by the diagonal matrix
\begin{align}\label{eq:DN-irrep-rotation}
\z^{\two{k}}(R_{\frac{2\pi}{N}00})=\begin{pmatrix}e^{i\frac{2\pi}{N}k} & 0\\
0 & e^{-i\frac{2\pi}{N}k}
\end{pmatrix}\,.
\end{align}
Thus, the two-dimensional irrep of $\DD_N$ decomposes as two one-dimensional irreps of $\Z_N$ which are interchanged by the reflection. When $N$ is even, there are $N/2 - 1$ two-dimensional irreps described by Eqs.~(\ref{eq:DN-irrep-reflection}-\ref{eq:DN-irrep-rotation}), and also two additional one-dimensional irreps (denoted by $\one^{\pm}$), representing the rotation $R_{\frac{2\pi}{N}00}$ by $-1$ and the reflection $R_{\frac{2\pi}{N}\pi0}$ by $\pm1$, respectively.

The irrep $D^\ell$ of $\SO_3$ decomposes into $2\ell + 1$ one-dimensional irreps of $\U_1$; these represent the rotation by angle $\phi$ about the $\zh$-axis  by $\{e^{im\phi}, m = -\ell, -\ell + 1, \cdots , \ell\}$. For $\ell < N/2$ and $m\ne 0$, the $\pm m$ irreps of $\U_1$ pair up to form a two-dimensional irrep of $\DD_N$, while the $m=0$ irrep of $\U_1$ provides a one-dimensional representation of $\DD_N$, either $\onep$ if $\ell$ is odd, or $\one$ if $\ell$ is even. From Eq.~(\ref{eq:DN-irrep-rotation}) we can infer the ``branching rules'' specifying how the irreps of $\DD_{2N}$ transform under the $\DD_N$ subgroup, namely $\one\to\one$, $\onep\to\onep$, and 
\begin{align}\label{eq:D2N-toDN}
\two{k} \to \two{k}, \quad \textrm{for } k < N/2\, .
\end{align}
This means that, of the irreps of $\DD_N$ that arise in the decomposition of $D^\ell$ for $\ell < N/2$, each is descended from a unique irrep of $\DD_{2N}$.

Suppose, now, that the momentum kick operator $\dd^\ell_{mn}$ acts on the codewords Eq.~(\ref{eq:DNcodewords}) of the code associated with $\DD_N\subset \DD_{2N}$, where $\ell < N/2$, and we are able to diagnose the irrep of $\DD_N$ according to which the damaged states transform. Because this irrep of $\DD_N$ points to a unique irrep of $\DD_{2N}$, the action of the rotation $R_{\frac{\pi}{N}00}$ on the states is unambiguously determined. This means it is possible to recover, mapping $\dd^\ell_{mn}|\overline 0\ket$ back to $|\overline 0\ket$, and $\dd^\ell_{mn}|\overline 1\ket$ back to $|\overline 1\ket$, without any logical error.

However, for $\ell \ge N/2$ the situation is different; the $\DD_{2N}$ irrep from which the $\DD_N$ irrep arises is no longer unique. Correspondingly, projecting onto an irrep of $\DD_N$ after $\dd^\ell_{mn}$ acts does not fix the relative phase of $|\overline 0\ket$ and $|\overline 1\ket$; therefore perfect recovery is not possible. 

To be concrete, consider the case $\DD_3\subset \DD_{6}\subset \SO_3$. Here, $\DD_3$ has just one two-dimensional irrep, which we will simply call $\twoo$. 
The $\ell=1$ irrep of $\SO_3$ decomposes as
\begin{align}
	D^{\ell = 1} \to \onep\oplus \two{1} \to \onep\oplus \twoo
\end{align}
under $\DD_6$ and $\DD_3$, respectively. 
The four matrix elements $D^1_{mn}$, for $m,n\in \{\pm 1\}$, constitute the two-dimensional irrep $\twoo$ of $\DD_3$ and $\two{1}$ of $\DD_6$; therefore, using the pairs of basis states (\ref{eq:DN-error-states}), we have
\begin{subequations}
\label{eq:DN-mom-kick}
\begin{align}
\dd_{mn}^{1}\kk{\overline{0}}&=\kk{\DD_{3};{}_{\m=m,\n=n}^{\twoo}}\\\dd_{mn}^{1}\kk{\overline{1}}&=\exp\left(i{\textstyle \frac{\pi}{3}}m\right)\kk{R_{\frac{\pi}{3}00}\DD_{3};{}_{\m=m,\n=n}^{\twoo}}\, ,\label{eq:DN-mom-kick-1}
\end{align}
\end{subequations}
for $m,n \in\{\pm 1\}$. The damage inflicted on the codewords by $\dd^{1}_{mn}$ can be reversed by applying $\dd^{1\dagger}_{mn}$. (The error operator $\dd^{1}_{00}$, realizing the representation $\onep$ of $\DD_3$ and $\DD_6$, is also easily reversed.)
Though $\dd^{1\dagger}_{mn}$ are not unitary operators acting on the full $\SO_3$ Hilbert space, a completely positive recovery map can be constructed which consists of projections onto $\DD_3$ irreps followed by appropriate momentum kicks. This map will successfully recover from any noise channel that can be expanded in $\{\dd^\ell_{mn}\}$ for $\ell \le 1$.

The $\SO_3$-irrep $\ell = 2$, on the other hand, branches as
\begin{align}
	D^{\ell = 2}\to \one \oplus \two{1}\oplus \two{2} \to \one \oplus \twoo \oplus \twoo
\end{align}
under $\DD_6$ and $\DD_3$ respectively. Now $D^2_{mn}$ constitutes the $\two{2} $ irrep of $\DD_6$ for $m, n \in \{\pm 2\}$ and the $\two{1} $ irrep of $\DD_6$ for $m, n \in \{\pm 1\}$; however these two distinct irreps of $\DD_6$ cannot be distinguished as irreps of $\DD_3$. Therefore diagnosing the irrep of $\DD_3$ according to which the damaged codewords transform does not suffice to determine the relative phase of the two code basis states; now 
\begin{subequations}
\label{eq:DN-mom-kick-ell2}
\begin{align}
\dd_{mn}^{2}\kk{\overline{0}}&=\kk{\DD_{3};{}_{\m=m,\n=n}^{\twoo}}\\\dd_{mn}^{2}\kk{\overline{1}}&=\exp\left(i{\textstyle \frac{2\pi}{3}}m\right)\kk{R_{\frac{\pi}{3}00}\DD_{3};{}_{\m=m,\n=n}^{\twoo}}\,,
\end{align}
\end{subequations}
for $m, n \in \{\pm 2\}$, in contrast to Eq.~(\ref{eq:DN-mom-kick}). If, say, a $\dd^2_{22}$ error were to occur, it could be mistaken for a $\dd^{1}_{-1,-1}$ error. An attempt to recover by applying  $\dd^{1\dg}_{-1,-1}$ would result in a logical phase error, with $|\overline 0\ket \to |\overline 0\ket$ and $|\overline 1\ket \to -|\overline 1\ket$. Thus, while the $\DD_3\subset \DD_6$ code can protect against angular momentum kicks with $\ell \le 1$, it does not protect against arbitrary kicks with $\ell \le 2$. In general, the observation (\ref{eq:D2N-toDN}) implies that the $\DD_N\subset \DD_{2N}$ code protects against all kicks with $\ell < N/2$.

An undetectable error corresponds to a nontrivial representation of $\DD_6$ that branches to the trivial representation of $\DD_3$. The lowest angular momentum at which this occurs is $\ell=3$, with branching rules
\begin{equation}
    D^{\ell=3} \to \one \oplus \one^{+} \oplus \one^{-} \oplus \two{1} \oplus \two{2} \to \one \oplus \one \oplus \onep \oplus \twoo \oplus \twoo\,.
\end{equation}
Here, the nontrivial irrep $\one^{+}$ of $\DD_6$ reduces to the trivial irrep $\one$ of $\DD_3$. Diagnosing this trivial irrep yields no information about its parent irrep of $\DD_6$, meaning that $\ell=3$ momentum kicks produce undetectable errors.

For $N=3$, the only nontrivial correction one needs while mapping the states back into the codespace is the correction of  the $\m$-dependent phase (\ref{eq:DN-mom-kick-1}) for $\l=\twoo$. Therefore, only knowledge of
$\l,\m$ is required, and the error syndrome can be obtained by performing a projective measurement onto a basis that resolves these indices without extracting the logical information. One such basis is the joint eigenbasis of the two commuting rotations from Table~\ref{fig:dihedral-syndromes}. Such a measurement will project the corrupted codewords onto eigenstates of these operators. A successful recovery operation, then, maps the resulting states back into the codespace, applying (in the case of $\l=\twoo$) a $\m$-dependent phase that undoes the action of the $\ell=1$ momentum kick from Eq.~(\ref{eq:DN-mom-kick-1}).

\begin{table}
\begin{tabular}{ccc}
\toprule 
 & $\lu_{0,\pi,0}$ & $\ru_{\frac{2\pi}{3},0,0}$\tabularnewline
\midrule
$|\cs\DD_{3};{}_{00}^{\one}\ket$ & $+1$ & $+1$\tabularnewline
$|\cs\DD_{3};{}_{00}^{\onep}\ket$ & $-1$ & $+1$\tabularnewline
$\frac{1}{\sqrt{2}}\big(|\cs\DD_{3};{}_{+1,+1}^{\twoo}\ket\pm|\cs\DD_{3};{}_{+1,-1}^{\twoo}\ket\big)$ & $\pm1$ & $e^{-i\frac{2\pi}{3}}$\tabularnewline
$\frac{1}{\sqrt{2}}\big(|\cs\DD_{3};{}_{-1,+1}^{\twoo}\ket\pm|\cs\DD_{3};{}_{-1,-1}^{\twoo}\ket\big)$ & $\pm1$ & $e^{i\frac{2\pi}{3}}$\tabularnewline
\bottomrule
\end{tabular}

\caption{\label{fig:dihedral-syndromes} \textsc{Dihedral codes.} Check operators for the $\protect\DD_{3}\subset\protect\DD_{6}$ codes
and their corresponding eigenvalues and eigenstates within the 12-dimensional
subspace $|R\protect\ket$ for $R\in\protect\DD_{6}$, where $\protect\cs\in\{I,R_{\frac{\pi}{N}00}\}$.
} 
\end{table}

\subsection{Other nonabelian molecular codes\label{subsec:Nonabelian-subgroup-codes}}

Other interesting codes can be constructed using the tetrahedral, octahedral, and icosahedral subgroups of the rotation group. All are finite nonabelian groups, denoted $\TT$, $\OO$ and $\II$, respectively, with order $|\TT|=12$, $|\OO|=24$, $|\II|=60$. $\TT$ is isomorphic to the alternating group $\A_4$,  $\OO$ is isomorphic to the permutation group $\S_4$, and $\II$ is isomorphic to the alternating group $\A_5$. Since $\TT$ is a subgroup of both $\OO$ and $\II$, codes can be constructed  based on the embedding $\TT \subset \OO$, with code dimension 2, or based on $\TT \subset \II$, with code dimension 5. 

The logical codewords for $\TT\subset\OO$ are uniform superpositions of $\SO_3$ elements, indicated as black/white balls in Fig.~\ref{fig:TinSO3}(b). That code can correct rotation errors in the fundamental Voronoi cell $\F_{\SO_3/\OO}$, the cube-like region bounded in blue in the figure. The $\TT\subset\II$ code can correct rotation errors in $\F_{\SO_3/\II}$, the dodecahedron bounded in blue in Fig.~\ref{fig:TinSO3}(c). In that figure, the balls of five different colors correspond to the $\SO_3$ elements making up this code's five logical codewords.

To investigate how well these codes protect against momentum kicks, we examine the branching rules for $\SO_3\to \K\to \H$, as in Sec.~\ref{subsec:dihedral-codes}, but where now $\H=\TT$ and $\K$ is either $\OO$ or $\II$. The group $\TT$ has four irreps labeled as $\{\one, \onep, \onepp, \three\}$ (with number denoting dimension), $\OO$ has five irreps $\{\one, \onep, \twoo, \three, \threep\}$, and $\II$ has five irreps $\{\one, \three, \threep, \four, \five\}$. We note that $D^{\ell=1}$, the defining 3-dimensional irrep of $\SO_3$, also provides defining irreps of the subgroups $\TT$, $\OO$, and $\II$; therefore the branching rule
\begin{align}\label{eq:branch-OI-to-T}
	D^{\ell = 1}\to \three \to \three
\end{align}
applies to both the $\TT\subset \OO$ and $\TT\subset \II$ codes. This means that projecting onto the basis of irreps of $\TT$ unambiguously identifies the error $\dd^{\ell}_{mn}$, which can therefore be corrected, assuming that $\ell \le 1$. Hence both codes protect against kicks with $\ell \le 1$.

Focusing on $\TT\subset\OO$ (the $\TT\subset\II$ code behaves similarly), the $\ell=2$ irrep of $\SO_{3}$ has branching rules \cite{Fallbacher2015}
\begin{align}
	D^{\ell = 2}\to \twoo \oplus \threep\to \onep\oplus \onepp \oplus \three \,.
\end{align}
Here, the $\TT$-irrep $\three$ is the same 3D irrep that appears in Eq.~(\ref{eq:branch-OI-to-T}), but the irrep $\threep$ of $\OO$ is different than the irrep $\three$ in Eq.~(\ref{eq:branch-OI-to-T}).
Therefore, the projection onto the basis of irreps of $\TT$ does not unambigously identify the irrep of $\OO$, and we conclude that the $\TT\subset \OO$ codes do not protect against arbitrary momentum kicks with $\ell \le 2$. 

Undetectable errors are associated with nontrivial irreps of $\OO$ which branch to trivial irreps of $\TT$. This occurs at $\ell=3$, due to the branching rules
\begin{align}
	D^{\ell = 3}\to \onep\oplus \three \oplus \threep\to \one\oplus \three \oplus \three\,.
\end{align}
Interestingly, the $\TT\subset \OO$ code can also detect all momentum kicks with $\ell = 4, 5$, because for these irreps of $\SO_3$, the trivial irrep of $\TT$ appears only as a descendant of the trivial irrep of $\OO$.

Momentum kick correction for the more general codes proceeds by measuring
a combination of left and right rotations that distinguishes the error
spaces sufficiently well for one to correct succesfully. As with the dihedral codes, momentum kicks produce $\m$-dependent relative phases between corrupted codewords, which need to be corrected upon recovery. Using $\TT\subset\OO$ as an example, successful reovery requires determining the irrep label $\l\in\{\one,\onep,\onepp,\three\}$ and the $\m$ label for $\l=\three$. After correcting position shifts, the corrupted states lie in the subspace $\{\kk{R}\,,\,R\in\OO\}$. The three check operators $\lu_{\frac{\pi}{2}\frac{\pi}{2}\pi}$, $\ru_{0\pi0}$, and $\ru_{\pi00}$ commute on this space, and measuring in their joint eigenbasis resolves $\l$ and $\m$.

Just like $\SO_{3}$ rotations permute cosets in the lens space $\SO_{3}/\Z_{N}$
in an induced representation (\ref{eq:inducedrepZninSO3}), elements
$\{\ru_{S}\}_{S\in\K}$ permute cosets in $\K/\H$, providing logical
$X$-type operators. For $\TT\subset\OO$, there are only two cosets,
so $\{\ru_{S}\}_{S\in\OO}$ either act trivially or exchange the two
codewords. For $\TT\subset\II$, the sixty rotations $\{\ru_{S}\}_{S\in\II}$
form a five-dimensional permutation representation of $\II$ when
acting on the five codewords. Since $\II=\A_{5}$ (the alternating
group), any permutations of the codewords in $\A_{5}$ are realized
by the unitary $\ru_{R}$'s. Moreover, these gates are fault tolerant. If there is a slight over- or under-rotation
$S^{\pr}\neq S$ and the rotated state $\ru_{S^{\pr}}|\overline{r}\ket$
is close to (but not quite equal to) the codeword $\ru_{S}|\overline{r}\ket$,
then the error-correction procedure will fix this by mapping $\ru_{S^{\pr}}|\overline{r}\ket$
to the closest codeword.

\section{Linear rotor codes\label{sec:Linear-rotor-codes}}

By a \textit{linear rotor} we mean a rigid body with a symmetry axis, such that rotations about that axis leave the orientation of the body invariant. The paradigmatic example is a diatomic molecule containing two distinct atoms; we discuss other manifestations in Sec~\ref{subsec:other-systems}. In contrast to an asymmetric body, for which orientations of the body are in one-to-one correspondence with elements of the rotation group $\SO_3$, the configuration space of the linear rotor is the coset space $\SO_3/\U_1= \S^2$, because the $\U_1$ rotations about the symmetry axis do not alter the orientation. This is equivalent to the configuration space of a particle moving on a two-sphere.

The position eigenstates $\{|\vh\ket\}$ provide an orthogonal basis for the Hilbert space of the linear rotor, with continuum normalization, where $\vh$ denotes a point on $\S^2$ (equivalently a unit 3-vector). It is convenient to parametrize points on the sphere using spherical coordinates $\vh= (\theta,\phi)$, where $\theta$ denotes the polar angle and $\phi$ is the aximuthal angle; thus $\theta \in [0,\pi]$ and $\phi\in [0,2\pi)$.

A rotation $R\in\SO_{3}$ rotates the linear rotor with orientation $\vh$
to a new orientation $R\vh$. It is represented by the
unitary operator $\r_{R}$, with action
\begin{equation}
\r_{R}|\vh\ket=|R\vh\ket\,.
\end{equation}
A rotation acting on $\S^2$, in contrast to a rotation acting on states of an asymmetric rigid rotor, has fixed points; the position eigenstate $|\vh\ket$ is left invariant by a rotation $R=(\o,\pm\vh)$ about the axis $\vh$ or the axis $-\vh$:
\begin{equation}
\r_{\o,\pm\vh}\kk{\vh}=\kk{\vh}\,.
\end{equation}
Any point $\vh$ on $\S^2$ can be obtained by applying a suitably chosen rotation $R$ to a fiducial initial point (for example the north pole $\vh_0$), but there are many more rotations than points, because two rotations in the same $\U_1$ coset map $\vh_0$ to the same point. 

Another relevant operation is inversion or parity, mapping $\vh$
to its antipode $-\vh$. The parity operation $P$ commutes with any rotation $R$; rotations together with inversions generate the group $\OO_{3}$ of proper and improper rotations in three dimensions, isomorphic to $\SO_3 \times \Z_2$. In Hilbert space, $P$ is represented by $\p$, with action
\begin{equation}
\p\kk{\vh}=\kk{-\vh}\,,
\end{equation}
which commutes with $\r_R$ for any $R$.

Dual to the continuous position basis is the discrete Fourier-conjugate basis,
defined on $\S^{2}$ by 
\begin{subequations}
\label{eq:S2-fourier}
\begin{align}
\,\,\,\,\,\,\,\,\,\,\,\,\,|\vh\ket & ={\displaystyle \sum_{\ell\geq0}\sum_{|m|\leq\ell}}Y_{m}^{\ell\star}(\vh)|{}_{m}^{\ell}\ket\label{eq:S2-fourier-0}\\
|_{m}^{\ell}\ket & ={\displaystyle \int_{\S^{2}}}\diff\vh Y_{m}^{\ell}(\vh)|\vh\ket\,,\label{eq:S2-fourier-1}
\end{align}
\end{subequations}
where $Y_{m}^{\ell}(\vh)$ is a spherical harmonic. The momentum
states satisfy the normalization
\begin{equation}
\bra_{m}^{\ell}|_{m^{\pr}}^{\ell^{\pr}}\ket={\displaystyle \int_{\S^{2}}}\diff\vh Y_{m}^{\ell\star}(\vh)Y_{m^{\pr}}^{\ell^{\pr}}(\vh)=\d_{\ell\ell^{\pr}}\d_{mm^{\pr}}\,,
\end{equation}
where $\diff\vh$ is the surface area element on the two-sphere. In the momentum basis,
\begin{subequations}
\label{eq:S2-position-shifts}
\begin{align}
\r_{R} & =\sum_{\ell\geq0}\sum_{|m|\leq\ell}D_{mn}^{\ell\star}(R)|_{m}^{\ell}\ket\bra_{n}^{\ell}|\\
\p & =\sum_{\ell\geq0}\sum_{|m|\leq\ell}\left(-1\right)^{\ell}|_{m}^{\ell}\ket\bra_{m}^{\ell}|\,.
\end{align}
\end{subequations}
Other relevant features of $\S^{2}$ are listed in the third column
of Table~\ref{t:GoverH}.

The spherical harmonics form a basis for functions on the sphere,
meaning that any operator on $\S^{2}$ that is diagonal in the position
basis can be expanded in $Z$-type operators
\begin{equation}
\y_{m}^{\ell}=\int_{\S^{2}}\diff\vh\,|\vh\ket Y_{m}^{\ell}\left(\vh\right)\bra\vh|\,.\label{eq:S2-momentum-shift}
\end{equation}
However, since there are more rotations than molecular orientations,
products of rotations and the above diagonal $Z$-type operators do
not form an orthonormal basis for operators on $\S^{2}$. They instead
form an overcomplete frame, satisfying 
the completeness relation
in Eqs.~(\ref{eq:operator-basis}),
\begin{equation}
{\textstyle \frac{1}{2\pi}}\int_{\SO_{3}}\diff R \sum_{\ell,m}\bra\vh|\r_{R}\y_{m}^{\ell}|\wh\ket\bra\wh^{\pr}|\y_{m}^{\ell\dg}\r_{R}^{\dg}|\vh^{\pr}\ket=\d_{\vh\vh^{\pr}}^{\S^{2}}\d_{\wh\wh^{\pr}}^{\S^{2}}\,.
\end{equation}
A similar relation holds for more general quotient spaces, as described
in Appx.~\ref{appx:GH}. 
Overcompleteness complicates the analysis of recovery from errors of the form
\begin{equation}
\rho\to\r_{R}\y_{m}^{\ell}\rho\y_{m}^{\ell\dg}\r_{R}^{\dg}\,,
\end{equation}
in which a momentum kick by $\ell,m$ is combined with a rotation $R$.

\subsection{Simplest linear rotor codes}

Here we embed the $\Z_{N}\subset\Z_{2N}$ code (\ref{eq:zakforZ3}) for
general $N$ into the linear rotor. While their $\SO_{3}$ counterparts
allowed protection against small momentum and position shifts, these
codes can correct \textit{either} against rotations around any axis
by sufficiently small angles \textit{or} against $O(N/2)$ angular
momentum kicks.

\prg{Codewords}

Constructing the simplest linear rotor codes is similar to
that for $\SO_{3}$ in Sec.~\ref{subsec:abelian-mol-codes}. Codewords
are equal superpositions of equatorial states $|\frac{\pi}{2},\phi\ket$,
whose azimuthal angle $\phi$ is every even or odd multiple of $\frac{\pi}{N}$,
\begin{subequations}
\label{eq:ZN-on-S2-codewords}
\begin{align}
\kk{\overline{0}} & ={\textstyle \frac{1}{\sqrt{N}}}\sum_{h\in\Z_{N}}\kk{{\textstyle \frac{\pi}{2},\frac{2\pi}{N}h}}\label{eq:ZN-on-S2-codewords-0}\\
\kk{\overline{1}} & ={\textstyle \frac{1}{\sqrt{N}}}\sum_{h\in\Z_{N}}\kk{{\textstyle \frac{\pi}{2},\frac{2\pi}{N}h+\frac{\pi}{N}}}\,.\label{eq:ZN-on-S2-codewords-1}
\end{align}
\end{subequations}
For the case $N=3$, these codewords are shown in Fig.~\ref{fig:Zd-rotors}(c). 
These codewords are not normalizable, but normalizable approximate codewords can be obtained by introducing a damping factor, just as we discussed for $\SO_{3}$ codes [see  Sec.~\ref{subsec:Approximate-codewords} and Appx.~\ref{appx:Microwave-dressing}].

Expressing the codewords in terms of angular momentum states $|_{m}^{\ell}\ket$
(\ref{eq:S2-fourier-1}) yields, for $r\in\{0,1\}$,
\begin{equation}
|\overline{r}\ket=\sqrt{N}\sum_{\ell\geq0}\sum_{|pN|\leq\ell}\left(-1\right)^{pr}Y_{pN}^{\ell}\left({\textstyle \frac{\pi}{2}},0\right)|_{pN}^{\ell}\ket\label{eq:ZN-on-S2-momentum-basis}\,.
\end{equation}
To derive Eq.~(\ref{eq:ZN-on-S2-momentum-basis}), it suffices to observe that
\begin{align}
	Y_{m}^{\ell}\left({\textstyle \frac{\pi}{2}},\phi\right)=Y_{m}^{\ell}\left({\textstyle \frac{\pi}{2}},0\right)e^{im\phi}\,.
\end{align}
Therefore,  the only terms that survive when we do the sum over $h\in\Z_N$ in Eq.~(\ref{eq:ZN-on-S2-momentum-basis}) are those in which $m$ is an integer multiple of $N$. 

The logical-$X$ codewords are 
\begin{align}
    \label{eq:S2-logical-X}
   \kk{\overline{0}}_X &=\sqrt{2N}\sum_{\ell\geq0}\sum_{|2pN|\leq\ell}Y_{2pN}^{\ell}\left({\textstyle \frac{\pi}{2}},0\right)|_{2pN}^{\ell}\ket, \\
   \kk{\overline{1}}_X &=\sqrt{2N}\sum_{\ell\geq0}\sum_{|(2p+1)N|\leq\ell}Y_{(2p+1)N}^{\ell}\left({\textstyle \frac{\pi}{2}},0\right)|_{(2p+1)N}^{\ell}\ket;\notag
\end{align}
that is, $|\overline 0\ket_X$ is a superposition of angular momentum eigenstates with $m$ an even multiple of $N$, and $|\overline 1\ket_X$ is a superposition of states with $m$ an odd multiple of $N$. In addition, because $Y^\ell_m(\frac{\pi}{2},\phi)=0$ whenever $\ell-m$ is odd \cite{VMH}, only every other value of $\ell$ appears in the superposition for each fixed value of $m$ [see Fig.~\ref{fig:Zd-rotors}(c)].

\prg{Position shifts}

We will use the error-correction conditions \cite{Bennett1996,Knill1997}
(see also \cite{nielsen_chuang}, Thm. 10.1) to determine which errors
can be handled by our code. To be able to correct against some subset
of rotations, one should satisfy for all such correctable rotations $R,R^{\pr}$,
\begin{subequations}
\label{eq:QEC}
\begin{align}
\bbra{\overline{0}}\r_{R}^{\dg}\r_{R^{\pr}}\kk{\overline{0}} & =\bbra{\overline{1}}\r_{R}^{\dg}\r_{R^{\pr}}\kk{\overline{1}}\label{eq:QEC-0}\\
\bbra{\overline{0}}\r_{R}^{\dg}\r_{R^{\pr}}\kk{\overline{1}} & =0\,.\label{eq:QEC-1}
\end{align}
\end{subequations}
This product of rotations is just another $\SO_{3}$ rotation, $\r_{R}^{\dg}\r_{R^{\pr}}$,
rotating the equatorial ``necklace'' of constituent orientations
of our codewords to another great circle.

To satisfy (\ref{eq:QEC-0}), notice that if $N$ is odd, the codeword
$|\overline 1\ket $ consists of superpositions of all points antipodal to
those of the codeword $|\overline 0\ket$,
\begin{equation}
\kk{\overline{1}}=\p\kk{\overline{0}}\,.\label{eq:antipodal}
\end{equation}
Therefore, assuming odd $N$ from now on and remembering that inversion
commutes with all rotations,
\begin{equation}
\bbra{\overline{1}}\r_{R}^{\dg}\r_{R^{\pr}}\kk{\overline{1}}=\bbra{\overline{0}}\p\r_{R}^{\dg}\r_{R^{\pr}}\p\kk{\overline{0}}=\bbra{\overline{0}}\r_{R}^{\dg}\r_{R^{\pr}}\kk{\overline{0}}.
\end{equation}
With the antipodal assumption (\ref{eq:antipodal}), the first condition
(\ref{eq:QEC-0}) is satisfied for all $R\in\SO_{3}$.

The second condition (\ref{eq:QEC-1}) puts restrictions on where
the rotations can map the codewords.
To be concrete, consider the case $N=3$, depicted in Fig.~\ref{fig:Zd-rotors}(c). The codeword $|\overline 0\ket$ is a uniform superposition of three ``constitutent'' points on the equator of $\S^2$, which are marked by black balls in the figure. The codeword $|\overline 1\ket$ is likewise a uniform superposition of three constituent points, marked by white balls. For each constituent point there is a corresponding Voronoi cell, containing all points on $\S^2$ which are closer to that point than to any other constituent point (see Appx.~\ref{appx:voronoi}). Each such Voronoi cell is a spherical lune, a sliver of the sphere bounded by two lines of longitude separated by angle $\frac{\pi}{3}$; one of these cells is colored blue in Fig.~\ref{fig:Zd-rotors}(c). The  condition Eq.~(\ref{eq:QEC-1}) will surely be satisfied as long as the rotation $R^{-1}R^\pr$ maps each constituent point to a point in its Voronoi cell. 

\prg{Momentum kicks}

Just like rigid-rotor codes, linear-rotor codes protect against sufficiently small momentum kicks. Selection rules for addition of angular momenta dictate that a momentum kick operator $\y^\ell_m$  
maps a momentum state $|_{m^{\pr}}^{\ell^{\pr}}\ket$ to states $|_{M}^{L}\ket$
that satisfy 
\begin{align}
|\ell - \ell^\pr| \le L \le |\ell + \ell^\pr|, \quad M = m + m^\pr .
\end{align}
Because the codewords have support on states such that $m$ is an integer multiple of $N$, the code can detect momentum kicks with $\ell\leq N-1$  and correct shifts with $\ell<N/2$. The procedure for diagnosing and correcting momentum kicks follows closely the corresponding discussion for rigid-rotor codes.

\prg{Combined shifts}

We have now seen that the code with basis states Eq.~(\ref{eq:ZN-on-S2-codewords}) can protect against both small rotations and small angular momentum kicks. But problems arise when we consider errors that combine a rotation and a kick. Suppose for example that $\vh$ is a constituent point of the codeword $|\overline 0\ket$, hence $-\vh$ is a constituent point of $|\overline 1\ket$, and consider a rotation $R_{\o, \vh}$ about the axis $\vh$ by a small \textit{nonzero} angle $\o$. Then, because one and only one constituent point of each codeword is preserved by the rotation, we have 
\begin{align}
	\bbra{\overline{0}}\y_{m}^{\ell}\r_{R}\kk{\overline{0}}&={\textstyle \frac{1}{N}}Y_{m}^{\ell}(\vh),\\\bbra{\overline{1}}\y_{m}^{\ell}\r_{R}\kk{\overline{1}}&={\textstyle \frac{1}{N}}Y_{m}^{\ell}(-\vh)={\textstyle \frac{1}{N}}(-1)^{\ell}Y_{m}^{\ell}(\vh)\,.\notag
\end{align}
To be specific, $Y^1_1(\theta,\phi) \propto e^{i\phi} \sin\theta$ is nonzero for $\theta = \frac{\pi}{2}$, and we therefore conclude that $\bra\overline 0| \y^1_1 \r_R|\overline 0\ket\ne \bra\overline 1| \y^1_1 \r_R|\overline 1\ket$. This means that the error-correction condition is not satisfied by this code for this error. 

More generally, suppose that $\vh_1$ is a constituent point of $|\overline 0\ket$, and that $R$, $R^\pr$ are two rotations both of which map $\vh_1$ to another point $\vh_2$. (There is a one-parameter family of such rotations.) Suppose in addition that $R\uh\ne R\uh^\pr$, where $\uh,\uh^\pr$ are any other constituent points of $|\overline 0\ket$. Then
\begin{align}
    \label{eq:S2-QEC}
	\bbra{\overline{0}}\r_{R}^{\dagger}\y_{m}^{\ell}\r_{R^{\pr}}\kk{\overline{0}}&={\textstyle \frac{1}{N}}Y_{m}^{\ell}(\vh_{2}),\\\bbra{\overline{1}}\r_{R}\y_{m}^{\ell}\r_{R}\kk{\overline{1}}&={\textstyle \frac{1}{N}}Y_{m}^{\ell}(-\vh_{2})={\textstyle \frac{1}{N}}(-1)^{\ell}Y_{m}^{\ell}(\vh_{2}).\notag
\end{align}
Again, because for odd $\ell$ and nonzero $Y^\ell_m(\vh_2)$ we find that $ \bra\overline 0| \r_R^\dagger \y^\ell_m \r_{R^\pr}|\overline 0\ket \ne  \bra\overline 1| \r_R^\dagger \y^\ell_m \r_{R^\pr}|\overline 1\ket$, the error-correction conditions are not satisfied. 

Given the above limitations, this code can protect against either (I) All rotations $R\in\SO_3$ that keep each constituent orientation in its corresponding Voronoi cell, or (II) All momentum kicks $\y^\ell_m$ with $0\leq m \leq \ell < N/2$. In addition, if we exclude from set (I) all rotations around axes corresponding to constituent points of our logical states, then the code can correct both the rotations remaining in (I) and momentum kicks (II). However, the code still cannot correct products of such rotations and kicks due to Eq.~(\ref{eq:S2-QEC}). This is in contrast to rigid rotor codes, which protect against any \textit{product} of a sufficiently small rotation and momentum kick.

The above diminished performance begs the question of whether such codes are of any use against realistic noise \cite{Ramakrishna2005,Zhong2016,Stickler2016,Schmidt2015,Papendell2017,Stickler2018,Stickler2018a}. Since the rotations themselves are \textit{overcomplete}, and since these codes protect against (virtually) all small rotations, such codes may be applicable to certain  environments, especially ones where the noise is biased \cite{Aliferis2008}. These codes can also be concatenated with other codes, whose purpose would be to provide a layer of protection against momentum kicks. It is likely that the formalism of approximate error-correction \cite{Leung1997,Crepeau2005,codecomp} may be required to study their applicability. 

In a sense, the continuous $\U_1$-symmetry of the linear rotor is too much symmetry for such GKP-type codes to perform well. However, the framework presented here can serve as a springboard to designing codes for the many molecules with discrete symmetries --- less symmetric than the linear rotor, but not completely asymmetric like the rigid rotor (see Sec.~\ref{subsec:other-systems}). Configuration spaces of less symmetric molecules should make it possible for codes of this type to perform better (see Sec.~\ref{sec:Conclusion}).

\prg{Partial Fourier transform}
To construct a recovery for the above error sets, we can once
again develop a partially Fourier-transformed basis. As before,
we use subgroups $\H\subset\K$ to split up our underlying space $\X=\S^{2}$
into various pieces as
\begin{equation}
\S^{2}=\bigcup_{\wh\in\S^{2}/\K}\K\wh=\bigcup_{\wh\in\S^{2}/\K}~\bigcup_{r\in\K/\H}r\H\wh\,.\label{eq:symmetric-space-formulation}
\end{equation}
But because $\S^{2}$ is not a group, the first quotient space does
not consist of cosets, but instead consists of orbits in $\S^{2}$
under $\K$. The \textit{orbit} $\K\wh$ of a point $\wh\equiv(\theta,\phi)$
under $\K$ is the set of points to which one can get to by applying
rotations in $\K$ to $(\theta,\phi)$. Identifying points in
$\S^{2}$ belonging to the same orbit, one comes up with the \textit{orbi}t-mani\textit{fold}
(i.e., orbifold) $\S^{2}/\K$ (see Table~\ref{t:quaotient_spaces}). We construct the partially Fourier-transformed basis on $\S^2$ and its corresponding recovery for the codes discussed above in Appx.~\ref{appx:S2-recovery}. 

\prg{Gates \& check operators}

Check operators for the $\Z_{N}$ code on $\S^{2}$ are similar to
its counterpart on $\SO_{3}$: the $Z$-type check operator $\sz$ selects
the $2N$ orientations $\{|\frac{\pi}{2},\frac{\pi}{N}h+\m\pi\ket\}_{h\in\Z_{N},\m\in\Z_{2}}$,
while the $X$-type check operator $\sx$ has maximal eigenvalue only at the
two particular superpositions of these orientations {[}corresponding
to the two codewords (\ref{eq:ZN-on-S2-codewords}){]}. 

The momentum shift $\y_{2N}^{2N}$ (\ref{eq:S2-momentum-shift}) acts as the identity on the codespace, while the shift $\y_{N}^{N}$ acts as a logical-$Z$ operator. As with the rigid rotor, neither of these are unitary on the full Hilbert space. We can obtain simpler versions by using operators of the form \cite{Lachieze-Rey2004}
\begin{equation}
    \left(\hat{\vh}\cdot\wh\right)^{p}\equiv\int_{\S^{2}}\diff\vh\left(\vh\cdot\wh\right)^{p}|\vh\ket\bra\vh|\,,
\end{equation}
where $\wh\in\S^2$ and $p$ is a nonnegative integer. Expressing the ``position operator'' $\hat \vh$ in spherical coordinates yields the $\sz$ below,
\begin{subequations}
\label{eq:ZN-in-S2-stab}
\begin{align}
\sz & =\cos(2N\hat{\phi})\sin^{2N}\hat{\theta}\label{eq:ZN-on-S2-Sz}\\
\sx & =\cos\left({\textstyle \frac{2\pi}{N}}\hat{L}_{z}\right)\,.\label{eq:ZN-on-S2-Sx}
\end{align}
\end{subequations}
$X$-type check operators include powers of the $\zh$-axis rotation $\r_{\frac{2\pi}{N},\zh}=e^{-i\frac{2\pi}{N}\zh\cdot\hat{\boldsymbol{L}}}$,
where $h\in\Z_{N}$, $\hat{\boldsymbol{L}}=(\hat{L}_{x},\hat{L}_{y},\hat{L}_{z})$
is the angular momentum operator, and $\hat{L}_{z}|_{m}^{\ell}\ket=m|_{m}^{\ell}\ket$.
A combination of such powers yields Eq.~(\ref{eq:ZN-on-S2-Sx}) above.
Inversion $\p$ is a logical-$X$ operator.

\subsection{Nonabelian subgroup codes\label{subsec:Nonabelian-subgroup-codes-S2}}

\begin{figure}
\includegraphics[width=1\columnwidth]{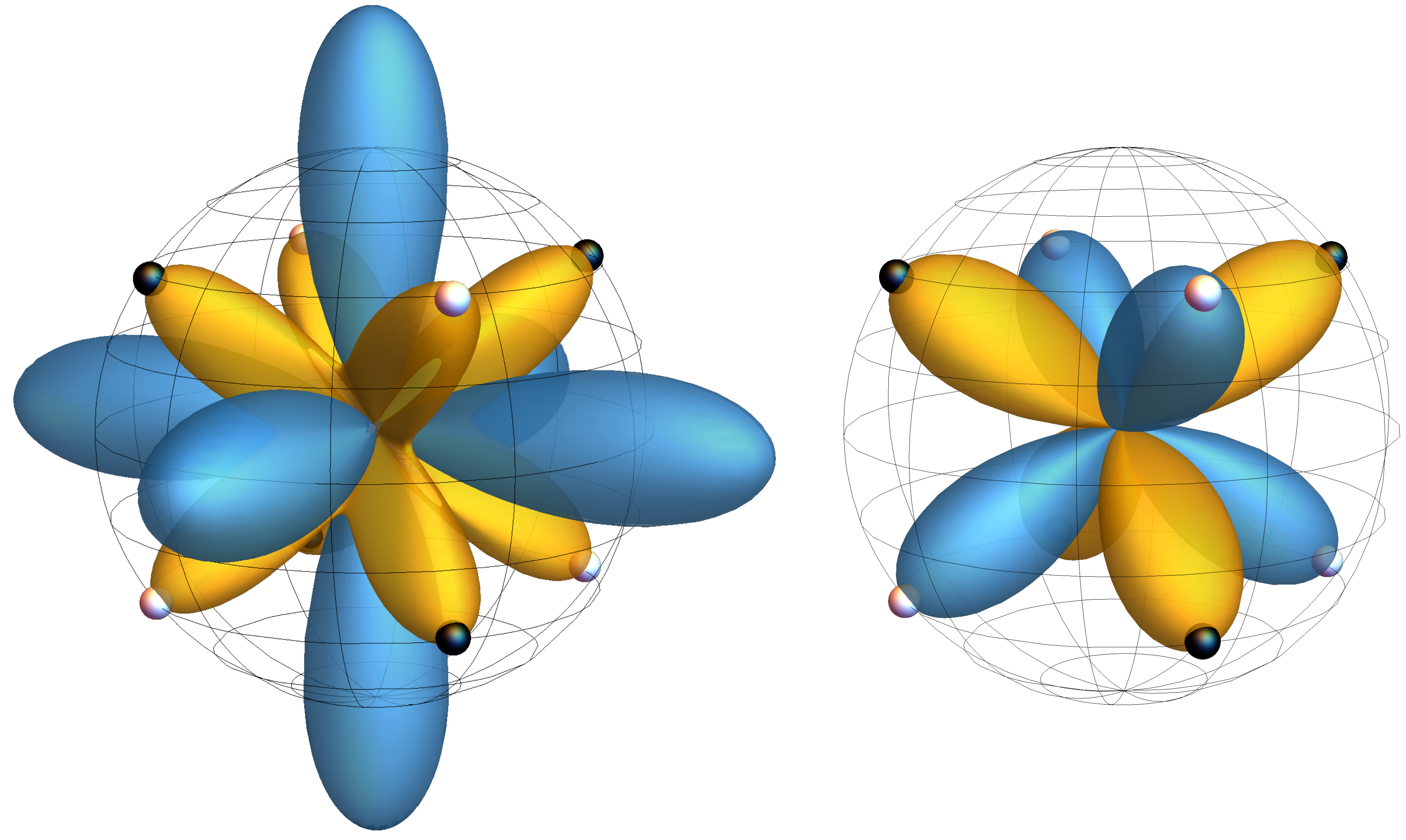}\caption{\label{fig:TinS2}
\textsc{Linear rotor codes. }Sketch of two polyhedral
harmonics for the $\protect\TT$ code on $\protect\S^{2}$, whose
two codewords are equal superpositions of the white and black points,
respectively. The left harmonic is the code's stabilizer $\protect\sz$
(\ref{eq:S2-stab-T-in-O}) and the right is the logical-$Z$ operator
(\ref{eq:S2-stab-T-in-O-1}). Positive (negative) values are in yellow
(blue), and the outlined spheres have radius 1.}
\end{figure}

Mimicking Sec.~\ref{subsec:Nonabelian-subgroup-codes}, we briefly discuss
more general codes based on nonabelian $\H\subset\K$. A simple example is $\H=\TT$ and $\K=\TT\times\Z_2^P$, where $\Z_2^P$ is the group generated by inversion $P$. Its codeword constituents lie on two antipodal tetrahedra that are invariant under $\TT$ (black and white points in Fig.~\ref{fig:TinS2}, respectively). Taken together, these tetrahedra make up a cube. Letting $\wh_{\text{cube}}$ be one of the vertices of the cube, we can express the codewords in terms of the orbit of $\wh_{\text{cube}}$ under $\TT$:
\begin{subequations}
\begin{align}
    \kk{\overline{0}}&=\frac{1}{6}\sum_{R\in\TT}\kk{R\wh_{\text{cube}}}\\\kk{\overline{1}}&=\frac{1}{6}\sum_{R\in\TT}\kk{-R\wh_{\text{cube}}}\,.
\end{align}
\end{subequations}
The normalization factor $1/6$ arises here because the 12 elements of $\TT$ map $\wh_{\text{cube}}$ to only 4 distinct constituents for each codeword. As before, these codewords are part of a partially Fourier-transformed basis associated with $\TT$, formulated in Appx.~\ref{appx:S2-recovery}. These $\TT$ codes correct against momentum shifts $\{\hat D^\ell_{mn}\}$ with $\ell \le 1$, and detect momentum shifts with $\ell \le 2$, like their counterparts on $\SO_3$ (see Sec.~\ref{subsec:Nonabelian-subgroup-codes}).

\prg{Check operators}

The $Z$-type check operators $\sz$ have the same eigenvalue at each of the constituent points of the code, which in the $\TT$ case means the corners of the cube in Fig.~\ref{fig:TinS2}. This condition is clearly satisfied by harmonics that are symmetric under $\K$, since that group leaves the cube invariant. In Appx.~\ref{appx:S2-recovery}, we describe how to obtain such harmonics by ``averaging'' or ``twirling'' the spherical harmonics over $\K$. Using this procedure, we obtain the $Z$-type check operator
\begin{align}
\sz & ={\textstyle \frac{3}{16}}\left(30\cos^{2}\hat{\theta}-35\cos^{4}\hat{\theta}-5\sin^{4}\hat{\theta}\cos4\hat{\phi}-3\right)\,,\label{eq:S2-stab-T-in-O}
\end{align}
corresponding to the $\K$-symmetric harmonic $Y_{0}^{4}$. Shown in the left panel of Fig.~\ref{fig:TinS2},
the above is normalized such that the constituent states $\{|\pm R\wh_{\text{cube}}\ket\}_{R\in\K}$
of the codewords are eigenstates with eigenvalue $+1$. 

Naturally, $\TT$-symmetric harmonics
can act as logical $Z$-type operators
within the codespace. The smallest-$\ell$ logical-$Z$ operator is
shown in the right panel of Fig.~\ref{fig:TinS2}, corresponding to the harmonic $Y_{2}^{3}$ averaged over $\TT$,
\begin{equation}
\zl={\textstyle \frac{3\sqrt{3}}{2}}\sin^{2}\hat{\theta}\cos\hat{\theta}\sin2\hat{\phi}\,.\label{eq:S2-stab-T-in-O-1}
\end{equation}

The $X$-type check operators $\sx$ consist of rotations $\{\r_{\o,\wh}\,,\,(\o,\wh)\in\TT\}$, commuting with all $\sz$ but not necessarily with each other
outside of the codespace. The group $\TT$ is generated by the rotations $(\frac{2\pi}{3},(\xh+\yh+\zh)/\sqrt{3})$ and $(\pi,\zh)$, so
\begin{subequations}
\label{eq:Stab-T-in-O}
\begin{align}
\sx^{(1)} & =\cos\left[{\textstyle \frac{2\pi}{3\sqrt{3}}}\left(\hat{L}_{x}+\hat{L}_{y}+\hat{L}_{z}\right)\right]\\
\sx^{(2)} & =\left(-1\right)^{\hat{L}_{z}}\,,
\end{align}
\end{subequations}
together with the $\sz$ check operator, are sufficient to identify the code space.

\prg{Relation to spherical designs}

There is a one-way connection between designs and momentum
kick detection. An $L$-design is a set of points $\PP\subset\S^2$
satisfying 
\begin{equation}
\int_{\S^{2}}\diff\vh\,f\left(\vh\right) = \frac{1}{|\PP|}\sum_{p\in\PP} f(p)\equiv f\left(\PP\right)\label{eq:design}
\end{equation}
for all polynomials $f$ of degree $\ell\leq L$. An $L$-design satisfies $f(\PP)=f(R\PP)$ for any rotation $R\in\SO_3$ [\citealp{Nebe2006}, Thm.~5.6.1] and any polynomial $f$ with degree $\le L$. Because of this property, and because spherical harmonics $\{Y_{m}^{\ell}\}$ are degree-$\ell$ polynomials restricted to the sphere, the states $|\PP\ket\propto\sum_{p\in\PP}|p\ket$ and $\r_R|\PP\ket$ form a code detecting $\leq L$ momentum kicks, where $R$ is any nontrivial rotation. 

Designs often arise as orbits of a group $\H$ acting on a particular point $\wh$, $\PP=\H\wh$. For example, the constituent orientations of each of the codewords of our $\TT\subset\TT\times\Z_2^{P}$ code form a two-design, and the $+1$ logical-$X$ state consists of all points on a cube and forms a 3-design. (Not all of our codes are designs: the equatorial sets of points making up our $\Z_{N}$ codewords detect momentum kicks, but do not form $N-1$-designs.) 

The connection to designs suggests a way to obtain other design-based codes, whose codewords are not based on a single orbit, or whose codewords make up more complicated polyhedra \cite{Hardin1996}. There is also a potentially interesting extension of oscillator-based error-correcting  codes based on designs \cite{Lacerda2016} to molecular state spaces.

\section{A qubit on a group\label{sec:A-qubit-on}}

In Sec.~\ref{sec:Rigid-rotor-codes} we described a family of quantum codes based on the nested subgroups $\H \subset \K\subset \SO_3$. In this section, we generalize this construction. The basic framework was already discussed in Sec.~\ref{subsec:dihedral-codes}.
We formulate quantum codes based on  $\H\subset \K\subset \G$  using a symbolic decomposition of $\G$ defined by a partial Fourier transform:
\begin{equation}
\G\cong\G/\K\times \K/\H \times  \widehat{\H}\,.
\end{equation}
We interpret elements of $\G/\K$ as correctable rotation errors and elements of $ \widehat{\H}$ as correctable momentum kick errors, while elements of $\K/\H$ correspond to basis states which span the code space. Data for these codes are summarized in Table~\ref{t:G-summary}. 

We consider error-correcting codes for quantum systems whose canonical
position basis $\{|g\ket\}$ corresponds to elements of a group, $g\in\G$.
Such spaces admit generalized versions of many of the features of
standard quantum mechanical spaces such as qubits or oscillators:
position and momentum bases, their corresponding shifts, orthogonality
relations, a Weyl-type relation, etc. We have collected these in Table~\ref{t:G},
intending it to be an extension of an analogous table {[}\citealp{Albert2017},
Table~1{]} for the standard spaces. 

The position/momentum bases for general $\G$ can be discrete/discrete (e.g., for qudit spaces $\G=\Z_D^{\times n}$), continuous/discrete (for rotors $\G\in\{\U_1,\SO_3\}$), or continuous/continuous (for oscillators $\G=\R$). These differences obscure the intuition we are trying to convey, so we keep $\G$ finite for clarity here. The caption of Table~\ref{t:G} adapts these discussions to other $\G$, and Ref.~\cite{Justel2018} rigorously formulates many of the required tools for type I unimodular second-countable groups.

Most of the structure for general $\G$ is already present for $\G=\SO_3$, which we outlined in Sec.~\ref{sec:Rigid-rotor-codes}. Position shifts $\ru$ for general $\G$ are represented by left multiplication, $\ru_h \kk{g}=\kk{hg}$, and analogous shifts $\lu$ exist for right multiplication. ``Momentum'' kick operators are diagonal in position space, acting as [\citealp{Brell2015}, Eq.~(5)]
\begin{equation}
    \zz^\ell_{mn}\kk{g}=Z^\ell_{mn}(g)\kk{g}\,,
\end{equation}
where $Z^\ell_{mn}(g)$ is the $m,n$th matrix element of the group element $g$ in the irrep $\ell$. These matrix elements are part of the momentum basis for $\G$:
\begin{equation}
    |_{mn}^{\ell}\ket={\displaystyle \sum_{g\in\G}}\textstyle{\sqrt{\frac{d_{\ell}}{|\G|}}}Z_{mn}^{\ell}(g)|g\ket\,.
\end{equation}
We collect all $^\ell_{mn}$ into $\widehat{\G}$, the ``dual space'' of $\G$.

Products of position shifts and momentum kicks, $\hat{B}_{g}^{\ell mn}=\sqrt{\frac{d_{\ell}}{|\G|}}\zz_{mn}^{\ell}\ru_{g}$,
form an orthonormal and complete basis for operators on $\G$,
\begin{subequations}
\label{eq:operator-basis}
\begin{align}
\tr\big(\hat{B}_{g}^{\ell mn\dg}\hat{B}_{g^{\pr}}^{\ell^{\pr}m^{\pr}n^{\pr}}\big)  =\d_{gg^{\pr}}^{\G}\d_{\ell\ell^{\pr}}\d_{mm^{\pr}}\d_{nn^{\pr}}\\
\!\!\!\!\!\!\!\!\!\!\!\sum_{g\in\G}\sum_{\ell mn\in\widehat{\G}}\bra h|\hat{B}_{g}^{\ell mn}|k\ket\bra k^{\pr}|\hat{B}_{g}^{\ell mn\dg}|h^{\pr}\ket  =\d_{hh^{\pr}}^{\G}\d_{kk^{\pr}}^{\G}\,,
\end{align}
\end{subequations}
where $\tr(\cdot)=\sum_{g\in\G}\bra g|(\cdot)|g\ket$ and $h,h^\pr,k,k^\pr\in\G$. Thus, any physical noise channel $\cal{E}$ acting on this space can be expanded in terms of this operator basis, as before (\ref{eq:U1-channel},\ref{eq:SO3-channel}). The purpose of our codes is to protect against ``small'' position shifts as well as certain momentum shifts.

\begin{table}
\begin{tabular}{lc}
\toprule 
Code & $\H\subset\K\subset\G$\tabularnewline
\midrule
Part. Fourier basis (\ref{eq:zakG})\phantom{$\{\ru_{k}\}$} & $\{|\cs\H;{}_{\m\n}^{\l}\ket\,,\,\cs\in\F_{\G/\H}\,,\,_{\m\n}^{\l}\in\widehat{\H}\}$\tabularnewline
Logicals (\ref{eq:codewords-G})\phantom{$\{\ru_{k}\}$} & $\{|r\H;{}_{00}^{\one}\ket\,,\,r\in\F_{\K/\H}\}$\tabularnewline
Corr. position shifts\phantom{$\{\ru_{k}$} & $\F_{\G/\K}$\tabularnewline
Corr. momentum kicks\phantom{$\{\ru_{k}$} & Use branching formulas\tabularnewline
Check operators $\sz$ (\ref{eq:G:check-ops})\phantom{$\{\ru_{k}$} & $\zz_{mn}^{\ell}\left(\K\right)$\tabularnewline
$Z$-type logicals\phantom{$\{\ru_{k}$} & $\zz_{mn}^{\ell}\left(\H\right)$\tabularnewline
Check operators $\sx$\phantom{$\{\ru_{k}$} & $\{\ru_{k}\lu_{h}\,,\,k\in\K^\pr\,,\,h\in\H\}$\tabularnewline
$X$-type logicals\phantom{$\{\ru_{k}$} & $\{\ru_{k}\,,\,k\in\K\}$\tabularnewline
\bottomrule
\end{tabular}

\caption{\label{t:G-summary}List of elements of a $\protect\H\subset\protect\K$
code on $\protect\G$ from Sec.~\ref{subsec:abelian-mol-codes}. The set $\widehat{\protect\H}$
consists of (equivalence classes) of all irreps of $\protect\H$,
$\protect\F_{\protect\G/\protect\K}$ is the Voronoi cell of the identity (see Appx.~\ref{appx:voronoi}), and $\protect\K^{\protect\pr}\subset\protect\K$ consists
of all elements of $\protect\K$ that map to identity when projected
onto the logical subspace.}
\end{table}

\prg{Partial Fourier transform}

Our code constructions make use of the partial Fourier transform on $\G$, whose states are parameterized by cosets in $\G/\H$ and $\H$-irreps,
\begin{equation}
\kk{\cs\H;{}_{\m\n}^{\l}}={\textstyle \sqrt{\frac{d_{\l}}{|\H|}}}\sum_{h\in\H}\z_{\m\n}^{\l}\left(h\right)\kk{\cs h}\,.\label{eq:zakG}
\end{equation}
Above, $\cs$ belongs to the coset space $\G/\H$, which we parameterize using $\F_{\G/\H}$, the Voronoi cell of the identity (see Appx.~\ref{appx:voronoi}). The coefficient $\z_{\m\n}^{\l}(h)$ is the $\m,\n$th matrix element of the $d_{\l}$-dimensional irrep $\l$ of $\H$, evaluated for the element $h\in\H$. We use the Greek letters $\lambda, \mu,\nu$ to label matrix elements of irreps of $\H$, and save the letters $\ell, m, n$ for labeling matrix elements of irreps of $\G$. 

The above basis interpolates
between the group's position states ($\H=\{1\}$) and momentum states ($\H=\G$). 
One can show that it is orthonormal and complete:\footnote{To show the above, we use $g=\cs h$ with $\cs\in\F_{\G/\H}$ and
$h\in\H$ for each $g\in\G$, $\bra\cs h|a^{\pr}h^{\pr}\ket=\d_{aa^{\pr}}^{\G/\H}\d_{hh^{\pr}}^{\H}$,
and Table~\ref{t:G}.E. We define generalized delta functions $\d_{xy}^{\X}$ for a space $\X$, satisfying $\frac{1}{|\X|}\sum_{y\in\X}f(y)\d_{xy}^{\X}=f(x)$ for $x\in\X$.}
\begin{subequations}
\begin{align}
~~~~~~~~~~~\big<\cs\H;{}_{\m\n}^{\l}\big|\cs^{\pr}\H;{}_{\m^{\pr}\n}^{\l^{\pr}}\big>  =\d_{aa^{\pr}}^{\G/\H}\d_{\l\l^{\pr}}\d_{\m\m^{\pr}}\d_{\n\n^{\pr}}\\
\!\!\!\!\!\!\!\!\!\!\!\!\!\sum_{a\in\G/\H}\sum_{\l\m\n\in\widehat{\H}}\big<g\big|\cs\H;{}_{\m\n}^{\l}\big>\big<\cs\H;{}_{\m\n}^{\l}\big|g^{\pr}\big>  =\d_{gg^{\pr}}^{\G}\,.
\end{align}
\end{subequations}
This basis arises in several other contexts in science and engineering, which we discuss in Appx.~\ref{appx:notable-examples}.

\prg{Codewords}

Our codewords correspond to cosets of $\H$ in $\K$. For $r\in\F_{\K/\H}$ and $\l=\one$ the trivial irrep,
\begin{equation}
\kk{\overline{r}}\equiv\kk{r\H;{}_{00}^{\one}}={\textstyle\frac{1}{\sqrt{|\H|}}}\sum_{h\in\H}|rh\ket~.
\label{eq:codewords-G}
\end{equation}
Expressing the $\G$ position states 
(\ref{eq:codewords-G}) in terms of momentum states yields 
\begin{equation}
\kk{\overline{r}}={\displaystyle \sum_{\ell mn\in\widehat{\G}}}{\textstyle \sqrt{\frac{d_{\ell}}{|\G/\H|}}}Z_{mn}^{\ell\star}(r\H)|_{mn}^{\ell}\ket\,.\label{eq:G-basis-dual}
\end{equation}
Here we have introduced the notation $f(\H)$ for the \textit{$\H$-average} (also called \textit{$\H$-twirl}) of a function $f$ on $\G$ over the subgroup $\H$, defined by 
\begin{equation}
f(\H)\equiv\frac{1}{|\H|}\sum_{h\in\H}f(h)\,.\label{eq:twirl-def}
\end{equation}

Observing that $Z_{mn}^{\ell\star}(r\H)=\left(\Z^{\ell\star}(r)Z^{\ell\star}(\H)\right)_{mn}$, we see that the momentum state $|_{mn}^{\ell}\ket$ ``participates'' in the expansion of $\kk{\overline{r}}$ (occurs with a nonzero coefficient) only if $Z^{\ell}(\H) \ne 0 $.
Irreps with this property make up the \textit{reciprocal space of} $\H$ \textit{in} $\G$,
\begin{equation}
\H^{\perp}\equiv\left\{ \ell\in\widehat{\G}\,,\,Z^{\ell}(\H)\neq0\right\} \,.\label{eq:reciprocal-space}
\end{equation}
For each $\ell$, we also have to determine the participating $m,n$ indices; these depend on the basis used for $Z^\ell$ (see Appx.~\ref{appx:GH}).

Denoting the set of participating momentum-state indices $^\ell_{mn}$ by $\widehat{\G/\H}$, Eq.~(\ref{eq:G-basis-dual}) becomes
\begin{equation}
\kk{\overline{r}}={\displaystyle \sum_{\ell mn\in\widehat{\G/\H}}}{\textstyle \sqrt{\frac{d_{\ell}}{|\G/\H|}}}Z_{mn}^{\ell\star}(r\H)|_{mn}^{\ell}\ket\,.\label{eq:codewords-G-dual}
\end{equation}
We have thus mapped the position degree of freedom $r$ from the ket $|rh\ket$ of the position-basis expansion (\ref{eq:codewords-G})
into the coefficient $Z_{mn}^{\ell\star}(r\H)$ of the momentum-basis expansion (\ref{eq:codewords-G-dual}).
This can be done for any coset state $|\cs\H;^{\one}_{00}\ket$ with $\cs\in\G/\H$ --- a manifestation of the Fourier transform on $\G/\H$ (see Appx.~\ref{appx:GH}). Analogously, one can develop a Fourier transform on the codespace $\K/\H$.

\prg{Position shifts}
Position shifts acting from the left realize an induced representation
for each $\l$, meaning that $\{\ru_{g}\}_{g\in\G}$ do not connect
different $\l$'s. The difference from the abelian
case (\ref{eq:inducedrepZninSO3}) is the behavior of the internal
indices $\m\n$,
\begin{equation}
\ru_{g}\kk{\cs\H;{}_{\m\n}^{\l}}=\sum_{\rho}\z_{\rho\m}^{\l\star}(k_{g})\kk{gak_{g}^{-1}\H;{}_{\rho\n}^{\l}}\,.\label{eq:position-shifts-acting-on-coset-states}
\end{equation}
Above, the compensating element $k_{g}\in\H$
is picked such that $gak_{g}^{-1}\in\F_{\G/\H}$.

Let us determine the set of
correctable position shifts $\ru_g$. First consider $g\in\F_{\G/\K}$, in which case there is no compensating element. Then, the error state obtained from applying a momentum kick and position shift to the  codeword $|\overline{r}\ket$ will consist of a superposition of the basis elements (\ref{eq:zakG}) with $\cs=gr$. By measuring a rotation in $\F_{\G/\K}$ and applying the corresponding position shift, the recovery will map each $gr$ to the element $r^\pr\in\K/\H$ whose Voronoi cell contains $gr$. (The partitioning into cosets ensures that the Voronoi cell of each $r^{\pr}\in\K/\H$ will contain only one $gr$.) Since $g\in\F_{\G/\K}$, $gr$ will be in the Voronoi cell of $r$, so $r^\pr=r$. After recovery, each $r$ will return to its original location.

Now consider $g\notin\F_{\G/\K}$. Now, the corrupted position label corresponding to codeword $r$ can stray into the the Voronoi cell of some other element $r^{\pr}\neq r$. The above recovery will snap such error words to the wrong codewords, leading
to logical errors. In the case of nonabelian codes, there may be errors due to the effect of the compensating element on $\m\n$ (\ref{eq:position-shifts-acting-on-coset-states}).

\prg{Momentum kicks}

Assuming we have exactly corrected a position shift, the resulting
state lies in the span of $\{|r\H;{}_{\m\n}^{\l}\ket\}$ for all $r\in\K/\H$
and $_{\m\n}^{\l}\in\widehat{\H}$.
Below, we show how to use the branching formulas for $\G$ restricted to $\K$, and then $\K$ restricted to $\H$, to determine detectable and correctable momentum kicks $\zz^\ell_{mn}$. We leave the precise formulation of a momentum-kick recovery for general $\G$ to future work.

Let $\ell\in\widehat{\G}$, $\kappa\in\widehat{\K}$
and $\l\in\widehat{\H}$, so that $\ell\to\kappa\to\l$ means that $\ell$ contains at least one copy of the $\K$-irrep $\kappa$ when restricted to $\K$, which in turns contains at least one copy of $\l$ when restricted to $\H$. We denote the trivial irrep by $\one$. For convenience,
we assume that $Z^{\ell}$ are written in a $\K$-admissible basis
(see Appx.~\ref{appx:GH}), meaning that $Z^{\ell}(k)$ for $k\in\K$
are block-diagonal with respect to the $\K$-irreps. Similarly, we
assume that those $Z^{\ell}(k)$ are in turn in an $\H$-admissible basis.

First consider detectable errors. Let $\ell$ be such that one of
its branches is $\ell\to\kappa\to\one$ with $\kappa\neq\one$.
Then, there exists a momentum kick $\zz_{mn}^{\ell}$ for some $m,n$
that is undetectable. To prove this, consider matrix-valued versions
of the error-correction criteria, projecting the matrix of $\G$-mometum
shifts $\zz^{\ell}$ into the codespace (\ref{eq:codewords-G}),
\begin{equation}
\bbra{\overline{r}}\zz^{\ell}\kk{\overline{r}}=Z^{\ell}\left(r\H\right)=Z^{\ell}\left(r\right)Z^{\ell}\left(\H\right)\,.\label{eq:G-detecting-momentum-shifts}
\end{equation}
We will show that the above
depends on $r$. 

Since $Z^{\ell}\left(h\right)$ is in an $\H$-admissible basis, we can express each $Z^{\ell}\left(h\right)$ in
$Z^{\ell}(\H)\propto\sum_{h\in\H} Z^{\ell}\left(h\right)$ as a direct sum of irreps of $\H$. By the group orthogonality relations on $\H$, $Z_{mn}^{\ell}\left(\H\right)=\d_{mn}$
only for those $n$ which correspond to matrix elements of $h$ in
the trivial $\H$-irrep. We have assumed that $\ell$ branches to
at least one trivial irrep of $\H$, so there exists such an $n$, which we call $n_\star$.
{[}This implies that $\ell\in\H^{\perp}$ (\ref{eq:reciprocal-space}), a necessary but not sufficient condition for undetectability.{]} Now, consider the
column $Z_{mn_\star}^{\ell}\left(r\H\right)$, with $m\in\{1,2,\cdots,d_{\kappa}\}$ and
$d_{\kappa}$ being the dimension of the $\kappa$ irrep that contains the trivial irrep $\l=\one$. 
When $r=1$ (the identity),
$Z_{mn}^{\ell}\left(r\H\right)=\d_{mn}$. But since $\kappa\neq\one$,
there exists another $r^{\pr}\neq1$ that is represented differently.
Thus, $Z_{mn}^{\ell}\left(r^{\pr}\H\right)\neq\d_{mn}$, and Eq.~(\ref{eq:G-detecting-momentum-shifts})
depends on $r$.

Now consider correctable errors. Let $\ell\neq\ell^{\pr}$ be such that
they branch to the same nontrivial $\H$-irrep via \textit{different} nontrivial $\K$-irreps
$\kappa\neq\kappa^{\pr}$, i.e., $\ell\to\kappa\to\l$
and $\ell^{\pr}\to\kappa^{\pr}\to\l$, respectively. Then, momentum kicks $\zz^{\ell}$ and $\zz^{\ell^{\pr}}$ will not
be simultaneously correctable. To prove this, let $m,n$ be the matrix
elements of the copy of $\l$ contained in $\kappa$, and $m^{\pr},n^{\pr}$
be those for the copy of $\l$ contained in $\kappa^{\pr}$. Then
\begin{subequations}
\begin{align}
\zz_{mn}^{\ell}\kk{\overline{r}} & ={\textstyle\frac{1}{\sqrt{|\H|}}} \sum_{h\in\H}\sum_{p=1}^{d_{\l}}\z_{mp}^{\kappa}(r)\z_{pn}^{\l}(h)|rh\ket\\
 & ={\textstyle\frac{1}{\sqrt{d_{\l}}}}\sum_{p=1}^{d_{\l}}\z_{mp}^{\kappa}(r)|r\H;{}_{pn}^{\l}\ket\,,
\end{align}
\end{subequations}
where we have used the basis (\ref{eq:zakG}) in the second line. 
Using orthogonality of this basis,
\begin{equation}
\bbra{\overline{r}}\zz_{m^{\pr},n^{\pr}}^{\ell^{\pr}\dg}\zz_{m,n}^{\ell}\kk{\overline{r}}=\frac{\d_{nn^{\pr}}}{d_{\l}}\sum_{p=1}^{d_{\l}}\z_{mp}^{\kappa}(r)\z_{pm^{\pr}}^{\kappa^{\pr}}(r^{-1})\,.
\end{equation}
When $r=1$, the sum over $p$ reduces to $\d_{mm^{\pr}}$. But since $\kappa\neq\kappa^{\pr}$, there exists an $r^{\pr}\neq1$ such that the above yields a different result. Therefore, one cannot
correct both $\ell$ and $\ell^{\pr}$ momentum kicks.

\prg{Gates \& check operators}

Logical $X$-type gates include all $\{\ru_{k}\}_{k\in\K}$,
which realize an induced representation on the logical subspace $\K/\H$. A subset of those, which we call $\K^{\pr}\subset\K$, acts as the identity in this induced representation; such operators can be used as check operators for momentum-kick syndrome measurement. Examples of such representations are discussed in Sec.~\ref{subsec:Nonabelian-subgroup-codes}.

The position shifts $\{\lu_h\}_{h\in\H}$ also act trivially on the codewords. These do not commute
with each other for nonabelian $\H$, but do commute with $\ru_k$ (since left- and right-multiplication commute).
These can also be used as check operators, and the resulting combined set of $X$-type check operators is listed in Table~\ref{t:G-summary}. 

If $\H$ is a normal subgroup of $\K$, then the shifts $\{\lu_k\}_{k\in\K}$ also realize logical gates; otherwise, such shifts may not preserve the code subspace (since left and right coset spaces, $\K/\H$ and $\H\textbackslash\K$, are not equal). For $\G=\SO_3$, such cases include $\Z_N\subset\Z_{2N}$ and $\TT\subset\OO$, but not $\TT\subset\II$.

Twirling momentum kick operators over $\K$, 
\begin{equation}
\label{eq:G:check-ops}
\zz_{mn}^{\ell}\left(\K\right)\equiv{\textstyle\frac{1}{|\K|}}\sum_{k\in\K}\lu_{k}\zz_{mn}^{\ell}\lu_{k}^{\dg}\,,
\end{equation}
offers a convenient method for generating $Z$-type check operators $\sz$. The above operators are functions on $\G/\K$:
\begin{equation}
\zz_{mn}^{\ell}\left(\K\right)\kk{wr\H;_{\m\n}^{\l}}=Z_{mn}^{\ell}\left(w\K\right)\kk{wr\H;_{\m\n}^{\l}}
\end{equation}
for $w\in\G/\K$ and $r\in\K/\H$. Thus, measuring them does not spoil the logical information. A projective measurement onto the basis of the joint eigenstates of these mutually commuting operators can be used to determine the syndrome $w$. (The scheme outlined below implicitly performs such a measurement.) Such $\sz$ commute with each other and all $\{\lu_k\}_{k\in\K}$, but only commute with $\{\ru_{k}\}_{k\in\K}$ on the subspace $\{|r\H;{}_{\m\n}^{\l}\ket\}$ with $r\in\K/\H$ and $_{\m\n}^{\l}\in\widehat{\H}$. 
Since they can be nonunitary, they do not in general form a group.

Twirling momentum kicks over $\H$ produces logical $Z$-type operators. A similar procedure yields $Z$-type check and logical operators for codes on $\S^2$ (see Appx.~\ref{appx:S2-recovery}).

\prg{Measurements}

Recall that, given a subgroup $\K$, each group element $g\in\G$
can be written as $g=\cs k$ for $\cs\in\F_{\G/\K}$ and $k\in\K$.
In order to diagnose which position shift occurred without destroying
the logical information, one needs to read off the coset label $\cs$
without obtaining information about $k$. 
Since there are only $|\G/\K|$ different
values one needs to distinguish, we can pick the ancillary space to be $\G/\K$ (instead of the larger $\G$) and still measure in one shot. 

Letting $\r_{g}$ be the induced representation of $\G$ on $\G/\K$, we apply the generalized $\crot$ gate (cf. \cite{Brell2015})
\begin{equation}
\crot_{\G/\K}=\sum_{g\in\G}|g\ket\bra g|\otimes\r_{g}
\end{equation}
onto the $\G$-space housing our logical information and an ancilla initialized in some state $|\K\ket$ (assumed invariant under $\{\r_{k}\}_{k\in\K}$). Since $\r_{g}=\r_{\cs}\r_{k}$ and since
the initial state is $\K$-invariant, the ancilla will obtain only
the coset label $\cs$, without destroying coherences between elements of the coset. This procedure can also be used for logical state initialization.

The space $\G/\K$ can be ``simulated''
by a finite-dimensional space spanned by generalized spin-coherent states \cite{perelomov_book}, similar to our construction from Sec.~\ref{subsec:msmnts}. 

\section{Conclusion \& future work\label{sec:Conclusion}}

We have developed error-correcting codes that protect against
small shifts in the position and momentum of a rigid body and, more
generally, of a state space $\{|g\ket\,,\,g\in\G\}$ where $\G$ is a group. Our treatment unifies CSS codes ($\G=\Z_{D}^{\times n}$)
with GKP codes for qudits ($\G=\Z_{N}=\C_N$), oscillators ($\G=\R$), and planar
rotors ($\G=\U_1=\SO_2=\C_{\infty}$ or $\G=\Z$). We propose using our rigid-body codes, for which $\G$ is the 3-dimensional proper rotation group $\SO_3$, to robustly encode quantum information in the rotational states of asymmetric molecules. 

We also constructed related codes that protect a linear rotor, whose configuration space $\S^2$ is a coset space rather than a group, and we formulated position and momentum bases, their associated shifts, and orthogonality relations for general coset spaces. 

A basis may be chosen for a rigid-body code space such that each basis state is a uniform superposition of a finite number of possible orientations for the body. Because position eigenstates in a continuous-variable system are not normalizable, the ideal codewords are likewise not normalizable and have infinite energy. But we may instead choose normalizable, finite-energy approximate codewords which retain good error-correction properties.

Our coding scheme has potential applications to polar molecules, certain spin systems, atomic ensembles, single atoms, and levitated nanoparticles. We now mention several possible topics for future investigation.

\prg{Physical noise}

Our codes are designed to protect against noise that acts ``locally in phase space.'' For CSS codes the correctable errors are low-weight Pauli operators acting on a few qubits. For GKP oscillator codes, the correctable errors are small shifts in the position or momentum of the oscillator. For rigid-rotor codes, the correctable errors are small shifts in the rotor's orientation or small kicks in its angular momentum.

Physical noise may act nonlocally in phase space. But it has recently been shown that the dominant noise in microwave cavities is sufficiently local for GKP codes to work effectively \cite{codecomp,Noh2018}. It remains to be seen whether the noise in realistic rigid rotors \cite{Ramakrishna2005,Zhong2016,Stickler2016,Schmidt2015,Papendell2017,Stickler2018,Stickler2018a} has similarly benign properties.

\prg{Symmetric molecules}

For a molecule with a symmetry group $\H\subset\SO_3$, the configuration space of molecular orientations is the coset space $\SO_3/\H$. A larger symmetry group means a smaller configuration space, and thus less room for diagnosing rotation errors. In the extreme case of a perfect sphere, invariant under any $\SO_3$ rotation, there is no room in the one-dimensional space $\SO_3/\SO_3$ for any logical information at all. 
It would be interesting to investigate further how the performance of our codes depends on the symmetry group. Of particular interest are $\Z_3$-symmetric molecules (such as monomethoxides \cite{Yu2019} or Posner molecules \cite{Fisher2015,YungerHalpern2019}), which are invariant under rotations by $\pm 120$ degrees. 

\prg{Nuclear motion}

Instead of considering rigid molecules that are assumed to be in a fixed vibrational state, one can also consider ``floppy'' molecules for which there is no clear separation between rotational and vibrational motion. Such motion ranges from small nuclear vibrations around equilibrium positions \cite{browncarrington} to larger-scale bending motion \cite{Iachello2003} and even nuclear permutations \cite{Schmiedt2016}. To devise codes that protect quantum information carried by (for example) floppy molecules, we will need to consider different configuration spaces than for the rigid-body codes described here. Nevertheless, some of the mathematical tools we have developed may be applicable in this broader context. 

\begin{acknowledgments}
We thank Stephen Bartlett, Juani Bermejo-Vega, Giacomo Bighin, Igor N. Cherepanov, David DeMille, Steven M.~Girvin, Alexey V.~Gorshkov, Nick Hutzler, Joe Iverson, Arian Jadbabaie, Dominik J\"{u}stel, Alexei Yu.~Kitaev, Roman Korol, Roman Krems, Richard Kueng, Mikhail Lemeshko, Angelo Lucia, Kang-Kuen Ni, Igor Pak, Hannes Pichler, Gil Refael, Grant Salton, Eugene Tang, and Jun Ye for useful suggestions and discussions. We gratefully acknowledge support from ARO-LPS, NSF, the Walter Burke Institute for Theoretical Physics, and the Division of Physics, Mathematics, and Astronomy (PMA) at Caltech.
The Institute for Quantum Information and Matter is an NSF Physics Frontiers Center. Our figures were drawn using \textsc{Mathematica} 12.
\end{acknowledgments}

\appendix

\begin{table*}
\begin{tabular}[t]{>{\raggedright}m{1.4in}cc}
\toprule 
 & Finite group $\mathscr{L}^{2}(\G)$ & Rigid rotor $\mathscr{L}^{2}(\SO_{3})$\tabularnewline
\midrule 
A. ``Phase space'' & $\left(g,\,{}_{mn}^{\ell}\right)\in\G\times\widehat{\G}$ & $\left(R,\,{}_{mn}^{\ell}\right)\in\SO_{3}\times\widehat{\SO_{3}}$\tabularnewline
\midrule 
\addlinespace
\multirow{2}{1.4in}{B. Conjugate bases} & $\,\,\,\,\,\,\,\,\,\,\,\,\,\,\,\,\,|g\ket={\displaystyle \sum_{\ell mn\in\widehat{\G}}}\sqrt{\frac{d_{\ell}}{|\G|}}Z_{mn}^{\ell\star}(g)|_{mn}^{\ell}\ket$ & $|R\ket={\displaystyle \sum_{\ell\geq0}\sum_{|m|,|n|\leq\ell}}\sqrt{\frac{2\ell+1}{8\pi^{2}}}D_{mn}^{\ell\star}(R)|{}_{mn}^{\ell}\ket$\tabularnewline
\addlinespace
 & $|_{mn}^{\ell}\ket={\displaystyle \sum_{g\in\G}}\sqrt{\frac{d_{\ell}}{|\G|}}Z_{mn}^{\ell}(g)|g\ket$ & $\!\!\!\!\!\!\!\!\!\!\!\!\!|_{mn}^{\ell}\ket={\displaystyle \int_{\SO_{3}}}\diff R \sqrt{\frac{2\ell+1}{8\pi^{2}}}D_{mn}^{\ell}(R)|R\ket$\tabularnewline
\midrule
\addlinespace
C. Overlap & $\bra g|{}_{mn}^{\ell}\ket=\sqrt{\frac{d_{\ell}}{|\G|}}Z_{mn}^{\ell}(g)$ & $\bra R|{}_{mn}^{\ell}\ket=\sqrt{\frac{2\ell+1}{8\pi^{2}}}D_{mn}^{\ell}(R)$\tabularnewline
\midrule 
\addlinespace
D. ``Resolution'' & ${\displaystyle \sum_{g\in\G}|g\ket\bra g|=\sum_{\ell mn\in\widehat{\G}}|_{mn}^{\ell}\ket\bra{}_{mn}^{\ell}|=\id_{\G}}$ & ${\displaystyle {\displaystyle \int_{\SO_{3}}}\diff R |R\ket\bra R|={\displaystyle \sum_{\ell\geq0}\sum_{|m|,|n|\leq\ell}}|{}_{mn}^{\ell}\ket\bra{}_{mn}^{\ell}|=\id_{\SO_3}}$\tabularnewline
\midrule 
\addlinespace
E. ``Orthocompleteness'' & ${\displaystyle \sum_{g\in\G}}Z_{mn}^{\ell\star}(g)Z_{m^{\pr}n^{\pr}}^{\ell^{\pr}}(g)=\frac{|\G|}{d_{\ell}}\d_{\ell\ell^{\pr}}\d_{mm^{\pr}}\d_{nn^{\pr}}$ & ~~~${\displaystyle \int_{\SO_{3}}}\diff R D_{mn}^{\ell\star}(R)D_{m^{\pr}n^{\pr}}^{\ell^{\pr}}(R)=\frac{8\pi^{2}}{2\ell+1}\d_{\ell\ell^{\pr}}\d_{mm^{\pr}}\d_{nn^{\pr}}$~~~\tabularnewline
\addlinespace
 & $\!\!\!\!\!\!\!\!\!\!\!\!\!\!\!\!\!\!\!\!\!\!\!\!\!\!\!\!\!\!\!\!\!\!\!\!\!\!\!\!\!\!\!\!\!\!\!\!\!\!{\displaystyle \sum_{\ell mn\in\widehat{\G}}}\frac{d_{\ell}}{|\G|}Z_{mn}^{\ell\star}(g)Z_{mn}^{\ell}(g^{\pr})=\d_{gg^{\pr}}^{\G}$ & ${\displaystyle \sum_{\ell\geq0}\sum_{|m|,|n|\leq\ell}}\frac{2\ell+1}{8\pi^{2}}D_{mn}^{\ell\star}(R)D_{mn}^{\ell}(R^{\pr})=\d_{RR^{\pr}}^{\SO_{3}}$\tabularnewline
\midrule 
\multirow{4}{1.4in}{F. Position shifts} & $\ru_{h}|g\ket=|hg\ket$ & $\ru_{S}|R\ket=|SR\ket$\tabularnewline
 & $\,\,\,\,\,\,\lu_{h}|g\ket=|gh^{-1}\ket$ & $\,\,\,\,\,\,\lu_{S}|R\ket=|RS^{-1}\ket$\tabularnewline
\cmidrule{2-3} \cmidrule{3-3} 
 & $\ru_{h}|{}_{mn}^{\ell}\ket={\displaystyle \sum_{p}}Z_{pm}^{\ell\star}(h)|{}_{pn}^{\ell}\ket$ & $\ru_{R}|{}_{mn}^{\ell}\ket={\displaystyle \sum_{p}}D_{pm}^{\ell\star}(R)|{}_{pn}^{\ell}\ket$\tabularnewline
 & $\,\lu_{h}|{}_{mn}^{\ell}\ket={\displaystyle \sum_{p}}Z_{pn}^{\ell}(h)|{}_{mp}^{\ell}\ket$ & $\,\lu_{R}|{}_{mn}^{\ell}\ket={\displaystyle \sum_{p}}D_{pn}^{\ell}(R)|{}_{mp}^{\ell}\ket$\tabularnewline
\midrule 
\multirow{3}{1.4in}{G. Momentum kicks} & $\zz_{mn}^{\ell}|g\ket=Z_{mn}^{\ell}(g)|g\ket$ & $\dd_{mn}^{\ell}|R\ket=D_{mn}^{\ell}(R)|R\ket$\tabularnewline
\cmidrule{2-3} \cmidrule{3-3} 
 & $\zz_{mn}^{\ell}|{}_{m^{\pr}n^{\pr}}^{\ell^{\pr}}\ket={\displaystyle \sum_{LMN\in\widehat{\G}}}c_{\ell mn,\ell^{\pr}m^{\pr}n^{\pr}}^{LMN}|{}_{MN}^{L}\ket$ & $\dd_{mn}^{\ell}|{}_{m^{\pr}n^{\pr}}^{\ell^{\pr}}\ket={\displaystyle \sum_{L\geq0}\sum_{|M|,|N|\leq L}}c_{\ell mn,\ell^{\pr}m^{\pr}n^{\pr}}^{LMN}|_{MN}^{L}\ket$\tabularnewline
 & $c_{\ell mn,\ell^{\pr}m^{\pr}n^{\pr}}^{LMN}=\frac{\sqrt{d_{\ell^{\pr}}d_{L}}}{|\G|}{\displaystyle \sum_{g\in\G}}Z_{MN}^{L\star}(g)Z_{mn}^{\ell}(g)Z_{m^{\pr}n^{\pr}}^{\ell^{\pr}}(g)$ & $c_{\ell mn,\ell^{\pr}m^{\pr}n^{\pr}}^{LMN}={\textstyle \sqrt{\frac{2\ell^{\pr}+1}{2L+1}}}C_{\ell m\ell^{\pr}m^{\pr}}^{LM}C_{\ell n\ell^{\pr}n^{\pr}}^{LN}$\tabularnewline
\midrule 
\multirow{2}{1.4in}{H. ``Weyl relation''} & ${\displaystyle \ru_{g}\zz_{mn}^{\ell}\ru_{g}^{\dg}=\sum_{p}Z_{pm}^{\ell\star}(g)\zz_{pn}^{\ell}}$ & ${\displaystyle \ru_{R}\dd_{mn}^{\ell}\ru_{R}^{\dg}=\sum_{p}D_{pm}^{\ell\star}(R)\dd_{pn}^{\ell}}$\tabularnewline
 & ${\displaystyle \lu_{g}\zz_{mn}^{\ell}\lu_{g}^{\dg}=\sum_{p}Z_{pn}^{\ell}(g)\zz_{mp}^{\ell}}$ & ${\displaystyle \lu_{R}\dd_{mn}^{\ell}\lu_{R}^{\dg}=\sum_{p}D_{pn}^{\ell}(R)\dd_{mp}^{\ell}}$\tabularnewline
\bottomrule
\end{tabular}

\caption{\label{t:G}Summary of relations for $\mathscr{L}^{2}(\protect\G)$ --- the space of $\mathscr{L}^{2}$-normalizable functions on a group $\G$ --- extending analogous summaries for ordinary qudit and oscillator state spaces [\citealp{Albert2017}, Table~1].
The $C_{\ell m\ell^{\protect\pr}m^{\protect\pr}}^{LM}$ are Clebsch-Gordan
coefficients \cite{Arovas,VMH}. 
When $\G=\Z_{D}^{\times n}$, the state space is that of $n$ qu$D$its, and the position states and their corresponding momentum states are both bona-fide (i.e., discrete) orthonormal bases.
The rotor state spaces $\U_{1}$ and $\SO_{3}$ and, more generally, any continuous compact $\G$ admit bases of position states in the continuous/Dirac sense {[}\citealp{Hall2013}, Sec.~6.6{]}. In those cases, $\frac{1}{|\G|}\sum_{g\in\G}$ is replaced by $\frac{1}{|\G|}\int_{\G}\diff g$, where $|\G|$ is the volume of $\G$ as a manifold and $\diff g$ is the Haar measure \cite{so3engbook}. However, since such spaces are compact, their corresponding momentum bases are still discrete. The oscillator $\G=\R$ is continuous and noncompact, meaning that both its position and momentum bases are continuous. For this group and others like it, $\frac{1}{|\G|}$ is omitted and the sum over $\ell$ turns into an integral (with respect to the Plancherel measure) \cite{Barut1986} (see also [\citealp{so3engbook}, Sec.~8.3.3]).
}
\end{table*}

\begin{table*}
\begin{tabular}[t]{>{\raggedright}m{1.4in}cc}
\toprule 
 & Coset space $\mathscr{L}^{2}(\G/\H)\subset \mathscr{L}^{2}(\G)$ & Linear rotor $\mathscr{L}^{2}(\S^{2})\subset \mathscr{L}^{2}(\SO_{3})$\tabularnewline
\midrule 
A. ``Phase space'' & $\left(\cs,\,{}_{mn}^{\ell}\right)\in\G/\H\times\widehat{\G/\H}$ & $\left(\vh,\,{}_{m}^{\ell}\right)\in\S^{2}\times\widehat{\S^{2}}$\tabularnewline
\midrule 
\addlinespace
\multirow{2}{1.4in}{B. Conjugate bases} & ${\displaystyle |\cs\H\ket={\displaystyle \sum_{\ell mn\in\widehat{\G/\H}}}}\sqrt{\frac{d_{\ell}}{|\G/\H|}}Z_{mn}^{\ell\star}\left(\cs\H\right)|_{mn}^{\ell}\ket$ & $\,\,\,\,\,\,\,\,\,\,\,\,\,|\vh\ket={\displaystyle \sum_{\ell\geq0}\sum_{|m|\leq\ell}}Y_{m}^{\ell\star}(\vh)|{}_{m}^{\ell}\ket$\tabularnewline
\addlinespace
 & $\!\!\!\!\!\!\!\!|_{mn}^{\ell}\ket={\displaystyle \sum_{\cs\in\G/\H}}\sqrt{\frac{d_{\ell}}{|\G/\H|}}Z_{mn}^{\ell}\left(\cs\H\right)|\cs\H\ket$ & $|_{m}^{\ell}\ket={\displaystyle \int_{\S^{2}}}\diff\vh Y_{m}^{\ell}(\vh)|\vh\ket$\tabularnewline
\midrule
\addlinespace
C. Overlap & $\bra\cs\H|{}_{mn}^{\ell}\ket=\sqrt{\frac{d_{\ell}}{|\G/\H|}}Z_{mn}^{\ell}\left(\cs\H\right)$ & $\bra\vh|_{m}^{\ell}\ket=Y_{m}^{\ell}(\vh)$\tabularnewline
\midrule 
\addlinespace
D. ``Resolution'' & ${\displaystyle \sum_{\cs\in\G/\H}|\cs\H\ket\bra\cs\H|={\displaystyle {\displaystyle \sum_{\ell mn\in\widehat{\G/\H}}}}|{}_{mn}^{\ell}\ket\bra{}_{mn}^{\ell}|=\id_{\G/\H}}$ & ${\displaystyle {\displaystyle \int_{\S^{2}}}\diff\vh|\vh\ket\bra\vh|=\sum_{\ell\geq0}\sum_{|m|\leq\ell}|{}_{m}^{\ell}\ket\bra{}_{m}^{\ell}|=\id_{\S^2}}$\tabularnewline
\midrule 
\addlinespace
E. ``Orthocompleteness'' & ${\displaystyle {\displaystyle \sum_{\cs\in\G/\H}}}Z_{mn}^{\ell\star}\left(\cs\H\right)Z_{m^{\pr}n^{\pr}}^{\ell^{\pr}}\left(\cs\H\right)=\frac{|\G/\H|}{d_{\ell}}\d_{\ell\ell^{\pr}}\d_{mm^{\pr}}\d_{nn^{\pr}}$ & ${\displaystyle {\displaystyle \int_{\S^{2}}}\diff\vh Y_{m}^{\ell\star}(\vh)Y_{m^{\pr}}^{\ell^{\pr}}(\vh)=\d_{\ell\ell^{\pr}}\d_{mm^{\pr}}}$\tabularnewline
\addlinespace
 & $\!\!\!\!\!\!\!\!\!\!\!\!\!\!\!\!\!\!\!\!\!\!\!\!\!\!\!\!\!\!\!\!\!\!\!\!\!\!\!\!\!\!\!\!\!\!\!\!\!\!\!\!\!\!\!{\displaystyle {\displaystyle \sum_{\ell mn\in\widehat{\G/\H}}}}\frac{d_{\ell}}{|\G/\H|}Z_{mn}^{\ell\star}\left(\cs\H\right)Z_{mn}^{\ell}\left(\cs^{\pr}\H\right)=\d_{\cs\cs^{\pr}}^{\G/\H}$ & $\!\!\!\!\!\!\!\!\!\!\!\!\!\!\!\!\!\!{\displaystyle {\displaystyle \sum_{\ell\geq0}\sum_{|m|\leq\ell}}Y_{m}^{\ell\star}(\vh)Y_{m}^{\ell}(\vh^{\pr})=\d_{\vh\vh^{\pr}}^{\S^{2}}}$\tabularnewline
\midrule 
\multirow{2}{1.4in}{F. Position shifts} & $\r_{h}|\cs\H\ket=|h\cs\H\ket$ & $\r_{R}|\vh\ket=|R\vh\ket$\tabularnewline
\cmidrule{2-3} \cmidrule{3-3} 
 & $\r_{h}|{}_{mn}^{\ell}\ket={\displaystyle \sum_{p}}Z_{pm}^{\ell\star}(h)|{}_{pn}^{\ell}\ket$ & $\r_{R}|{}_{m}^{\ell}\ket={\displaystyle \sum_{|p|\le\ell}}D_{pm}^{\ell\star}(R)|{}_{p}^{\ell}\ket$\tabularnewline
\midrule 
\multirow{3}{1.4in}{G. Phase shifts} & ${\displaystyle \zz_{mn}^{\ell}(\H)|\cs\H\ket=Z_{mn}^{\ell}(\cs\H)|\cs\H\ket}$ & ${\displaystyle \y_{m}^{\ell}|\vh\ket=Y_{m}^{\ell}(\vh)|\vh\ket}$\tabularnewline
\cmidrule{2-3} \cmidrule{3-3} 
 & ${\displaystyle \zz_{mn}^{\ell}(\H)|{}_{m^{\pr}n^{\pr}}^{\ell^{\pr}}\ket=\sum_{LMN\in\widehat{\G/\H}}c_{\ell mn,\ell^{\pr}m^{\pr}n^{\pr}}^{LMN}|{}_{MN}^{L}\ket}$ & $\y_{m}^{\ell}|{}_{m^{\pr}}^{\ell^{\pr}}\ket={\displaystyle \sum_{L\geq0}\sum_{|M|\leq L}}c_{\ell m,\ell^{\pr}m^{\pr}}^{LM}|{}_{M}^{L}\ket$\tabularnewline
 & $\,\,c_{\ell mn,\ell^{\pr}m^{\pr}n^{\pr}}^{LMN}=\frac{\sqrt{d_{\ell^{\pr}}d_{L}}}{|\G/\H|}{\displaystyle \sum_{\cs\in\G/\H}}Z_{MN}^{L\star}\left(\cs\H\right)Z_{m^{\pr}n^{\pr}}^{\ell^{\pr}}\left(\cs\H\right)Z_{mn}^{\ell}(\cs\H)\,\,$ & $\,\,c_{\ell m,\ell^{\pr}m^{\pr}}^{LM}=\sqrt{\frac{(2l+1)(2l^{\pr}+1)}{4\pi(2L+1)}}C_{\ell m\ell^{\pr}m^{\pr}}^{LM}C_{\ell0\ell^{\pr}0^{\pr}}^{L0}$\tabularnewline
\midrule 
H. ``Weyl relation'' & ${\displaystyle \r_{h}\zz_{mn}^{\ell}(\H)\r_{h}^{\dg}=\sum_{p}Z_{pm}^{\ell\star}(h)\zz_{pn}^{\ell}(\H)}$ & ${\displaystyle \r_{R}\y_{m}^{\ell}\r_{R}^{\dg}{\displaystyle =\sum_{|p|\leq\ell}D_{pm}^{\ell\star}(R)\y_{p}^{\ell}}}$\tabularnewline
\bottomrule
\end{tabular}

\caption{\label{t:GoverH}Summary of relations for coset spaces $\mathscr{L}^{2}(\protect\G/\protect\H)$,
treated as subspaces of group spaces $\mathscr{L}^{2}(\protect\G)$; see Appx.~\ref{appx:GH}.
The $C_{\ell m\ell^{\protect\pr}m^{\protect\pr}}^{LM}$ are Clebsch-Gordan
coefficients \cite{Arovas,VMH}. The operator $\protect\r_h$ is $\protect\ru_h$ projected onto the quotient space.}
\end{table*}

\clearpage

\section{Building codes via microwave dressing\label{appx:Microwave-dressing}}

As discussed in Sec.~\ref{sec:Realizations}, the most versatile approach to generating the approximate codewords is to build a linear combination of angular momentum eigenstates with the proper weights using an array of microwave couplings. For this preliminary scheme, we neglect hyperfine structure and mixing of momentum states with nuclear quadrupole moments \cite{Ospelkaus2010}. As a concrete physical platform, we focus on molecules composed of bosonic isotopes of alkaline-earth(-like) atoms~\cite{Stellmer2014} as well as $^{12}\text{C}$ and $^{16}\text{O}$, which have zero nuclear spin. A small electric field lifts the degeneracy of all $|^\ell_{|m|,|n|}\ket\leftrightarrow|^{\ell^\pr}_{|m^\pr|,|n^\pr|}\ket$ transitions, but does not significantly mix eigenstates. Similarly, a magnetic field can split the $\pm m$ and $\pm n$ degeneracies. Recall that the eigenenergies for a given $\ell$ are proportional to $B\ell(\ell+1)$.$^{\ref{fn:spherical-top}}$ Hence any transition $\ell-\ell^\pr=\pm1$ has a unique energy.

\begin{figure}[t]
\includegraphics[width=1\columnwidth]{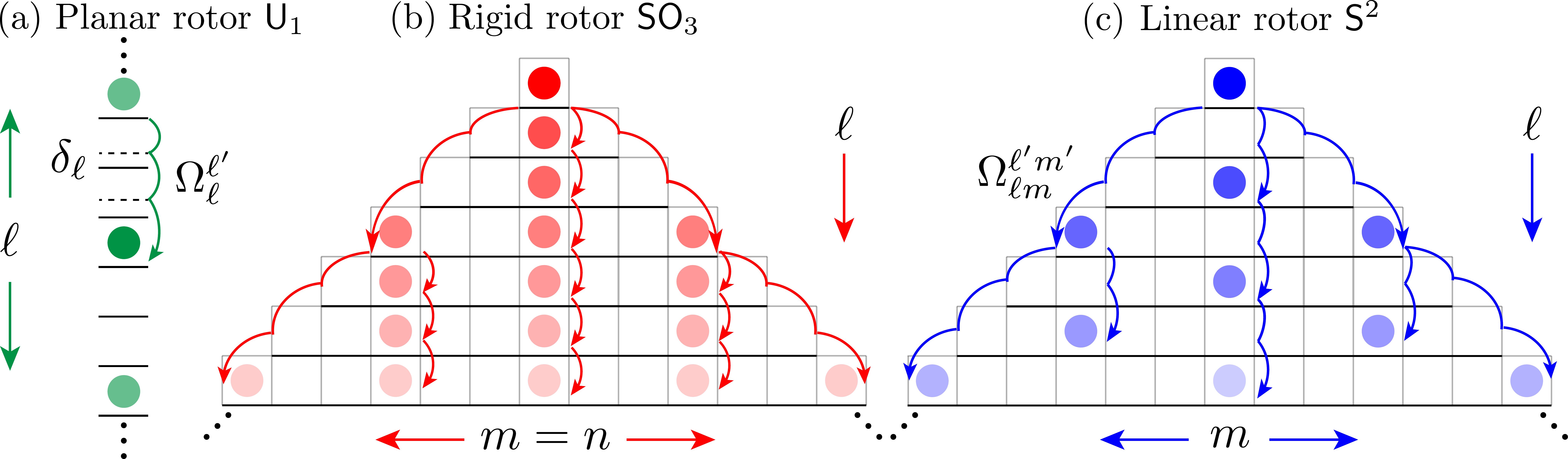}
\caption{
\label{fig:appmicdress}\textsc{Microwave dressing.} Sketch of the microwave tones required to build the approximate codewords out of angular momentum eigenstates up to $|\lmax|=3$ for $\U_1$ {\bf (a)} and $\lmax=6$ for $\SO_3$ {\bf (b)} and $\S^2$ {\bf (c)}. We assume the molecule is initialized in $|\ell=0\ket$ for $\U_1$, $|^0_{00}\ket$ for $\SO_3$, and $|^0_0\ket$ for $\S^2$. An array of microwave tones is applied to construct the codewords. Any transition allowed by dipole selection rules can be driven with a Rabi frequency $\Omega_{\ell m}^{\ell^\pr m^\pr}$, phase $\phi_{\ell m}^{\ell^{\pr}m^{\pr}}$, and detuning $\delta_{\ell m}^{\ell^{\pr}m^{\pr}}$. (For $\SO_3$ we utilize only the $n=m$ states, so no explicit mention of $n$ is necessary for those cases. For $\U_1$, there are only states with integer $\ell$.) Curved lines without arrows illustrate vastly off-resonant coupling to a state for which the population is effectively zero, and the state can be adiabatically eliminated. Curved lines with arrows illustrate couplings with finite population of the state. To prepare approximate codewords in the subspaces shown, at most 12, 24, and 22 unique microwave tones are required for $\U_1$, $\SO_3$, and $\S^2$, respectively.
}
\end{figure}

Since $B\sim1$ GHz for most molecules, $|^\ell_{|m|,|n|}\ket\leftrightarrow|^{\ell^\pr}_{|m^\pr|,|n^\pr|}\ket$ transitions can conveniently be driven with microwave fields. Note also that the dipole matrix elements (DMEs) of such transitions are relatively large for polar molecules, often $\mu\approx1$ atomic unit (a.u.). Therefore, multiphoton processes which are off-resonant from intermediate states can still achieve sufficient couplings. Figure~\ref{fig:appmicdress} shows the codewords for $\U_1$ (a), $\SO_3$ (b), and $\S^2$ (c) in the angular momentum basis, where the opacity qualitatively indicates the fractional population of the normalized state. Recall that, for $\SO_3$, we require only the $n=m$ states, so no explicit discussion of $n$ is necessary in any of these cases.

Consider, for example, $^1\Sigma$-type molecules, such as bi-akalis. Their orientations correspond to the state space $\S^2$, and we consider realizing the codewords (\ref{eq:ZN-on-S2-momentum-basis}) with $N=3$. We apply the damping function $e^{-\half\D^2\lh}$ from Sec.~\ref{subsec:Approximate-codewords} to normalize the codewords. The parameter $\D>0$ depends on how many momentum states $\lmax$ we want to consider, while $\lh$ is the total angular momentum operator. The resulting logical zero approximate codeword, up to normalization, is
\begin{equation}
|\tilde{0}\ket\propto\sum_{\ell=0}^{\lmax}\sum_{|3p|\leq\ell}e^{-\half\D^2\ell\left(\ell+1\right)}Y_{3p}^{\ell}\left({\textstyle \frac{\pi}{2}},0\right)|_{3p}^{\ell}\ket.\label{eq:S2-approximate}
\end{equation}

We assume the molecule is initially in the rigid-rotor ground state $|^\ell_m\ket=|^0_0\ket$, and assemble the state (\ref{eq:S2-approximate}) using microwave tones. All couplings ``cascade'' down from $|^0_0\ket$ as shown in Fig.~\ref{fig:appmicdress}(c). Control over the frequency, power, polarization, and phase of a microwave tone can readily be achieved in the laboratory, and is essential to build the state (\ref{eq:S2-approximate}). The microwave couplings are all adiabatic with respect to the timescales of molecular rotation, and we expect that continous-wave (CW) microwave pulses will be possible to build the codewords as the steady state solution of the Hamiltonian with the microwave driving terms of the form
\begin{equation}
  \Omega_{\ell m}^{\ell^{\pr}m^{\pr}}\exp\left[-i\left(\delta_{\ell m}^{\ell^{\pr}m^{\pr}}t+\phi_{\ell m}^{\ell^{\pr}m^{\pr}}\right)\right]|_{m^{\pr}}^{\ell^{\pr}}\ket\bra_{m}^{\ell}|,\label{eq:tone}
\end{equation} 
with Rabi frequency $\Omega_{\ell m}^{\ell^{\pr}m^{\pr}}$, detuning $\delta_{\ell m}^{\ell^{\pr}m^{\pr}}$, and phase $\phi_{\ell m}^{\ell^{\pr}m^{\pr}}$
in the rotating frame of the original rigid-rotor Hamiltonian $B \lh$. 

In the perturbative limit $\O/\d\ll1$, where we have suppressed the decorations on $\O,\d$ for clarity, the population $P_{\ell^{\pr}m^{\pr}}$ can be expressed as
$P_{\ell^{\pr}m^{\pr}}=(\O/\delta)^{2}P_{\ell m}$, where $P_{\ell m}$ is the population in $|^\ell_m\ket$. We ignore the Stark shift of $|^{\ell^\pr}_{m^\pr}\ket$, given by $\delta^S=\Omega^2/\delta$, since $\delta^S\ll\delta$. Applying the same analysis to multiphoton processes, now consider the 3-photon pulses $|^{\ell_0}_{m_0}\ket\to|^{\ell_1}_{m_1}\ket\to|^{\ell_2}_{m_2}\ket\to|^{\ell_3}_{m_3}\ket$. Assuming vastly off-resonant drives for the intermediate states, we can adiabatically eliminate said states. The population in $|^{\ell_3}_{m_3}\ket$ is given by
\begin{equation}
	P_{\ell_{3},m_{3}}=\left(\frac{\Omega_{\ell_{0}m_{0}}^{\ell_{1}m_{1}}~\Omega_{\ell_{1}m_{1}}^{\ell_{2}m_{2}}~\Omega_{\ell_{2}m_{2}}^{\ell_{3}m_{3}}}{\delta_{\ell_{0}m_{0}}^{\ell_{1}m_{1}}~\delta_{\ell_{1}m_{1}}^{\ell_{2}m_{2}}~\delta_{\ell_{2}m_{2}}^{\ell_{3}m_{3}}}\right)^{2}P_{\ell_{0}m_{0}}.
\end{equation}
Thus, $|^{l_3}_{m_3}\ket$ can effectively be attained from $|^{l_0}_{m_0}\ket$ using the above combination of tones. 

The above scheme can be used to transfer population in each momentum state present in the approximate state (\ref{eq:S2-approximate}) to neighboring states. That way, the population in $|^0_0\ket$ cascades down the angular momentum pyramid. The required tones are shown in Fig.~\ref{fig:appmicdress}. Curved lines with arrows illustrate couplings with finite population of the state. For the $\S^2$ state (\ref{eq:S2-approximate}) with $\lmax=6$, we require 22 unique microwave tones. Similar schemes for $\U_1$ and $\SO_3$, shown in Fig.~\ref{fig:appmicdress}(a-b), require 12 and 24 tones, respectively.

The above is just a sketch. A detailed analysis based on, e.g., exact diagonization of the coupling Hamiltonian or a master equation would be required to choose the frequency, power, polarization, and phase of each microwave tone (\ref{eq:tone}). One must be aware of the formation of dark states as well as potentially dynamic evolution of the populations. Advanced modeling is particularly important outside the perturbative limit when $\O_{\ell m}^{\ell^{\pr}m^{\pr}}/\delta_{\ell m}^{\ell^{\pr}m^{\pr}}\sim1$, for which population dynamics must be addressed. Time-dependent pulses can also be employed for which power and phase can be adjusted on timescales comparable to the Rabi frequencies (but still slow compared to rotational timescales). Such analysis is outside the scope of this work. 

An electric field may be helpful for building the codewords, and could potentially reduce the number of required microwave tones. In our notation, an electric field is represented by operator $\cos \hat \phi$ for $\U_1$, $\dd_{00}^{1}$ for $\SO_3$, and $\y_{0}^{1}$ for $\S^2$. For the latter two spaces, a large electric field mixes $\ell$-states with the same $m$ [see Eq.~(\ref{eq:CG})]. This point is particularly salient in the case of $\SO_3$, where couplings down the columns for $|m|=0,3,6$ are required. A large electric field will naturally create these couplings down each column of Fig.~\ref{fig:appmicdress}(b).

\section{Voronoi cells\label{appx:voronoi}}

We determine the Voronoi-Dirichlet cells of the quotient spaces $\SO_{3}/\H$
from Table~\ref{t:quaotient_spaces}. This is strictly an adaptation
of the work of Postnikov \cite{postnikov} (see also \cite{Wulker2019}).

Let $\X$ be a metric space with distance function $\mathtt{d}$ and distinguished origin point $x_0$. Let $\H$ be a discrete group whose elements $R\in\H$ map points $x\in\X$ as $x\to R x$. This group maps the origin to the orbit $\{Rx_0\}_{R\in\H}$. Each such point $Rx_0$ has its own \textit{Voronoi cell} --- a region consisting of points that are closer (or as close) to $Rx_0$ than to $R^\pr x_0$ for any $R^\pr \neq R$. When $R$ is the identity, we call the corresponding cell the \textit{fundamental Voronoi cell},
\begin{equation}
    \F_{\X/\H} = \{x\in\X~|~\forall R\in\H,~\mathtt{d}\left(x,x_0\right) \leq \mathtt{d}\left(x,Rx_0\right)\}~.
\end{equation}
When the space is a group ($\X=\G$), $x_0$ is the identity, and the above is thus the Voronoi cell of the identity.

To properly account for distances in $\SO_3$, we recall that it is equivalent to the 3-sphere $\S^3=\SU_2$ with opposite points identified, $\SO_{3}=\S^{3}/\Z_{2}$ \cite{Arovas}. The 3-sphere can be parameterized by either a 4-dimensional unit vector or by a quaternion.
Just like a semicircle parameterizes a circle with opposite points
identified, hemispherical quanternions,
\begin{equation}
\eta(\o,\vh)=\left(\cos\frac{\o}{2},\,\vh\sin\frac{\o}{2}\right)\,,
\end{equation}
are an equivalent way to parameterize $R=(\o,\vh)\in\SO_{3}$. The distance function $\mathtt{d}$ we use is the dot product of the above 4-vectors.

The Voronoi cell of a point $p$
is bounded by the intersection of the interiors of the \textit{mediatrices}
of $p$ and all $q$ that are in the orbit of $p$ under $\H$. A
mediatrix of points $p$ and $q$ is the set of all points that lie
the same distance from $p$ and $q$. For a 2D square lattice, the boundary of $\F_{\R^{\times2}/\Z^{\times2}}$ consists of segments that bisect lines that connect
the center of the cell with centers of neighboring cells. The boundaries of $\F_{\SO_3/\H}$ are hyperplanes that go through the midpoint $(\O/2,\wh)$
of the geodesic connecting the center with each group element $(\O,\wh)\in\H$.
Points $\eta$ on such hyperplanes satisfy
\begin{equation}
\dot{\eta}\left(\O/2,\wh\right)\cdot\eta\left(\o,\vh\right)=0\,,
\end{equation}
where $\dot{\eta}\equiv\partial\eta/\partial\o$. Solving for
$\o$ yields
\begin{equation}
\o=\left|2\cot^{-1}\left(\vh\cdot\wh\cot\frac{\O}{4}\right)\right|\,.
\end{equation}
We use this to plot the various manifolds in Figs.~\ref{fig:Zd-rotors}-\ref{fig:TinSO3}.
Letting $\H$ be a $\zh$-axis rotation subgroup yields Eq.~(\ref{eq:omega-max}).

\section{Normalizable codewords\label{appx:Approximate-states}}

\prg{Observables}

For the calculation of $\lb$ (\ref{eq:lbar}), in the $\D\to0$
limit
\begin{equation}
\bbra{\tilde{r}}\lh\kk{\tilde{r}}=\frac{\bra\overline{r}|\lh e^{-\D^{2}\lh}|\overline{r}\ket}{\bra\overline{r}|e^{-\D^{2}\lh}|\overline{r}\ket}\sim\frac{\bra I|\lh e^{-\D^{2}\lh}|I\ket}{\bra I|e^{-\D^{2}\lh}|I\ket}\,,
\end{equation}
where $I$ is the identity rotation. To obtain the above, recall that
each $|\overline{r}\ket$ is a superposition of position states $|R_{\frac{2\pi}{N}h+\frac{\pi}{N}r,\zh}\ket$
for $h\in\Z_{N}$. The state $e^{-\D^{2}\lh}|R_{\frac{2\pi}{N}h+\frac{\pi}{N}r,\zh}\ket$
can be thought of as a Gaussian distribution of orientations centered
at $|R_{\frac{2\pi}{N}h+\frac{\pi}{N}r,\zh}\ket$, overlapping with
other states for different $h$. However, since the overlap is exponentially
suppressed with $h$, we ignore such contributions. We then use the
fact that all rotations commute with $\lh$, allowing us to remove
$h,r$-dependence:
\begin{equation}
\bbra{R_{\o,\vh}}f(\lh)\kk{R_{\o,\vh}}=\bbra If(\lh)\kk I\label{eq:invariance-under-rotations}
\end{equation}
for any $(\o,\vh)$ and function $f(\lh)$.

Having used $\SO_{3}$-symmetry to remove dependence of $(\o,\vh)$,
we now express the identity state in the momentum basis, yielding
\begin{equation}
\bbra I f(\lh)\kk I=\frac{1}{8\pi^{2}}\sum_{\ell\geq0}\left(2\ell+1\right)^{2}f\left(\ell\left(\ell+1\right)\right)\,.
\end{equation}
The identity state is supported only on momentum states $|_{mm}^{\ell}\ket$
with amplitude $\frac{2\ell+1}{8\pi^{2}}$, and we have performed
the sum over $m$ to obtain the extra $2\ell+1$ factor. The function $f(\lh)$ becomes as such due to Eq.~(\ref{eq:L2-action-on-lmn-states}). We then rearrange
the above sum to obtain a sum over integers, which can then be approximated
using Poisson summation (\ref{eq:poisson-normal}) for the relevant
$f$,
\begin{align}
\bbra I f(\lh)\kk I & =\frac{1}{16\pi^{2}}\sum_{\ell\in\Z}\left(2\ell+1\right)^{2}f\left(\ell\left(\ell+1\right)\right)\\
 & \sim\frac{1}{16\pi^{2}}\int_{\R}\diff x\left(2x+1\right)^{2}f\left(x\left(x+1\right)\right)\,.\nonumber 
\end{align}
Plugging in explicit forms for $f$ yields Eq.~(\ref{eq:lbar}).

\prg{Leakage error}

To evaluate $\pl$, we first start with its complement $\pk=1-\pl$, the projection of $|\tilde{0}\ket$ onto its own Voronoi cells
{[}for $N=3$, those in the left panel of Fig.~\ref{fig:ladder}(c){]}:
\begin{equation}
\pk=\sum_{h\in\Z_{N}}\int_{\F_{\SO_{3}/\Z_{2N}}}\diff S_{\o,\vh}\left|\bra S_{\o,\vh}R_{\frac{2\pi}{N}h,\zh}|\tilde{0}\ket\right|^{2}\,.
\end{equation}
Recall that $|\tilde{0}\ket$ is a superposition of smeared group
elements $|R_{\frac{2\pi}{N}k,\zh}\ket$ for $k\in\Z_{N}$. Inserting
this expansion for $|\tilde{0}\ket$, we estimate $\pk$ in the $\D\to0$
limit by ignoring contributions from elements leaking outside of their
own Voronoi cells,
\begin{equation}
\pk\sim\sum_{h\in\Z_{N}}\int_{\F_{\SO_{3}/\Z_{2N}}}\!\!\!\!\!\!\diff S_{\o,\vh}\frac{\left|\bra S_{\o,\vh}R_{\frac{2\pi}{N}h,\zh}|e^{-\half\D^{2}\lh}|R_{\frac{2\pi}{N}h,\zh}\ket\right|^{2}}{N\bra\overline{0}|e^{-\D^{2}\lh}|\overline{0}\ket}.
\end{equation}
Inserting the asymptotic expression for $\bra\overline{0}|e^{-\D^{2}\lh}|\overline{0}\ket$
and using invariance under rotations (\ref{eq:invariance-under-rotations})
brings us to
\begin{equation}
\pk\sim \frac{8\left(\sqrt{\pi}\D\right)^{3}}{e^{\D^{2}/4}}\int_{\F_{\SO_{3}/\Z_{2N}}}\!\!\!\!\!\!\diff S_{\o,\vh}\left|\bra S_{\o,\vh}|e^{-\half\D^{2}\lh}|I\ket\right|^{2}.\label{eq:interim}
\end{equation}

Now we get rid of the $\vh$-dependence of the absolute value. Recall
that rotations conjugate each other as
\begin{equation}
R_{\O,\wh}R_{\o,\vh}R_{\O,\wh}^{\dg}=R_{\o,R_{\O,\wh}\vh}\,.
\end{equation}
For each $\vh$,
we pick $\left(\O,\wh\right)$ such that $R_{\O,\wh}\vh=\zh$.
We then ``create'' that rotation inside the ket and commute it through
to the bra,
\begin{subequations}
\begin{align}
\!\!\!\!\!\!\!\bra S_{\o,\vh}|f(\lh)|I\ket & =\bra S_{\o,\vh}|f(\lh)|R_{\O,\wh}^{\dg}R_{\O,\wh}\ket\\
 & =\bra R_{\O,\wh}S_{\o,\vh}R_{\O,\wh}^{\dg}|f(\lh) |I\ket\\
 & =\bra S_{\o,\zh}|f(\lh)|I\ket\,.
\end{align}
\end{subequations}
We further estimate this matrix element via the same procedure as
that for the average angular momentum,
\begin{align}
\bra S_{\o,\zh}|e^{-\half\D^{2}\lh}|I\ket & \sim\int_{\R}\diff x\frac{2x+1}{e^{\frac{\D^{2}}{2}x\left(x+1\right)}}\frac{\sin\left[\left(2x+1\right)\o/2\right]}{16\pi^{2}\sin\left(\o/2\right)}\nonumber \\
 & =\frac{\sqrt{2}}{8}\frac{\o~ e^{\frac{\D^{2}}{8}-\frac{\o^{2}}{2\D^{2}}}}{\left(\sqrt{\pi}\D\right)^{3}\sin\left(\o/2\right)}\,.
\end{align}

Plugging all of the above into $\pl=1-\pk$ yields
\begin{equation}
\pl\sim1-\frac{1}{4\left(\sqrt{\pi}\D\right)^{3}}\int_{\F_{\SO_{3}/\Z_{2N}}}\!\!\!\!\!\!\diff S_{\o,\vh}\frac{\o^{2}e^{-\o^{2}/\D^{2}}}{\sin^{2}\left(\o/2\right)}.
\end{equation}
The integration measure {[}\citealp{VMH}, Sec.~4.5.4{]} is
\begin{equation}
\diff S_{\o,\vh}=4\sin\T\sin^{2}(\o/2)\diff\T \diff\P \diff\o\,,
\end{equation}
and for $\F_{\SO_{3}/\Z_{2N}}$, the integration of $(\T,\P)=\vh$
is over $\S^{2}$ and $\o\in[0,\om(\T)]$ {[}see Eq.~(\ref{eq:omega-max}){]}.
To absorb the ``$1-$'' part, we integrate over the complementary
region, for which $\o\in[\om(\T),\pi]$. Trivially integrating
over the azimuthal angle and simplifying yields
\begin{align*}
\pl\sim & \frac{2}{\sqrt{\pi}\D^{3}}\int_{0}^{\pi}\sin\T \diff\T\int_{\om\left(\T\right)}^{\pi}\diff\o\,\o^{2}e^{-\o^{2}/\D^{2}}\,.
\end{align*}

To simplify the integral, we first use invariance under $\T\to\pi-\T$
to write
\begin{equation}
\pl\sim\frac{4}{\sqrt{\pi}\D^{3}}\int_{0}^{\pi/2}\sin\T \diff\T\int_{\o_{\star}}^{\pi}\diff\o\,\o^{2}\,e^{-\o^{2}/\D^{2}}\,,
\end{equation}
where, due to the new integration domain, we can remove the absolute
value in $\om$ and define $\o_{\star}\equiv2\cot^{-1}\left(\cos\T~\cot\frac{\pi}{4N}\right)$.
Now we apply Laplace's method \cite{vaughn_book}: in the $\D\to0$ limit, the
leading-order contribution to the $\o$-integral is around $\o=\o_\star$,
since the exponential $\o^{2}/\D^{2}$ is minimized there.  Thus,
we can increase the upper $\o$-bound to infinity without losing accuracy
and perform the resulting Gaussian-type integral. Plugging the result into
the remaining integral yields
\begin{equation}
\pl\sim\frac{2}{\sqrt{\pi}\D}\int_{0}^{\pi/2}\diff\T\,\o_{\star}\sin\T\,e^{-\o_{\star}^{2}/\D^{2}}\,.
\end{equation}
Now the dominant contribution is around $\T=0$. Expressing $\o_{\star}$
in terms of $\T$, we expand both $\o_{\star}\sin\T$ and the exponential
around zero,
\begin{subequations}
\begin{align}
\o_{\star}^{2} & \approx\left(\frac{\pi}{2N}\right)^{2}+\frac{\pi}{2N}\sin\left(\frac{\pi}{2N}\right)\T^{2}\\
\o_{\star}\sin\T & \approx\frac{\pi}{2N}\T\,.
\end{align}
\end{subequations}
Plugging this in, extending the upper bound to infinity, evaluating
the resulting Gaussian-type integral, and simplifying yields the result
(\ref{eq:prob-of-leakage}).

\prg{Momentum kick distortion}

A way of understanding why detection of $\ell<N$ momentum kicks implies
correction of $\ell<N/2$ kicks stems from the fact that products
of $D_{mn}^{\ell}(R)D_{m^{\pr}n^{\pr}}^{\ell^{\pr}}(R)$ can be expanded
in terms of a sum of single $D$-matrices {[}\citealp{VMH}, Sec.~4.6.1{]}.
Upgrading this to operators and using selection rules (\ref{eq:CG})
yields
\begin{equation}
\dd_{mn}^{\ell}\dd_{m^{\pr}n^{\pr}}^{\ell^{\pr}}=\sum_{L=\left|\ell^{\pr}-\ell\right|}^{\ell^{\pr}+\ell}C_{\ell m\ell^{\pr}m^{\pr}}^{L,m^{\pr}+m}C_{\ell n\ell^{\pr}n^{\pr}}^{L,n^{\pr}+n}\dd_{m^{\pr}+m,n^{\pr}+n}^{L},\label{eq:products-of-D-matrices}
\end{equation}
where the Clebsch-Gordan coefficients $C_{\ell m\ell^{\pr}m^{\pr}}^{L,m^{\pr}+m}=0$
when $|m^{\pr}+m|,|n^{\pr}+n|>L$. Per the Knill-Laflamme conditions, in order to
correct against kicks by angular momentum $\ell<N/2$, we need to
detect products of kicks $(\dd_{mn}^{\ell})^{\dg}\dd_{m^{\pr}n^{\pr}}^{\ell^{\pr}}$ with $\ell,\ell^\pr<N/2$.
Using the above equation and
\begin{equation}
(\dd_{mn}^{\ell})^{\dg}=\left(-1\right)^{m+n}\dd_{-m,-n}^{\ell}
\end{equation}
{[}\citealp{VMH}, Sec.~4.5{]}, such products can be expanded in
terms of single kicks with momentum $<N$. Detection of such single
kicks thus implies correction of kicks with $\ell<N/2$.

For the normalizable codewords, the problem comes from violations
of the Knill-Laflamme conditions stemming from $\dd_{00}^{\ell}$,
namely, $\bra\tilde{0}|\dd_{00}^{\ell}|\tilde{1}\ket\neq0$. Using
Eq.~(\ref{eq:products-of-D-matrices}), this translates to not being
able to perfectly resolve kicks $\dd_{mm}^{\ell}$ with fixed $m$
but different $\ell$, as the distortion caused by such kicks depends
on $\ell$. Working in the $\D\to0$ limit, plugging in the
approximate codeword normalization, using Table \ref{t:G}.G, and
using selection rules (\ref{eq:CG}),
\begin{align}
\bra\tilde{0}|\dd_{00}^{\ell}|\tilde{1}\ket & \sim\frac{\D^{3}N}{e^{\D^{2}/4}\sqrt{\pi}}\sum_{\ell^{\pr}\geq0}\left(2\ell^{\pr}+1\right)e^{-\frac{\D^{2}}{2}\ell^{\pr}(\ell^{\pr}+1)}\label{eq:sums}\\
 & \!\!\!\!\!\!\!\!\!\!\!\!\!\!\!\!\!\!\!\!\!\!\!\!\!\!\times\sum_{\left|\ell^{\pr}-\ell\right|\leq L\leq\ell^{\pr}+\ell}e^{-\frac{\D^{2}}{2}L(L+1)}\sum_{|PN|\leq L}\left(-1\right)^{P}(C_{\ell,0,\ell^{\pr},PN}^{L,PN})^{2}\,.\nonumber 
\end{align}
Numerically, thus sum is exponentially suppressed with $1/\D^{2}$
for all $\ell<N$ that we have tested. Below, we estimate its asymptotic
behavior, showing that its dependence on $\D$ is surprisingly similar
to $\pl$ (\ref{eq:prob-of-leakage}).

As $\D\to0$, ever increasing values of $L,\ell^{\pr}$ contribute
to the sums (\ref{eq:sums}). For $\ell\ll L,\ell^{\pr}$, the Clebsch-Gordan
coefficients $C_{\ell,0,\ell^{\pr},PN}^{L,PN}$ approach their semiclassical
limit --- a particular $D$-matrix element {[}\citealp{VMH}, Sec.~8.9.1{]}.
Since $D$-matrices are unitary, this element is bounded by one. Setting
all $C$'s to one and evaluating the innermost sum over $P$ yields
\begin{align}
\bra\tilde{0}|\dd_{00}^{\ell}|\tilde{1}\ket & \approx\frac{\D^{3}N}{e^{\D^{2}/4}\sqrt{\pi}}\sum_{\ell^{\pr}\geq0}\left(2\ell^{\pr}+1\right)e^{-\frac{\D^{2}}{2}\ell^{\pr}(\ell^{\pr}+1)}\nonumber \\
 & \!\!\!\!\!\!\times\sum_{\left|\ell^{\pr}-\ell\right|\leq L\leq\ell^{\pr}+\ell}e^{-\frac{\D^{2}}{2}L(L+1)}\left(-1\right)^{\left\lfloor \frac{L}{N}\right\rfloor }\,.
\end{align}
For all $\ell^{\pr}>\ell$, there are $2\ell+1$ different values
of $L$. For each value, we approximate the resulting sum over $\ell^{\pr}$
by the sum associated with $L=\ell^{\pr}$. This yields
\begin{align}
\bra\tilde{0}|\dd_{00}^{\ell}|\tilde{1}\ket & \approx{\textstyle \frac{\left(2\ell+1\right)\D^{3}N}{e^{\D^{2}/4}\sqrt{\pi}}}\sum_{\ell^{\pr}\geq0}\left(2\ell^{\pr}+1\right)e^{-\D^{2}\ell^{\pr}(\ell^{\pr}+1)}\left(-1\right)^{\left\lfloor \frac{\ell^{\pr}}{N}\right\rfloor }.
\end{align}
The sign of the summand oscillates with $\ell^{\pr}$, with periodicity
$N$. Numerically, we observe that this expression scales as the quoted estimate (\ref{eq:momentum-distortion}).

\section{Coset spaces $\mathscr{L}^{2}(\protect\G/\protect\H)$}
\label{appx:GH}

In close analogy to group spaces $\G$, here we construct Table~\ref{t:GoverH}
--- an analogue to Table~\ref{t:G} for coset spaces $\G/\H$
with $\H\subseteq\G$. The key idea is to treat these spaces as subspaces
of the group space $\G$. This allows us to develop position/momentum shift operators and orthogonality relations. This treatment is intended for molecular state spaces $\SO_{3}/\H$, but it also provides a framework for qudit-type spaces $\G/\H$ for finite $\G$ as well as symmetric spaces for $\G$ a Lie group. However, as with Sec.~\ref{sec:A-qubit-on}, we consider
finite $\G$ to better flesh out the key intuition.

Mathematically, our result is a ``coordinates statement'' of the Peter-Weyl theorem for homogeneous spaces {[}\citealp{Carter1995},
Corr.~9.14{]}\cite{farashahi}{[}\citealp{Childs2010}, Eq.~(116){]}[\citealp{Vilenkin1991}, Sec.~2.3.9]. The ability to make such a statement stems from a particular choice of the ``coordinates'', namely, bases for the irreps of $\G$ that are block-diagonal when restricted to $\H$. 

Recalling that a space $\G$ consists of states $\{|g\ket\,,\,g\in\G\}$,
the defining position states of the space $\G/\H$ are equal superpositions
of elements of cosets of $\H$ in $\G$,
\begin{equation}
\kk{\cs\H}\equiv\frac{1}{\sqrt{|\H|}}\sum_{h\in\H}\kk{\cs h}\,,
\end{equation}
where $\cs$ is any element of the coset $\cs\H$. In effect, projecting into the subspace spanned by the above states is equivalent to performing the quotient map on the level of the group. Some applications require unique choices of coset reprensetatives, so we pick $\cs\in\F_{\G/\H}$ --- the Voronoi cell of the identity (see Appx.~\ref{appx:voronoi}). We abuse notation and use $\G/\H$ and $\F_{\G/\H}$ interchangeably throughout the paper.

Expressing these
in terms of the momentum states of $\G$ from Table~\ref{t:G}.B,
switching sums, and using definition (\ref{eq:twirl-def}) yields
a generalization of Eq.~(\ref{eq:G-basis-dual}) to all $\G/\H$,
\begin{equation}
\kk{\cs\H}={\displaystyle \sum_{\ell mn\in\widehat{\G}}}{\textstyle \sqrt{\frac{d_{\ell}}{|\G/\H|}}}Z_{mn}^{\ell\star}\left(\cs\H\right)|_{mn}^{\ell}\ket\,.
\end{equation}
We observe that this is zero unless $Z^{\ell}\left(\H\right)\neq0$.
The participating $\ell$ thus form the reciprocal space \cite{Justel2018}
\begin{equation}
\H^{\perp}=\left\{ \ell\in\widehat{\G}\,,\,Z^{\ell}\left(\H\right)\neq0\right\} \,,
\end{equation}
a notion that generalizes the dual/reciprocal lattice for $\H=\Z^{\times d}$ \cite{tinkhambook} and the dual code for CSS codes $\H=\Z_{2}^{\times n}$
{[}\citealp{grasslbook}, Lemma~7.1{]}. 

The reciprocal space depends significantly
on $\G$ and $\H$. For example, the $\Z_{3}\subset\Z_{6}\subset\U_{1}$
codewords (\ref{eq:Z3}) contain every third $\ell$, $\Z_{3}^{\perp}=\{\ell\in\Z\,,\,e^{i\frac{2\pi}{3}\ell}=1\}$.
For the $\TT\subset\OO\subset\SO_{3}$ code from Sec.~\ref{subsec:Nonabelian-subgroup-codes},
$\ell\in\{1,2,5\}$ do not participate. For $\Z_{3}\subset\Z_{6}\subset\SO_{3}$
(\ref{eq:Z3inSO3inMomentumBasis}), all $\ell\in\widehat{\SO_{3}}$
participate, but the internal indices $m,n$ are restricted. 

While
the surviving $\ell$ are established by $\H^{\perp}$, surviving
$m,n$ depend on the choice of basis used for the irreps $Z^{\ell}$. We pick an
\textit{$\H$-admissible basis} \cite{farashahi}, for which each $Z^{\ell}$ decomposes into blocks of some irreps $\l(\ell)\in\widehat{\H}$ when restricted
to $h\in\H$,
\begin{equation}
Z^{\ell}(h)=\bigoplus_{\l(\ell)\in\widehat{\H}}\z^{\l(\ell)}(h)\,.\label{eq:decomp-into-irreps-of-h}
\end{equation}
Picking a particular matrix element $p,n$ selects one of three cases:
\begin{enumerate}
\item a matrix element that is outside of the above block decomposition
and hence zero,
\item a matrix element of a nontrivial irrep $\l(\ell)\neq\one$,
\item a matrix element of the trivial irrep $\l(\ell)=\one$.
\end{enumerate}
In case (2), the group orthogonality relations (Table~\ref{t:G}.E)
for $\H$ tell us that such a matrix element averaged over $\H$ will
be zero (since we assumed the irrep is nontrivial and $\z_{00}^{\one}(h)=1$).
Thus, only case (3) survives, and we see that $\H^{\perp}$ consists
of only those irreps $\ell$ that contain a trivial irrep of $\H$.
Since the trivial irrep is one-dimensional, since we are using an
$\H$-admissible basis, and since our average has a $\frac{1}{|\H|}$
factor, one must have $Z_{pn}^{\ell}\left(\H\right)=\d_{pn}$ for
all $n$ corresponding to trivial $\l(\ell)=\one$ in decomposition (\ref{eq:decomp-into-irreps-of-h}).
Plugging this in yields
\begin{subequations}
\begin{align}
\kk{\cs\H} & ={\displaystyle \sum_{\ell mn\in\widehat{\G}}}{\textstyle \sqrt{\frac{d_{\ell}}{|\G/\H|}}}\sum_{p=1}^{d_{\ell}}Z_{mp}^{\ell\star}\left(\cs\right)Z_{pn}^{\ell\star}\left(\H\right)|_{mn}^{\ell}\ket\\
 & ={\displaystyle \sum_{\ell\in\H^{\perp}}}\sum_{n=1}^{d_{\ell}}\sum_{m=1}^{d_{\ell}}{\textstyle \sqrt{\frac{d_{\ell}}{|\G/\H|}}}Z_{mn}^{\ell\star}\left(\cs\right)Z_{nn}^{\ell\star}\left(\H\right)|_{mn}^{\ell}\ket\\
 & ={\displaystyle \sum_{\ell mn\in\widehat{\G/\H}}}{\textstyle \sqrt{\frac{d_{\ell}}{|\G/\H|}}}Z_{mn}^{\ell\star}\left(\cs\H\right)|_{mn}^{\ell}\ket\,,\label{eq:coset-in-terms-of-momentum-states}
\end{align}
\end{subequations}
where where we collect the surviving $\ell\in\H^{\perp}$, the
surviving $n$ designated by our $\H$-admissible basis, and all $m\in\{1,\cdots,d_{\ell}\}$
into $\widehat{\G/\H}$. This set determines the $\G$-momentum states
$|_{mn}^{\ell}\ket$ that form the momentum basis for $\G/\H$.

Since the position states $|\cs\H\ket$ each consist of different
group elements, they are orthogonal. Inserting a resolution of identity
in terms of $\G$-momentum states into $\bra\cs\H|\cs^{\pr}\H\ket$
yields completeness relations
\begin{subequations}
\begin{align}
\d_{\cs\cs^{\pr}}^{\G/\H} & =\bra\cs\H|\cs^{\pr}\H\ket\\
 & =\sum_{\ell mn\in\widehat{\G}}\bra\cs\H|_{mn}^{\ell}\ket\bra_{mn}^{\ell}|\cs^{\pr}\H\ket\\
 & =\sum_{\ell mn\in\widehat{\G/\H}}{\textstyle \frac{d_{\ell}}{|\G/\H|}}Z_{mn}^{\ell}\left(\cs\H\right)Z_{mn}^{\ell\star}\left(\cs^{\pr}\H\right)\,.
\end{align}
\end{subequations}
Taking two triples from $\widehat{\G/\H}$ and inserting the position-state
identity resolution on $\G/\H$ yields the completeness relation
\begin{subequations}
\begin{align}
\d_{\ell\ell^{\pr}}\d_{mm^{\pr}}\d_{nn^{\pr}} & =\bra_{mn}^{\ell}|_{m^{\pr}n^{\pr}}^{\ell^{\pr}}\ket\\
 & =\sum_{\cs\in\G/\H}\bra_{mn}^{\ell}|\cs\H\ket\bra\cs\H|_{m^{\pr}n^{\pr}}^{\ell^{\pr}}\ket\\
 & ={\textstyle \frac{d_{\ell}}{|\G/\H|}}{\displaystyle {\displaystyle \sum_{\cs\in\G/\H}}}Z_{mn}^{\ell\star}\left(\cs\H\right)Z_{m^{\pr}n^{\pr}}^{\ell^{\pr}}\left(\cs\H\right)\,.
\end{align}
\end{subequations}
Using Eq.~(\ref{eq:coset-in-terms-of-momentum-states}) and applying
the completeness relation yields the Fourier transfrom on $\G/\H$,
\begin{equation}
{\displaystyle \sum_{\cs\in\G/\H}}{\textstyle \sqrt{\frac{d_{\ell}}{|\G/\H|}}}Z_{mn}^{\ell}\left(\cs\H\right)\kk{\cs\H}=\kk{_{mn}^{\ell}}\,.
\end{equation}

All position shifts $\ru_g$ act in an induced representation (\ref{eq:position-shifts-acting-on-coset-states}),
but only $\H$-twirled momentum kicks (\ref{eq:G:check-ops}) keep one inside $\G/\H$,
\begin{equation}
\zz_{mn}^{\ell}\left(\H\right)\kk{\cs\H}=Z_{mn}^{\ell}\left(\cs\H\right)\kk{\cs\H}\,.
\end{equation}
The remaining identities in the 2nd column of Table~\ref{t:GoverH}
are determined from the above and Table~\ref{t:G}.

Denoting $\ru_g$ projeted onto $\G/\H$ as $\r_g$, the products
\begin{equation}
\hat{B}_{g}^{\ell mn}={\textstyle \sqrt{\frac{d_{\ell}}{|\G|}}}\zz_{mn}^{\ell}\left(\H\right)\r_g
\end{equation}
 for $g\in\G$ and $_{mn}^{\ell}\in\widehat{\G/\H}$ form an overcomplete
frame for operators acting on $\G/\H$,
\begin{equation}
\!\!\!\!\sum_{g\in\G}\sum_{\ell mn\in\widehat{\G/\H}}\bra\cs\H|\hat{B}_{g}^{\ell mn}|b\H\ket\bra b^{\pr}\H|\hat{B}_{g}^{\ell mn\dg}|a^{\pr}\H\ket=\d_{\cs\cs^{\pr}}^{\G/\H}\d_{bb^{\pr}}^{\G/\H},
\end{equation}
for all $\cs,\cs^{\pr},b,b^{\pr}\in\F_{\G/\H}$. This formula can
be obtained by using the group orthogonality relations and noting
that
\begin{equation}
\sum_{g\in\G}\r_g|a\H\ket\bra a\H|\r_g^{\dg}=\r_a\sum_{g\in\G}|g\H\ket\bra g\H|\r_a^\dg=\left|\H\right|\id_{\G/\H}\,.
\end{equation}

Some of the other position shifts $\lu_g$ remain unitary when projected onto $\G/\H$, as we will shortly see in the examples below.

\prg{Example: rigid rotor}

Recall from Appx.~\ref{appx:voronoi} that the rigid rotor is itself a quotient space, $\SO_{3}=\SU_{2}/\Z_{2}$, where $\Z_2=\{I,-I\}$. We can interpret it as a subspace of $\SU_2$, with position states
\begin{equation}
	\kk R=\frac{1}{\sqrt{2}}\left(\kk{+R}_{\SU_2}+\kk{-R}_{\SU_2}\right)~,
\end{equation}
where $\R\in\SO_3$ and $\kk{\pm R}_{\SU_2}$ are position states in $\SU_2$. To determine the $\SU_2$ momentum states $|^\ell_{mn}\ket$ (with $\ell\in\Z/2$ now being integer or half-integer) participating in the momentum basis of $\SO_3$, we can expand $\kk{\pm R}_{\SU_2}$ in terms of momentum states and simplify the sum (i.e., $\Z_2$-twirl). The irreps of $\SU_2$ are also expressible in terms of Wigner $D$-matrices, and such $\Z_{2}$-twirls are simple:
\begin{align}
{\textstyle \sqrt{\frac{d_{\ell}}{|\SU_{2}|/|\Z_{2}|}}}D^{\ell}\left(R\Z_{2}\right) & ={\textstyle \sqrt{\frac{2\ell+1}{8\pi^{2}}}}{\textstyle \half}\left(D^{\ell}\left(R\right)+D^{\ell}\left(-R\right)\right)\nonumber \\
 & =\d_{\ell\in\Z}{\textstyle \sqrt{\frac{2\ell+1}{8\pi^{2}}}}D^{\ell}\left(R\right)\,,\label{eq:O3-rotor}
\end{align}
since $D^{\ell}(-R)=-D^{\ell}(R)$ for the half-integer irreps.
This shows that only the integer $\SU_{2}$ irreps participate in $\SO_{3}$. The momentum states of $\SO_3$ are thus $|^\ell_{mn}\ket$ with $0\leq|m|,|n|\leq\ell$.

The $\ru_R$, $\lu_R$, and $\dd^\ell_{mn}$ of the rigid rotor (see Sec.~\ref{sec:Rigid-rotor-codes}) are inherited directly from their analogues on $\SU_2$. The $\ru_R$ and $\lu_R$ operators together from the group $\SO_3\times\SO_3$, the joint group of lab-frame and molecule-frame transformations for an asymmetric molecule.

\prg{Example: linear rotor}

A canonical example of a coset space is the two-sphere, where $\G=\SO_{3}$, $\H=\U_{1}$. Here we show that this space is equivalent to an appropriately chosen subspace of the rigid rotor $\SO_3$. Picking $\U_{1}$ to be $\zh$-axis $\SO_3$-rotations $R_{00\g}$ (in the
Euler angle $\phi\theta\g$ parameterization), $\vh=(\theta,\phi)\in\S^{2}$,
and the Wigner $D$-matrices$^{\ref{fn:Z-gauge}}$ are already in a $\U_{1}$-admissible basis. The position states are then
\begin{equation}
	\kk{\vh}\equiv\frac{1}{\sqrt{2\pi}}\int_{\U_{1}}\diff\g\kk{R_{\phi\theta\g}}~.\label{eq:S2-pos-states}
\end{equation}
To determine the momentum states $|^\ell_{mn}\ket$ participating in the momentum basis of $\S^2$, we can expand $|R_{\phi\theta\g}\ket$ in terms of momentum states and perform the $\U_{1}$-integral (i.e., twirl) of the corresponding coefficients $D^\ell_{mn}$. Using {[}\citealp{VMH}, Sec.~5.2.7{]} yields
\begin{align}
{\textstyle \sqrt{\frac{d_{\ell}}{|\SO_{3}|/|\U_{1}|}}}D_{mn}^{\ell}\left(\vh\U_{1}\right) & ={\textstyle \sqrt{\frac{2\ell+1}{4\pi}}\frac{1}{2\pi}}\int_{\U_{1}}\diff\g D_{mn}^{\ell}\left(\phi,\theta,\g\right)\nonumber \\
 & =\d_{n0}Y_{m}^{\ell}\left(\theta,\phi\right)\,.
\end{align}
Thus, the $\S^2$ momentum states are $|^\ell_m\ket\equiv|^\ell_{m0}\ket$. Further applying this machinery works out the rest of the third column of Table~\ref{t:GoverH}.

In this framework, the $X$- (\ref{eq:S2-position-shifts}) and $Z$-type (\ref{eq:S2-momentum-shift}) operators of $\S^2$ can be viewed as projections of the $X$- and $Z$-type operators of $\SO_3$ onto $\S^2$. From Table \ref{t:G}.F, we see that all $\ru_R$ for $R\in\SO_3$ are also operators on $\S^2$, acting on the $m$ indices of the $\S^2$ momentum states and yielding the position shifts $\r_R$. Projecting the $\lu_R$ operators, on the other hand, retains only certain matrix elements,
\begin{equation}
	\bra_{m}^{\ell}|\lu_{\a\b\g}|_{m^{\pr}}^{\ell^{\pr}}\ket=\d_{\ell\ell^{\pr}}\d_{mm^{\pr}}D_{00}^{\ell}\left(0\b0\right)~.
\end{equation}
Only two values $\b\in\{0,\pi\}$ yield unitary operators on $\S^2$, with $\b=0$ being the identity and $\b=\pi$ being the $\S^2$ inversion operation $\p$. These form the group $\Z_2^P$, which together with the projected $\ru_R$'s forms the group $\SO_3\times\Z_2^P=\OO_3$ of proper and improper rotations on $\S^2$. 

\prg{Example: $\Z_N$-symmetric rotor}

Another space of interest is $\SO_{3}/\Z_{N}$, the orientation space
of a $\Z_{N}$-symmetric molecule. Using $\zh$-axis rotations for
$\Z_{N}$, this is a subspace of $\SO_{3}$ with position states 
\begin{equation}
\kk{\cs}=\frac{1}{\sqrt{N}}\sum_{h\in\Z_{N}}\kk{R_{\phi,\theta,\xi+\frac{2\pi}{N}h}}\,,
\end{equation}
where $\cs=(\phi,\theta,\xi)\in\F_{\SO_{3}/\Z_{N}}$, $(\theta,\phi)\in\S^{2}$,
and $\xi\in[0,\frac{2\pi}{N})$. To determine the participating momentum states,
we perform the $\Z_{N}$-twirl
\begin{equation}
{\textstyle \sqrt{\frac{d_{\ell}}{\left|\SO_{3}\right|/\left|\Z_{N}\right|}}D_{mn}^{\ell}\left(\cs\Z_{N}\right)=\d_{n0}^{\Z_{N}}\sqrt{\frac{2\ell+1}{8\pi^{2}/N}}D_{mn}^{\ell}\left(\cs\right)}\,,
\end{equation}
where $\d_{nm}^{\Z_{N}}=1$ if $m=n$ modulo $N$. Thus, the set of
momentum states is
\begin{equation}
\left\{ \kk{_{m,Np}^{\ell}}\,,0\leq|Np|,|m|\leq\ell\right\} \,.
\end{equation}

As with the linear rotor, this space inherits all $\ru_{R}$ rotations.
However, projecting the $\lu_{R}$ rotations yields
\begin{equation}
\bra{}_{m,Np}^{\ell}|\lu_{\a\b\g}|{}_{m^{\pr},Np^{\pr}}^{\ell^{\pr}}\ket=\d_{\ell\ell^{\pr}}\d_{mm^{\pr}}D_{Np,Np^{\pr}}^{\ell}\left(\a\b\g\right)\,.
\end{equation}
As opposed to the case of the linear rotor, $\lu_{R}$ is not diagonal
in the momentum basis, and so there are more unitary operators inherited
from such rotations. For example, all triples $(\a0\g)$ and $(\a\pi\g)$
yield unitary operators, forming the group $\OO_{2}$. Together with
all $\ru_{R}$'s, these form the group $\SO_{3}\times\OO_{2}$.

As $N\to\infty$, $\xi\to0$ and this space approaches
$\S^{2}$. In pictures, the saucer-like space {[}see Fig.~\ref{fig:Zd-rotors}(b){]}
compresses to a flat pancake with all of its boundary points identified,
which is equivalent to the two-sphere.

\prg{Poisson summation}
A final interesting note is the presence of a Poisson summation formula
on these spaces. Recall the standard formula for functions $f\in \mathscr{L}^{2}(\R)$,
\begin{equation}
\sum_{h\in\Z}f\left(h\right)=\sum_{\ell\in\Z}\int_{\R}\diff x e^{i\ell x}f(x)\,.\label{eq:poisson-normal}
\end{equation}
Oftentimes, the first term in the sum over $\ell$ is sufficient asymptotically
with some parameter (see Sec.~\ref{subsec:Approximate-codewords}),
so this formula is useful to approximate sums with integrals. A closer
inspection reveals that this is a special case ($\H=\Z$ and $\G=\R$)
of the more general formula for evaluating the sum of a function
$f\in\mathscr{L}^{2}(\G)$ over $\H$ \cite{Justel2018}:
\begin{subequations}
\begin{align}
{\textstyle \frac{1}{|\H|}}\sum_{h\in \H}f\left(h\right) & ={\textstyle \frac{1}{\sqrt{|\H|}}}\bra\H|f\ket\\
 & =\sum_{\ell m\in\widehat{\G/\H}}{\textstyle \sqrt{\frac{d_{\ell}}{|\G|}}}\sum_{g\in\G}\bra_{mm}^{\ell}|g\ket\bra g|f\ket\\
 & =\sum_{\ell m\in\widehat{\G/\H}}{\textstyle \frac{d_{\ell}}{|\G|}}\sum_{g\in\G}Z_{mm}^{\ell\star}\left(g\right)f\left(g\right)\,.
\end{align}
\end{subequations}
This can be generalized to sums over cosets $\cs\H$.

\section{Partial Fourier transform on $\S^2$}
\label{appx:S2-recovery}

Our focus here is on protection against small rotations, which necessitates the use of coherences between antipodal orientations to store the logical information [i.e., satisfaction of Eq.~(\ref{eq:antipodal})]. Given this condition, it is convenient to let the subgroup $\K\supset\H$ be $\H\times\Z_2^{P}$, where $\Z_2^P$ is the group generated by inversion $P$.  Since there are only two cosets $\{\H,P\H\}$ of $\H$ in $\K$, our choice of $\K$ restricts us to only qubit codes. (Other choices for $\K$ are of course possible, but we do not expound on them here.)

\prg{Abelian subgroup codes}

Let us pick $\H=\Z_{N}$ (for odd $N$) and $\K=\Z_{N}\times\Z_2^{P}$,
where $\Z_{N}$ corresponds to the group of $\zh$-axis rotations. The orbit of a point
$(\vartheta,\varphi)$ (with $\vartheta\notin\{0,\pi\}$) under $\K$ consists of the
$2N$ points $(\vartheta,\varphi+{\textstyle \frac{\pi}{N}}h)$ and
$(\pi-\vartheta,\varphi+{\textstyle \frac{\pi}{N}}h+\pi)$ with $h\in\{0,1,\cdots,N-1\}$.
To form our basis, these points are then split up into two sets of
$N$ points, each set corresponding to one of the two cosets labeled by $r\in\{0,1\}$.
The two cosets are then Fourier-transformed to construct the respective
basis states $\{r,\l\}$ (with $\l\in\{0,1,\cdots,N-1\}$) for each
orbit $(\vartheta,\varphi)$, yielding the basis
\begin{subequations}
\begin{align}
\kk{0\Z_{N}(\vartheta,\varphi);\l} & ={\textstyle \frac{1}{\sqrt{N}}}{\displaystyle \sum_{h\in\Z_{N}}}e^{i\frac{2\pi}{N}\l h}\kk{{\textstyle \vartheta,\varphi+\frac{2\pi}{N}h}}\\
\kk{1\Z_{N}(\vartheta,\varphi);\l} & ={\textstyle \frac{1}{\sqrt{N}}}{\displaystyle \sum_{h\in\Z_{N}}}e^{i\frac{2\pi}{N}\l h}\kk{\pi-{\textstyle \vartheta,\varphi+\frac{2\pi}{N}h+\pi}}\,.
\end{align}
\end{subequations}
These states are defined for all $(\vartheta,\varphi)$ belonging
to the Voronoi cell of $|\frac{\pi}{2},0\ket$,
\begin{align}
\F_{\S^{2}/(\Z_{N}\times\Z_2^{P})} & =\left\{ (\vartheta,\varphi)\,\big|\,\vartheta\in[0,\pi],\,\varphi\in\left(-{\textstyle \frac{\pi}{2N},\frac{\pi}{2N}}\right]\right\} \,,\label{eq:lune}
\end{align}
except at the points $\vartheta\notin\{0,\pi\}$. This cell is depicted by the blue
spherical lune in Fig.~\ref{fig:Zd-rotors}(c). The codewords (\ref{eq:ZN-on-S2-codewords})
correspond to $|0\Z_{N}(\frac{\pi}{2},0);0\ket$ and $|1\Z_{N}(\frac{\pi}{2},0);0\ket$,
respectively. The \textit{cone points} $\vartheta\in\{0,\pi\}$ are
special in that they are invariant under any $\Z_{N}$ rotations around
the $\zh$-axis. For such points, $\l=0$, and their orbits under $\Z_N$ are simply
the points themselves, $|0\Z_{N}(0,0);\l\ket=|0,0\ket$ and $|1\Z_{N}(\pi,0);\l\ket=|\pi,0\ket$.

The above basis is orthonormal and complete due to Eq.~(\ref{eq:symmetric-space-formulation}),
and with this basis at hand, we can devise a recovery map for our
code. A simple map consists of isometries mapping the subspace $\{|r\Z_{N}(\vartheta,\varphi);\l\ket\}_{r\in\{0,1\}}$
for each $(\vartheta,\varphi)$ and $\l$ (with $\vartheta\notin\{0,\pi\}$)
into the codespace $\{|\overline{r}\ket\}_{r\in\{0,1\}}$. The remaining cone points $\vartheta=0,\pi$ can be mapped to any state in the codespace, and we choose
to map them to $\frac{1}{\sqrt{2}}(|\overline{0}\ket+|\overline{1}\ket)$.

Because it preserves coherences between antipodal points, this recovery protects from all rotations $R$ that keep each orientation
in its Voronoi cell. To see this, consider $N=3$ and write a general code state
(with $|c_{0}|^{2}+|c_{1}|^{2}=1$) as
\begin{equation}
|\psi\ket=c_{0}|\overline{0}\ket+c_{1}|\overline{1}\ket=\frac{1}{\sqrt{3}}\left(|\psi_{0}\ket+|\psi_{1}\ket+|\psi_{2}\ket\right)\,,
\end{equation}
where for $h\in\{0,1,2\}$, the states
\begin{equation}
|\psi_{h}\ket=c_{0}\kk{{\textstyle \frac{\pi}{2},\frac{2\pi}{N}h}}+c_{1}\kk{{\textstyle \frac{\pi}{2},\frac{2\pi}{N}h+\pi}}
\end{equation}
are superpositions of a pair of antipodal points. These states are
mapped to $\r_{R}|\psi_{h}\ket\bra\psi_{h}|\r_{R}^{\dg}$ upon a rotation
$\r_{R}$, with each constituent orientation $|{\textstyle \frac{\pi}{2},\frac{2\pi}{N}h}\ket$
being mapped to some point $|\vh_{h}\ket$, and its antipode to $\kk{-\vh_{h}}$.
Each $|\vh_{h}\ket$ is supported on $|0\Z_{N}(\vartheta_{h},\varphi_{h});\l\ket$
for all $\l$ and some $\vartheta_{h},\varphi_{h}$, and similarly
$\kk{-\vh_{h}}$ overlaps with $|1\Z_{N}(\vartheta_{h},\varphi_{h});\l\ket$
for all $\l$. Our recovery maps each $\r_{R}|\psi_{h}\ket\bra\psi_{h}|\r_{R}^{\dg}$
back into the codespace, preserving the logical information. Coherences
$|\psi_{h}\ket\bra\psi_{h^{\pr}\neq h}|$ are not preserved, but this
is not detrimental since the logical information is already inside
each $|\psi_{h}\ket$.

\prg{Nonabelian subgroups}

Here we construct the partial Fourier-transformed basis for $\H\subset\K$, where for simplicity we assume $\H$ to be the maximal subgroup of $\K$. By identifying points connected by actions of rotations in $\K$, $\S^{2}$ can be partitioned into orbits $\K\wh=\{|k\wh\ket\}_{k\in\K}$ with $\wh\in \S^2/\K$.

Since $\S^2$ is not a group, the number of points in an orbit $\K\wh$ depends on the starting point $\wh\in\S^2$. Generically, each rotation $R\in\K$ maps $\wh$ to a distinct point $R\wh$, but there exist special points (e.g., the aforementioned cone points), invariant under some (or even all) $R$, for which the size of the orbit is $<|\K|$. We have to consider such complications when designing the partial Fourier-transformed basis on $\S^2$.

We now further partition each $\K\wh$ into one or more parts, corresponding to cosets of $\H$ in $\K$. To do so, we apply the orbit-stabilizer theorem for each orbit. Consider the subset $\H\wh$ for each orbit
$\K\wh$, whose maximal invariant group is either $\H$ or $\K$
(as there are no subgroups
inbetween). If this group is $\H$, then by the theorem there is a one-to-one-correspondence between elements of the orbit of $\H\wh$ under $\K$ and cosets $\cs\in\K/\H$. If the group is $\K$, then $\H\wh=\K\wh$. 

Applying
the $\H$ Fourier transform on each subset $\cs\H\wh=\{|\cs R\wh\ket\}_{R\in\H}$
yields (cf.~\cite{Justel2018})
\begin{equation}
\kk{\cs\H\wh;{}_{\m\n}^{\l(\wh)}}=\frac{\sqrt{|\H\wh|}}{|\H|}\sum_{R\in\H}\z_{\m\n}^{\l(\wh)}\left(R\right)\kk{\cs R\wh}\,,
\label{eq:Zak-S2-over-H}
\end{equation}
indexed by orbits $\wh\in\S^{2}/\K$, cosets $\cs\in\K/\H$, and
irrep elements $_{\m\n}^{\l(\wh)}\in\widehat{\H}$. The irrep elements
depend on $\wh$, since the size $|\H\wh|$ of each orbit, and therefore the number of states in the subspace $\{| R\wh\ket\}_{R\in\H}$, depends on $\wh$. 
Likewise, the coset index $\cs$ is used only when $|\K\wh|\neq|\H\wh|$; otherwise, $\H\wh$ is invariant under $\K$, and no $\cs$ index is needed.

A simple example is $\H=\TT$ and $\K=\TT\times\Z_2^{P}$. A generic
orbit $\cs\TT\wh$ is a tetrahedrally symmetric set of 12 points,
and $\K\wh$ is an octahedrally symmetric set of 24 points. At a special
point $\wh_{\text{cube}}$, $\TT\wh_{\text{cube}}$ form the vertices
of a tetrahedron, and $\K\wh_{\text{cube}}$ is a cube {[}Fig.~\ref{fig:TinS2}{]}.
We pick the corresponding two states $\{|\cs\TT\wh_{\text{cube}};{}_{00}^{\one}\ket\}_{\cs\in\Z_2^{P}}$
to be the codewords (where $\one$ is the trivial irrep of $\TT$).

\prg{Symmetric harmonics}

A simple way to obtain $\K$-symmetric harmonics from the spherical harmonics is to average or twirl them over $\K$, 
\begin{equation}
\y_{m}^{\ell}\left(\K\right) \equiv\frac{1}{|\K|}\sum_{k\in\K}\r_{k}\y_{m}^{\ell}\r_{k}^{\dg}={\displaystyle \sum_{|p|\leq\ell}}D_{pm}^{\ell\star}\left(\K\right)\y_{p}^{\ell}\,.
\end{equation}
Above, we have used Eq.~(\ref{eq:twirl-def}), the ``Weyl relation''
(Table~\ref{t:GoverH}.H), and {[}\citealp{VMH}, Sec.~5.5.2, Eq.~(1){]}
to express the twirl in terms of Wigner $D$-matrices.$^{\ref{fn:Z-gauge}}$
The above is nonzero only for those $\ell$ admitting $\K$-symmetric
harmonics.

\section{Broader context}
\label{appx:notable-examples}

The partially Fourier-transformed basis (\ref{eq:zakforZ3}) for $\Z_3\subset\U_1$ and its
generalization (\ref{eq:zakG}) for subgroups $\H\subset\G$ are prominent in many areas
of science and engineering. We list notable examples
and three interpretations below.

\prg{Notable examples}

A particularly famous example is lattice systems $\{\G,\H\}=\{\R^{\times d},\Z^{\times d}\}$,
where
\begin{equation}
\R^{\times d}\cong(\R/\Z)^{\times d}\times\widehat{\Z^{\times d}}\cong\U_{1}^{\times d}\times\U_{1}^{\times d}\,.
\end{equation}
The corresponding basis (\ref{eq:zakG}) is called the Weil-Brezin
transform {[}\citealp{follandharm}, Eq.~(1.112){]} or, in the solid-state
context, the Zak or $kq$-basis \cite{Zak1967}: the first $\U_{1}^{\times d}$
factor is parameterized by angles replacing the discrete band index
$n$ in the standard Bloch functions, while the second factor is simply
$k$-space. This basis has seen applications in signal processing
\cite{fstorm}, where it is useful for resolving signals from noise.
This basis has also been studied in quantum foundations \cite{Aharonov1969},
and its constituent states can be grouped to form a codespace and
error spaces for GKP codes \cite{Gottesman2001} (a motivation for
this work).

Another interesting example is $\{\SO_{3},\U_{1}\}$. Its corresponding
decomposition
\begin{equation}
\SO_{3}\cong\SO_{3}/\U_{1}\times\widehat{\U_{1}}\cong\S^{2}\times\Z
\end{equation}
has been useful for expressing vector fields on the sphere {[}\citealp{marpec},
Sec.~12.3{]}.

\prg{Symmetry-adapted bases}
Such bases are used to block-diagonalize $\H$-symmetric Hamiltonians
acting on a space $\X$ into blocks corresponding to irreps
$\l\in\widehat{\H}$. For the $\{\U_{1},\Z_{3}\}$ example, a $\Z_{3}$-symmetric
Hamiltonian written in the basis (\ref{eq:zakforZ3}) will not have
any matrix elements connecting different values of $\l$. This diagonalization
procedure is ubiquitous in physics and chemistry {[}\citealp{tinkhambook},
Sec.~3-8{]} (see also \cite{Justel2018}).

\prg{Coherent states}
The states $\left\{ \kk{\cs\Z_{3};\l}\right\} _{\cs\in\U_{1}/\Z_{3}}$
for fixed $\l$ can then be obtained by applying position shifts on
the fiducial state $\kk{0\Z_{3};\l}$, making them similar to Perelomov
coherent states {[}\citealp{perelomov_book}, Sec.~2.1{]} (with the
caveat that the representation of the group of shifts is reducible).
A key difference between such states for $\{\R,\Z\}$ and the conventional
oscillator coherent states is the choice of fiducial state: a GKP
state for the former and the vacuum Fock state for the latter.

\prg{Fiber bundles}
If we instead take a look at $\left\{ \kk{\cs\Z_{3};\l}\right\} _{\l\in\Z_{3}}$,
we have a 3D space for each $\cs\in\U_{1}/\Z_{3}$. For this and any
$\{\G,\H\}$ case where $\G$ is a Lie group, the states form a fiber
bundle with base space $\G/\H$, fiber $\widehat{\H}$, and cross-section
$\{\cs\}$ \cite{perelomov_book}.

\bibliography{C:/Users/russi/Documents/library}

\end{document}